%% file: paper.tex
\documentclass[a4paper,11pt]{article}
\pdfoutput=1

\usepackage{jheppub}
\usepackage{graphicx}
\usepackage{amsmath}
\usepackage{amssymb}
\usepackage{bbm}
\usepackage{tensor}
\usepackage{subfigure}
\usepackage{slashed}
\usepackage[english]{babel}
\newcommand{\op}{{\cal O}}

\newcommand{\C}{{\cal C}}
\newcommand{\Q}{{\cal Q}}
\newcommand{\todo}[1]{{\color{red} \ifmmode\else[todo]\fi #1}}
\usepackage[usenames,dvipsnames]{xcolor}
     \definecolor{hgreen}{rgb}{0,.3,0}
     \definecolor{hred}{rgb}{.3,0,0}
     \definecolor{hblue}{rgb}{0,0,.3}
     \definecolor{LightGray}{gray}{0.95}
\newcommand{\beq}{\begin{equation} }
\newcommand{\eeq}{\end{equation}} 
\newcommand{\bi}{\begin{itemize} }
\newcommand{\ei}{\end{itemize} }

\definecolor{Red}{rgb}{1.,0.,0.}
\definecolor{Grn}{rgb}{0.,0.75,0.}
\definecolor{Blu}{rgb}{0.,0.,1.}
\definecolor{Bro}{rgb}{0.6,0.3,0.1}

\DeclareMathOperator{\diag}{diag}

\newcommand{\lrpartial}{\negthickspace\stackrel{\leftrightarrow}{\partial}\negthickspace{}}
\newcommand{\lrD}{\negthickspace\stackrel{\leftrightarrow}{D}\negthickspace{}}
\newcommand{\ddm}{\texttt{DirectDM}}
\DeclareRobustCommand{\SkipTocEntry}[5]{}
\newcommand{\ncdot}{\negthinspace \cdot \negthinspace}
\newcommand{\lDslashed}{\thinspace\slash\negthickspace\negthickspace\negthickspace\negthinspace\stackrel{\leftarrow}{D}\negthickspace}

\graphicspath{{./figs/}}

\begin{document}

\title{Renormalization Group Effects in Dark Matter Interactions}

\affiliation[a]{Deutsches Elektronen-Synchrotron (DESY), D-22607 Hamburg, Germany}
\affiliation[b]{Department of Physics, University of Cincinnati, Cincinnati, Ohio 45221, USA}
\affiliation[c]{Department of Physics, University of California-San Diego, La Jolla, CA 92093, USA}
%\affiliation[d]{Fakult\"at f\"ur Physik, TU Dortmund, D-44221 Dortmund, Germany} 

\author[a]{Fady Bishara,}
\author[b]{Joachim Brod,}
\author[c]{Benjamin Grinstein,}
\author[b]{and Jure Zupan}

\emailAdd{fady.bishara AT desy.de}
\emailAdd{brodjm AT ucmail.uc.edu}
\emailAdd{bgrinstein AT ucsd.edu}
\emailAdd{zupanje AT ucmail.uc.edu}

\abstract{
We present a renormalization-group (RG) analysis of dark matter
interactions with the standard model, where dark matter is allowed to
be a component of an electroweak multiplet, and has a mass at or below
the electroweak scale. We consider, in addition to the gauge
interactions, the complete set of effective operators for dark matter
interactions with the standard model above the weak scale, up to and
including mass dimension six. We calculate the RG evolution of these
operators from the high scale $\Lambda$ down to the weak scale, and
perform the matching to the tower of effective theories below the weak
scale. We also summarize the RG evolution below the weak scale and the
matching to the nonrelativistic nuclear interactions. We present
several numerical examples and show that in certain cases the dark
matter -- nucleus scattering rate can change by orders of magnitude
when the electroweak running is included.
}

\preprint{DO-TH 18/01}
\arxivnumber{1809.03506}

\maketitle

%%%%%%
%Body of the text
%%%%%%

\input{introduction}

\input{RGrunning-EWmatching}
\input{EffectOfRG}

\input{summary}

%%%%%%%%
% Appendices
%%%%%%%%
\appendix
\input{conventions}

\input{appcNR}

\input{appWarsaw}

\input{appDM}
\input{appUnphys}

%%%%%%%%%

\bibliographystyle{JHEP}
\bibliography{paper}

\end{document}

%% file: introduction.tex
%!TEX root = paper.tex

%%%%%%%%%%%%%
%%%% Introduction  %%
%%%%%%%%%%%%%
\section{Introduction}
For a large class of dark matter (DM) models, the physics of direct
detection experiments can be described using Effective Field Theory
(EFT)~\cite{Bagnasco:1993st, Pospelov:2000bq, Kurylov:2003ra,
  Kopp:2009qt, Fan:2010gt, Cirigliano:2012pq,
  Hill:2011be,Hill:2013hoa, Fitzpatrick:2012ix, Fitzpatrick:2012ib,
  Menendez:2012tm, Anand:2013yka, Klos:2013rwa, DelNobile:2013sia,
  Barello:2014uda, Hill:2014yxa, Catena:2014uqa, Hoferichter:2015ipa,
  Hoferichter:2016nvd, Bishara:2016hek, Bishara:2017pfq,
  DEramo:2016gos, Bishara:2017nnn, DEramo:2017zqw, Brod:2018ust,
  Brod:2017bsw, Chen:2018uqz}.  There are several scales that enter
the problem: the DM mass, $m_\chi$, the scale of the mediators,
$\Lambda$, through which the DM interacts with the visible sector,
and, finally, the standard model (SM) scales -- the masses of the SM
particles and the scale of strong interactions, $\Lambda_{\rm
  QCD}$. The EFT description of DM direct detection is appropriate as
long as the mediators are heavier than a few hundred MeV, i.e., above
the typical momentum exchange in direct detection
experiments. Furthermore, the EFT description is necessary in order to
consistently treat the hadronic physics in the scattering of DM on
nuclei.

The EFT approach is especially appealing if one does not want to
commit to a particular DM model when interpreting the results of
direct detection experiments.  The direct detection bounds can be
expressed as the bounds on the coefficients of local operators, which
can then be compared between different direct detection experiments in
a model-independent manner~\cite{Hill:2014yxa, Bishara:2016hek,
  Bishara:2017nnn, Bishara:2017pfq, Chen:2018uqz}. If the mediator
scale is above the DM mass, $\Lambda\gtrsim m_\chi$, they can also be
compared to indirect detection bounds~\cite{Kumar:2013iva, Cao:2009uw,
  Goodman:2010qn, Ciafaloni:2011sa, Cheung:2011nt,
  Cheung:2012gi,Vietze:2014vsa}, and to colliders searches if
$\Lambda$ is above the typical partonic momentum exchange in the
collision \cite{Haisch:2015ioa, Cotta:2012nj, Busoni:2013lha,
  Fox:2011fx, Rajaraman:2011wf, Fox:2011pm, Racco:2015dxa,
  Jacques:2015zha,Bauer:2016pug}. At the LHC the typical partonic
momentum often does exceed the mediator scale, $\Lambda$, in which
case one needs to resort to simplified models~\cite{Bauer:2016gys,
  Abdallah:2015ter, Bruggisser:2016nzw, DeSimone:2016fbz,
  Kahlhoefer:2015bea, Boveia:2016mrp, Goncalves:2016iyg, Bell:2016ekl,
  Duerr:2016tmh, Englert:2016joy, Brennan:2016xjh, Alanne:2017oqj,
  Bertone:2017adx, Bernreuther:2018nat}.
  
\begin{figure}[t]\centering
\includegraphics[scale=1]{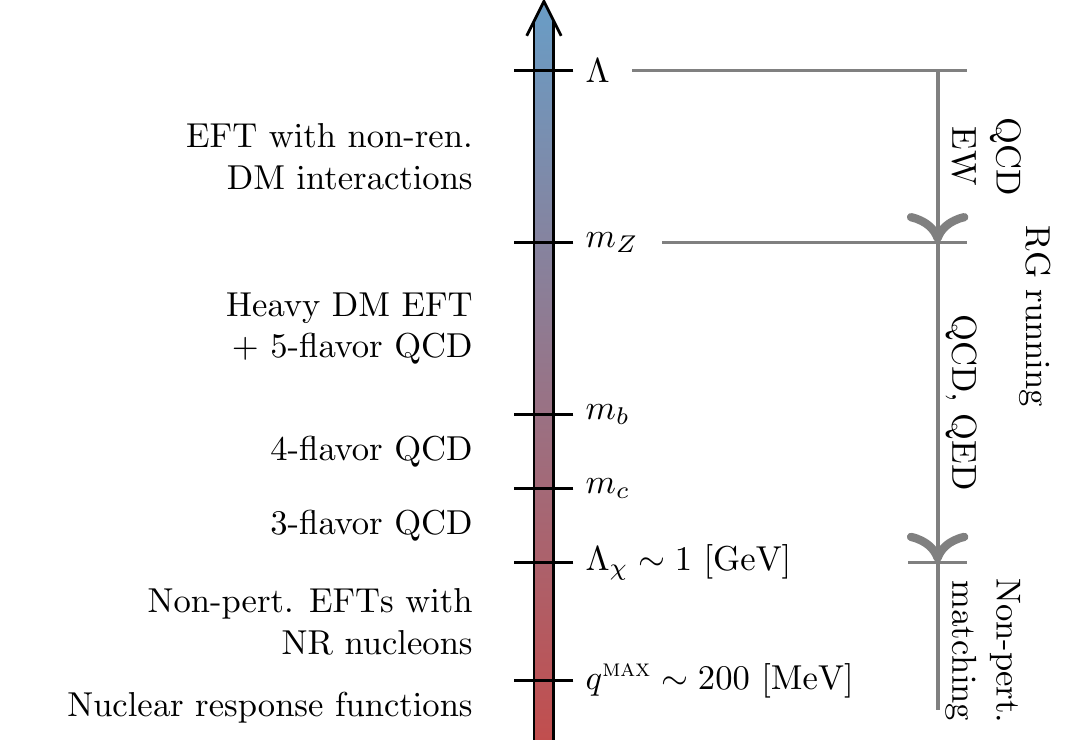}
\caption{The tower of EFTs linking the UV scale $\Lambda$ to the scale
  of interactions between the nucleons and the DM.}
\label{fig:eft-tower}
\end{figure}

In the present manuscript we are interested in the connection between
the DM theory at the mediator scale, $\Lambda$, and the EFT describing
DM direct detection.  To do so one needs to run through a tower of
EFTs that connects the UV scale $\Lambda$ with the nuclear scale. We
assume that\footnote{For $\Lambda\sim m_\chi$ one needs to match onto
  an EFT with non-relativistic DM, the Heavy Dark Matter EFT (HDMET),
  already at the scale $\Lambda$.}
\begin{equation}
\Lambda\gg m_\chi \sim m_Z\,,
\end{equation}
where $m_Z=91.1876\,$GeV is the $Z$-boson
mass. Fig.~\ref{fig:eft-tower} depicts the resulting tower of EFTs.
At a particular scale $\mu$ the appropriate EFT is constructed from
the relevant propagating degrees of freedom.

At $\mu\sim \Lambda$ the propagating degrees of freedom are either the
full theory of DM interactions, presumably renormalizable, or a
simplified model of DM interactions, including the mediators. For
$\mu< \Lambda$ the mediators are integrated out, leading to an EFT
with nonrenormalizable interactions between DM and the visible sector.
At $\mu\sim m_Z$ the top quark, the Higgs, and the $W,Z$ bosons are
integrated out. For $\mu<m_Z$ the DM interactions are therefore
described by nonrenormalizable operators in an EFT that contains only
DM (which, for $m_\chi \sim m_Z$, is now described by a
nonrelativistic field), and the bottom-, charm-, strange-, down-, and
up-quark, as well as the leptons, gluons and photons. At $\mu\sim m_b$
one integrates out the bottom quark, and at $\mu \sim m_c$ the charm
quark. Finally, at $\mu\sim {\mathcal O}(1{\rm~GeV})$ a
nonperturbative matching to an EFT with pions and nucleons, i.e., a
chiral effective theory, is performed~\cite{Hoferichter:2015ipa,
  Bishara:2016hek, Bishara:2017pfq, Hoferichter:2016nvd}. This is then
used in a chiral EFT approach to nuclear forces together with the
nuclear response functions to obtain the hadronic matrix element for
each of the DM-nucleon interaction operators~\cite{Klos:2013rwa,
  Vietze:2014vsa, Menendez:2012tm, Anand:2013yka, Fitzpatrick:2012ix,
  Fitzpatrick:2012ib, Bishara:2017nnn, Gazda:2016mrp, Korber:2017ery}.

The EFT operators mix under the renormalization-group (RG) evolution
when going from $\Lambda$ to $m_Z$, from $m_Z$ to $m_b$, etc. The
primary purpose of this paper is to calculate the anomalous dimensions
for the RG running from $\Lambda$ to $m_Z$ for the case of Dirac
fermion DM in an arbitrary electroweak multiplet. This RG running can
be phenomenologically important since it can mix operators that are
velocity suppressed in the nonrelativistic limit with operators that
are not velocity suppressed (see Refs.~\cite{Crivellin:2014qxa,
  D'Eramo:2014aba, DEramo:2016gos, Crivellin:2015wva, Haisch:2013ata,
  Haisch:2013uaa, Haisch:2012kf, Frandsen:2012db, Freytsis:2010ne} for
further examples of relevant loop corrections in DM interactions). In
addition, we also perform the rest of the running and matching down to
the nuclear level and give several numerical examples.

\begin{figure}\centering
\includegraphics[scale=0.9]{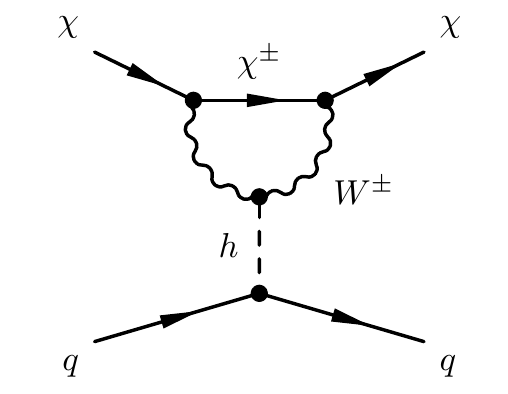}
\includegraphics[scale=0.9]{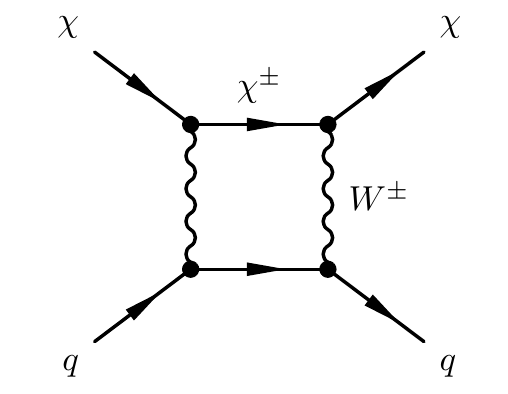}
\includegraphics[scale=0.9]{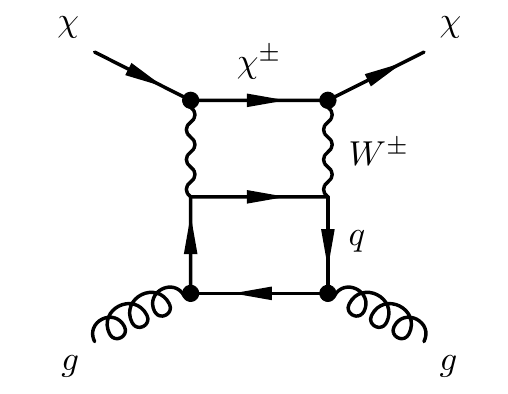}
\caption{Representative one-loop (left and middle) and two-loop
  (right) diagrams contributing to the direct detection scattering of
  DM that is part of an electroweak multiplet.}
         \label{fig:DD-Higgs_loop}
\end{figure}

\begin{figure}\centering
\includegraphics[width=0.45\textwidth]{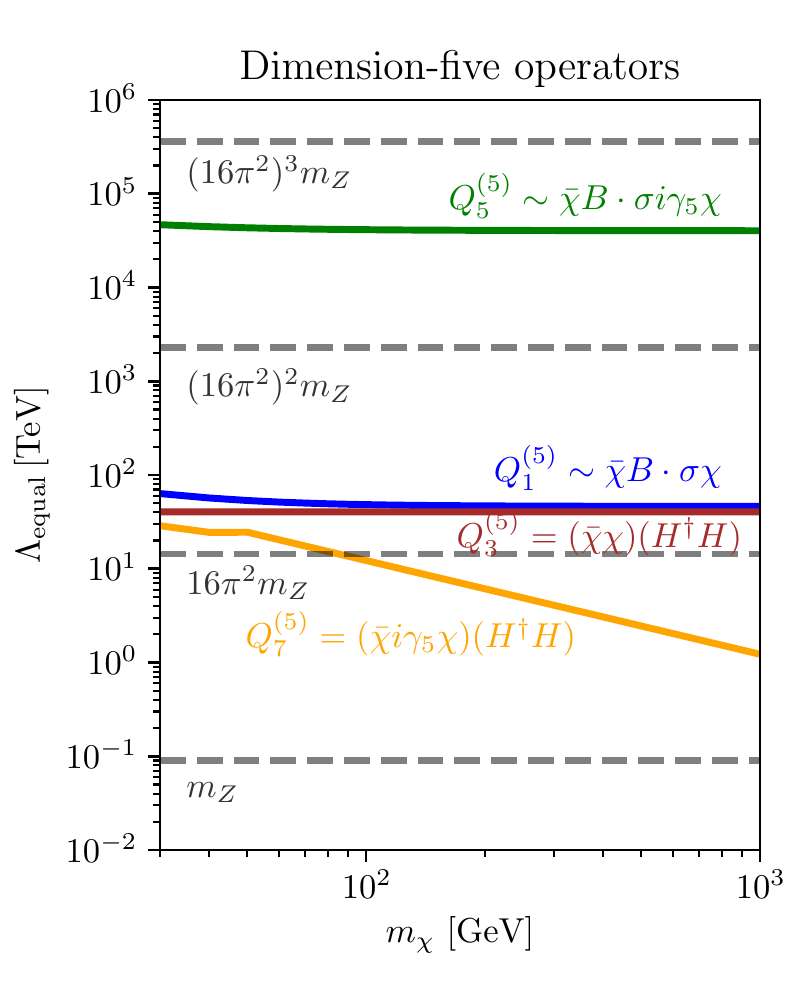}
~~\includegraphics[width=0.45\textwidth]{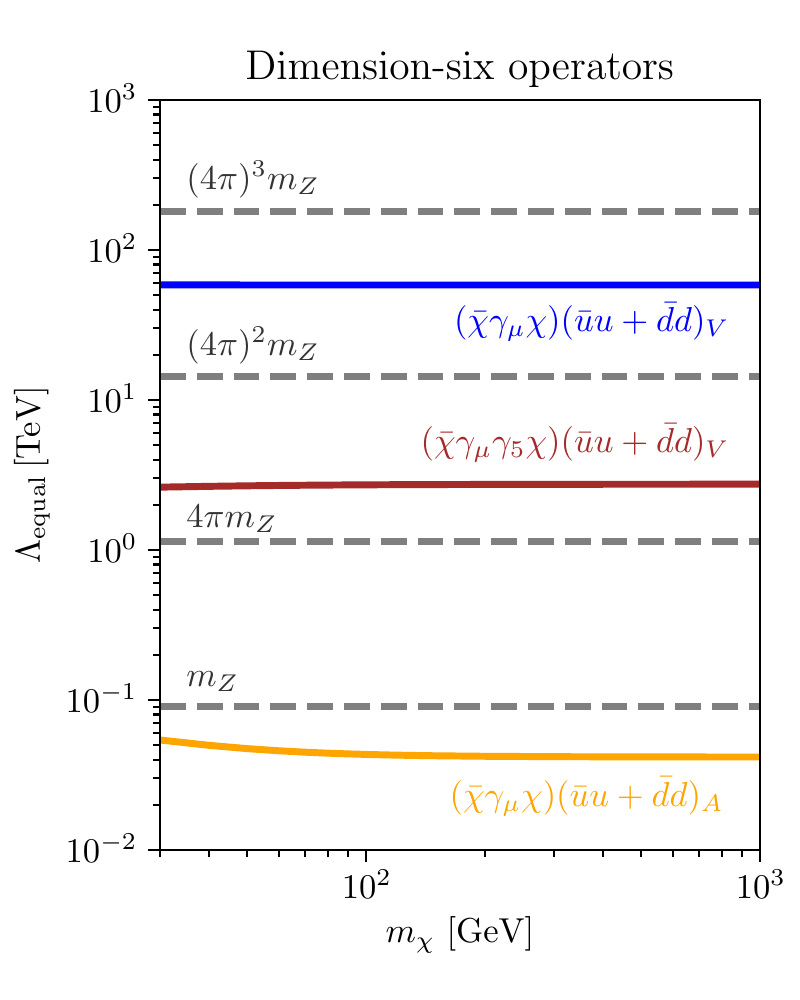}
\caption{The contribution from the non-renormalizable operator
  $Q_i^{(d)}$ dominates in direct detection scattering over the
  renormalizable contributions, if the suppression scale $\Lambda$ is
  below the corresponding solid line (i.e., for $\Lambda=\Lambda_{\rm
    equal}$ the non-renormalizable and the renormalizable
  contributions are of the same size). Examples shown are for triplet
  Dirac fermion DM with $Y_\chi=0$ scattering on a Xenon target, so
  that the contributions from renormalizable interactions start at
  one-loop. The dashed gray lines denote the electroweak scale, $m_Z$,
  and the scales roughly $n$ loop factors above it, $(4\pi)^{2n} m_Z$
  on the left and $(4\pi)^{n} m_Z$ on the right. Dimension-five
  operators are shown in the left panel, dimension-six operators in
  the right panel.}
          \label{fig:scales-DDexample}
\end{figure}

The possibility that DM is part of an electroweak multiplet is allowed
by direct detection constraints as long as DM does not couple to the
$Z$ boson at tree level (for instance, this is the case if the DM
multiplet has odd dimensionality and does not carry hypercharge). The
exchanges of $W,Z,h$ bosons with a quark current then generate a
contribution to DM--nucleon scattering at one-loop and two-loop level,
see Fig.~\ref{fig:DD-Higgs_loop} and Refs.~\cite{Hisano:2011cs,
  Hisano:2010ct}. Since these contributions are loop-suppressed and
result in either a chirality flip or spin-dependent scattering, it is
quite possible that the leading contribution is due to exchanges of
heavy mediators.  This is illustrated in
Fig.~\ref{fig:scales-DDexample} where we show for several
non-renormalizable interactions at which values of the mediator mass,
$\Lambda_{\rm equal}$, the non-renormalizable and renormalizable
contributions to scattering on Xenon are equal. For mediators lighter
than $\Lambda_{\rm equal}$ the scattering rates are dominated by the
non-renormalizable interactions. Even if the mediators are very heavy,
many orders of magnitude heavier than the weak scale, they can still
give the leading effect in spin-independent scattering.  Furthermore,
the operators that lead to velocity-suppressed contributions, such as
vector-axial interactions, are only poorly constrained. A mixing into
velocity unsuppressed, coherently enhanced operators at one-loop,
two-loop, or potentially even three-loop can therefore still be the
leading contribution to the scattering rate. This motivates both the
use of the complete tower of EFTs and the calculation of the
leading-logarithmic effects captured by RG running.

In our analysis we cover both the case of DM with electroweak-scale
mass, $m_\chi \sim m_Z$, and light DM, $m_\chi \ll m_Z$. Note that we
do not require DM to be a thermal relic, and therefore allow for a
large range of DM masses and interactions.  Above the electroweak
scale we limit our analysis to a basis of operators with mass
dimension five and six, and work to one-loop order for the anomalous
dimensions. The matching corrections are calculated at tree level,
except for the cases where one-loop contributions can be numerically
important, for instance, if the matching generates gluonic
operators. The subsequent RG evolution below the electroweak scale has
been described in detail in Refs.~\cite{Hill:2014yxa,
  Bishara:2017pfq}; see also Ref.~\cite{Bishara:2017nnn} for a
computer code that implements the running numerically.  Several
interesting cases are left for future work, such as the case of
several DM multiplets, the case of scalar DM, the case of very heavy
DM, $m_\chi \gg m_Z$, as well as the analysis of higher dimension
operators.

The paper is organized as follows. In Section~\ref{sec:UV} we give the
complete basis of dimension-five and dimension-six operators for DM
interacting with the SM, valid above the electroweak scale. The
anomalous dimensions describing the mixing of these operators are
presented in Section~\ref{sec:RGrunning}, while in
Section~\ref{sec:match-ew} we give the matchings to the tower of EFTs
below the electroweak symmetry breaking scale, and collect the results
on the running down to the hadronic scale, along with the subsequent
nonperturbative matching to the chiral EFT and the nuclear
responses. Section~\ref{sec:RG-run} contains illustrative examples
showcasing the effects of operator mixing on DM direct detection
phenomenology. The conclusions are given in
Section~\ref{sec:summary}. Appendix~\ref{app:not-conv} contains our
notation and conventions, Appendix~\ref{app:SMEFT} the mixing with the
pure SM operators, Appendix~\ref{app:DM} the mixing with the pure dark
sector, and Appendix~\ref{sec:unphys} a list of unphysical operators
used in the calculation.

\section{Effective Lagrangian above the electroweak scale}\label{sec:UV}
We extend the SM by a single Dirac fermion, a $Z_2$-odd electroweak
multiplet of dimension $d_\chi=2 I_\chi+1$, hypercharge $Y_\chi$, and
mass $m_\chi$, whose electrically neutral component is the DM. Here
$I_\chi$ is the weak isospin of the DM multiplet (see
App.~\ref{app:not-conv} for our conventions). One-loop electroweak
corrections split the multiplet components, so that the charged
particles are heavier than DM and decay in the early universe (see
Refs.~\cite{Cirelli:2005uq, Hill:2011be}).  In the numerical examples
in Section~\ref{sec:RG-run} we set $Y_\chi=0$, so that the
phenomenologically dangerous tree-level vectorial $Z$ couplings are
absent.  In the calculation of anomalous dimension in this section we
do, however, keep the $Y_\chi$ dependence, 
so that the results can be more generally applied. \footnote{This added generality is needed in the case of light DM. Collider experiments require that the charged components of the multiplet that contains the DM have electroweak scale masses. The required large  splitting of the spectrum arises from mixing with additional multiplets that have to be properly accounted for in the running of Wilson coefficients below the NP scale,  $\Lambda$. In our sample applications below, the light DM case is unrealistic -- and given for illustration purposes only -- because the effect of the required additional multiplets is ignored. Analysis of realistic examples is left for future work.}
%as this may prove useful in
%other applications of our results. \jz{We also keep the discussion 
%completely general when running below the electroweak scale. This includes
%the case of light DM, even though phenomenologically viable models
%inevitably require
%more than one multiplet so that charged components are heavy enough, 
%a possibility we postpone to future work. }

Within our set-up there are two types of DM interactions with the
visible sector: either through the exchanges of SM particles, or
through new states -- the mediators. In general both of these
contributions are present. Our default assumption is that DM has
electroweak scale mass, while the mediators are much heavier, with
masses of order $\Lambda \gg m_Z$. We thus have the following
hierarchy of scales,
\begin{equation} 
  \Lambda \gg m_\chi \sim m_Z \gg \Lambda_\text{QCD}\gtrsim q \,,
\label{eq:scales} 
\end{equation} 
where $q\sim {\mathcal O}(100~{\rm MeV})$ is the typical momentum
transfer in DM scattering on nuclei. We will also discuss the case of
light DM, $m_\chi \ll m_Z$.

When one considers processes at energy scales below the mass of the
mediators, $\mu<\Lambda$, the mediators can be integrated out. The
effective DM Lagrangian, valid for $\Lambda>\mu>m_Z$, is given by
\begin{equation}
\label{eq:Lchi:Lambda}
{\cal L}_\chi={\cal L}_\chi^{(4)}+{\cal L}_\chi^{(5)}+{\cal L}_\chi^{(6)}+\cdots, 
\end{equation}
where the superscripts denote the dimensionality of the operators in
the Lagrangian. The renormalizable part of the effective Lagrangian
is, for a Dirac-fermion DM multiplet,
\begin{equation}\begin{split}
\label{eq:gauge-int}
\mathcal{L}_\chi^{(4)} = \overline{\chi} i \gamma^\mu D_\mu\chi -
m_\chi \bar \chi \chi\,.
\end{split}\end{equation}
The covariant derivative comprises the interactions with the
electroweak gauge bosons $W^a_\mu$ and $B_\mu$; see
App.~\ref{app:not-conv} for further details on our notation.
For $\mu \gg m_Z \sim m_\chi$ the mass parameter $m_\chi$ can
effectively be set to zero.

The non-renormalizable terms in the effective
Lagrangian~\eqref{eq:Lchi:Lambda},
\begin{equation} 
{\cal L}_{\chi}^{(5)} = \sum_a \frac{C_a^{(5)}}{\Lambda} Q_a^{(5)}, \qquad {\cal L}_\chi^{(6)} =\sum_a
\frac{C_a^{(6)}}{\Lambda^2} Q_a^{(6)} \,, \quad \ldots
\label{sumC}
\end{equation}
arise from integrating out the mediators. Depending on the mediator
model it is possible that only ${\cal L}_\chi^{(5)}$ or only ${\cal
  L}_\chi^{(6)}$ are generated, but in general both will be
present. We truncate the expansion at dimension six since most
mediator models generate nonzero Wilson coefficients, $C_a^{(d)}$, in
at least one of the two effective Lagrangians, ${\cal L}_\chi^{(5)},
{\cal L}_{\chi}^{(6)}$ (for exceptions where the first contributions
arise only at dimension seven, see, e.g.,
Ref.~\cite{Bai:2015swa,Weiner:2012gm}; the complete basis at dimension
seven has been presented in Ref.~\cite{Brod:2017bsw}).  When writing
the basis we assume that there is a conserved global dark $U(1)_D$
quantum number, which forbids currents of the form $\bar \chi^c \Gamma
\chi$, where $\chi^c$ is the charge-conjugated DM field, and $\Gamma$
denotes a generic string of Dirac matrices. (This assumption is to be
relaxed in a follow-up work, where we plan to extend our analysis to
the case of Majorana fermions and more than one multiplet.)

\subsection{Dimension-five operator basis}\label{sec:UVdim5}
The CP-conserving dimension-five operators are
\begin{align}
Q_{1}^{(5)} &= \frac{g_1}{8\pi^2} (\bar\chi\sigma^{\mu\nu}\chi)B_{\mu\nu}\,,
&Q_{2}^{(5)} &= \frac{g_2}{8\pi^2} (\bar\chi\sigma^{\mu\nu}\tilde\tau^a\chi)W_{\mu\nu}^a\,, \label{Q12}\\
Q_{3}^{(5)} &= (\bar\chi\chi)(H^\dagger H)\,,
&Q_{4}^{(5)} &= (\bar\chi\tilde\tau^a\chi)(H^\dagger \tau^a H)\,,\label{Q34}
\end{align}
while the CP-odd operators have an extra insertion of $\gamma_5$, 
\begin{align}
Q_{5}^{(5)} &= \frac{g_1}{8\pi^2} (\bar\chi\sigma^{\mu\nu}i\gamma_5\chi)B_{\mu\nu}\,,
&Q_{6}^{(5)} &= \frac{g_2}{8\pi^2} (\bar\chi\sigma^{\mu\nu}\tilde\tau^ai\gamma_5 \chi)W_{\mu\nu}^a\,,\label{Q56}\\
Q_{7}^{(5)} &= (\bar\chi i\gamma_5 \chi)(H^\dagger H)\,,
&Q_{8}^{(5)} &= (\bar\chi\tilde\tau^a i\gamma_5 \chi)(H^\dagger \tau^a H)\,.\label{Q78}
\end{align}
Here and below, $H$ is the SM Higgs doublet, and the $SU(2)$
generators $\tilde\tau^a$, $\tau^a$ are defined in
App.~\ref{app:not-conv}. All non-displayed $SU(2)$ (and, below, also
color) indices are assumed to be contracted within the brackets. Note
that if $\chi$ is a $SU(2)$ singlet, the operators $Q_{2}^{(5)}$,
$Q_{4}^{(5)}$, $Q_{6}^{(5)}$, and $Q_{8}^{(5)}$ are absent.  In a
perturbative UV theory the operators $Q_{1,2}^{(5)}$ and
$Q_{5,6}^{(5)}$ are generated at one loop, while the operators
$Q_{3,4}^{(5)}$ and $Q_{7,8}^{(5)}$ are typically generated at tree
level. This expectation is reflected in our normalization of the
operators.

\subsection{Dimension-six operator basis}\label{sec:UVdim6}
At dimension six there are many more operators. We do not consider
flavor-violating operators, keeping our discussion minimal.  For each
SM fermion generation, $i=1,2,3$, there are then eight operators that
are products of DM currents and quark currents,
\begin{align}
Q_{1,i}^{(6)} &= (\bar\chi\gamma_\mu \tilde\tau^a\chi)(\bar Q_L^i \gamma^\mu \tau^a Q_L^i)\,, & Q_{5,i}^{(6)} &= (\bar\chi\gamma_\mu \gamma_5 \tilde\tau^a\chi)(\bar Q_L^i \gamma^\mu \tau^a Q_L^i)\,, \label{eq:dim6:Q15}\\
Q_{2,i}^{(6)} &= (\bar\chi\gamma_\mu \chi)(\bar Q_L^i \gamma^\mu Q_L^i)\,, & Q_{6,i}^{(6)} &= (\bar\chi\gamma_\mu \gamma_5 \chi)(\bar Q_L^i \gamma^\mu Q_L^i)\,,\\
Q_{3,i}^{(6)} &= (\bar\chi\gamma_\mu \chi)(\bar u_R^i \gamma^\mu u_R^i)\,, & Q_{7,i}^{(6)} &= (\bar\chi\gamma_\mu \gamma_5 \chi)(\bar u_R^i \gamma^\mu u_R^i)\,,\\
Q_{4,i}^{(6)} &= (\bar\chi\gamma_\mu \chi)(\bar d_R^i \gamma^\mu d_R^i)\,, & Q_{8,i}^{(6)} &= (\bar\chi\gamma_\mu \gamma_5 \chi)(\bar d_R^i \gamma^\mu d_R^i)\,. \label{eq:dim6:Q48}
\end{align}
Here $Q_L$ denotes the left-handed quark doublet, and $u_R$, $d_R$ the
right-handed up- and down-type quark singlets, respectively. The
analogous operators involving lepton currents are
\begin{align}
Q_{9,i}^{(6)} &= (\bar\chi\gamma_\mu \tilde\tau^a\chi)(\bar L_L^i \gamma^\mu \tau^a L_L^i)\,, & Q_{12,i}^{(6)} &= (\bar\chi\gamma_\mu \gamma_5 \tilde\tau^a\chi)(\bar L_L^i \gamma^\mu \tau^a L_L^i)\,,\label{eq:dim6:Q9Q12}\\
Q_{10,i}^{(6)} &= (\bar\chi\gamma_\mu \chi)(\bar L_L^i \gamma^\mu L_L^i)\,, & Q_{13,i}^{(6)} &= (\bar\chi\gamma_\mu \gamma_5 \chi)(\bar L_L^i \gamma^\mu L_L^i)\,,\\
Q_{11,i}^{(6)} &= (\bar\chi\gamma_\mu \chi)(\bar \ell_R^i \gamma^\mu \ell_R^i)\,, & Q_{14,i}^{(6)} &= (\bar\chi\gamma_\mu \gamma_5 \chi)(\bar \ell_R^i \gamma^\mu \ell_R^i)\,,\label{eq:dim6:Q11Q14}
\end{align}
where $L_L$ denotes the left-handed lepton doublet, and $\ell_R$ the
right-handed down-type lepton singlet. Finally, there are four
dimension-six operators involving Higgs currents,
\begin{align}
Q_{15}^{(6)} &= (\bar\chi\gamma^\mu \tilde\tau^a\chi)(H^\dagger
i\stackrel{\leftrightarrow}{D^a}_\mu H)\,, & Q_{17}^{(6)} &=
(\bar\chi\gamma^\mu \gamma_5 \tilde\tau^a\chi)(H^\dagger
i\stackrel{\leftrightarrow}{D^a}_\mu H)\,,
\label{eq:dim6:Q15Q17}
\\ 
Q_{16}^{(6)} &= (\bar\chi\gamma^\mu \chi)(H^\dagger
i\stackrel{\leftrightarrow}{D}_\mu H)\,, & Q_{18}^{(6)} &=
(\bar\chi\gamma^\mu \gamma_5 \chi)(H^\dagger
i\stackrel{\leftrightarrow}{D}_\mu H)\,. 
\label{eq:dim6:Q16Q18} 
\end{align}
The Higgs currents are defined in terms of hermitian combinations of
the covariant derivatives, $\stackrel{\leftrightarrow}{D}_\mu
\,\equiv\, D_\mu - \stackrel{\leftarrow}{D}_\mu^\dagger$ and
$\stackrel{\leftrightarrow}{D^a}_\mu \,\equiv\, \tau^a D_\mu -
\stackrel{\leftarrow}{D}_\mu^\dagger \tau^a$. Additional operators
with covariant derivatives acting on the DM fields vanish via the DM
equations of motion, up to total derivatives.  As in the case of
dimension-five operators, the basis simplifies if DM is a $SU(2)$
singlet. In this case, the operators $Q_{1,i}^{(6)}$, $Q_{5,i}^{(6)}$,
$Q_{9,i}^{(6)}$, $Q_{12,i}^{(6)}$, $Q_{15}^{(6)}$, and $Q_{17}^{(6)}$
vanish and should be dropped from the basis.

While the operators \eqref{Q12}-\eqref{Q78} and
\eqref{eq:dim6:Q15}-\eqref{eq:dim6:Q16Q18} mix under RG running, they
do not yet form a closed set under renormalization; for this we also
need to include the pure SM operators (see App.~\ref{app:SMEFT}) and
the operators with only DM fields.
To the extent that we neglect the mixing of the pure DM operators
among themselves, we need only four operators for our purposes, which
we can choose as
\begin{align}\label{eq:op:DM-DM}
D_{1}^{(6)} & = (\bar\chi\gamma_\mu \chi)(\bar \chi \gamma^\mu \chi)\,, 
&D_{2}^{(6)} & = (\bar\chi\gamma_\mu \gamma_5 \chi)(\bar \chi \gamma^\mu \chi)\,,\\
D_{3}^{(6)} & = (\bar\chi\gamma_\mu \tilde\tau^a \chi)(\bar \chi \gamma^\mu \tilde\tau^a \chi)\,, 
&D_{4}^{(6)} & = (\bar\chi\gamma_\mu \gamma_5 \tilde\tau^a \chi)(\bar \chi \gamma^\mu \tilde\tau^a \chi)\,.
\end{align}

%% file: RGrunning-EWmatching.tex
%!TEX root = paper.tex

%%%%%%%%%%%%%%%%%%%%%
\section{Renormalization group running}
\label{sec:RGrunning}
%%%%%%%%%%%%%%%%%%%%%
The RG running proceeds through several sequential steps, $\Lambda \to
\mu_{\rm EW} \to \mu_b \to \mu_c$, with matching thresholds at the
electroweak scale, $\mu_{\rm EW}$, the bottom-quark mass scale,
$\mu_b$, and the charm-quark mass scale, $\mu_c$.  We first review
briefly each of the steps, and then give the details in this and the
subsequent section.

Running from the mediator scale, $\Lambda$, to the EW scale, $\mu_{\rm
  EW} \sim m_Z \sim m_\chi$, results in the mixing of the operators in
the effective DM Lagrangian, Eq.~\eqref{eq:Lchi:Lambda}. We perform
the calculation of the RG running using dimensional regularization in
$d=4-2\epsilon$ dimensions. Following the conventions in
Ref.~\cite{Buchalla:1995vs}, we define the anomalous dimension matrix
$\gamma$ by
\begin{equation}\label{eq:ADMdef}
\mu\frac{d}{d\mu} \vec C(\mu) = \gamma^{T} \vec C(\mu) \,,
\end{equation}
where $\vec C$ is a vector of Wilson coefficients\footnote{For
  dimension-five operators some Wilson coefficients need to be
  redefined to have simple forms of anomalous dimensions, see
  Eq.~\eqref{eq:vecC:define}.}, and the superscript $T$ denotes matrix
transposition. The anomalous dimension matrix receives a number of
different contributions that we treat separately, so that
\begin{equation}\label{eq:ADMnotation}
\gamma = \frac{\alpha_s}{4\pi} \gamma_s^{(0)} + 
\left(\frac{\alpha_s}{4\pi}\right)^2 \gamma_s^{(1)} + 
\frac{\alpha_1}{4\pi}
\gamma_1^{(0)} + \frac{\alpha_2}{4\pi} \gamma_2^{(0)}
+\sum_{f=t,b,c,\tau} \frac{\alpha_f}{4\pi} \gamma_{y_f}^{(0)} +
\frac{\alpha_\lambda}{4\pi} \gamma_\lambda^{(0)} +\cdots \, .
\end{equation}
Here, we defined $\alpha_{f} \equiv y_{f}^2/4\pi$ and $\alpha_\lambda
\equiv \lambda/4\pi$, where $y_f$ is the Yukawa coupling of the
fermion $f$ and $\lambda$ the Higgs quartic coupling (for
normalizations see Appendix \ref{app:not-conv}), while the other
parameters are defined in terms of the gauge couplings in the usual
way, $\alpha_i \equiv g_i^2/4\pi$.  The ellipsis denotes higher-order
contributions. Note that the anomalous dimension above the EW scale
does not depend on the QCD coupling constant, since DM does not carry
color, while all the DM--quark operators in
\eqref{eq:dim6:Q15}-\eqref{eq:dim6:Q48} contain conserved quark
currents in the limit of zero quark masses. The situation is different
below the EW scale.

The solution to the RG evolution equation \eqref{eq:ADMdef} gives the
Wilson coefficients at any scale $\mu_{\rm EW}<\mu<\Lambda$,
\begin{equation}\label{eq:RGevol}
\vec C(\mu)=U(\mu, \Lambda)\vec C(\Lambda),
\end{equation}
where $U(\mu, \Lambda)$ is the evolution operator from $\Lambda$ to
$\mu$, obtained by solving \eqref{eq:ADMdef}, or equivalently
\begin{equation}
\frac{d}{d\ln\mu}U(\mu, \Lambda)=\gamma^T U(\mu, \Lambda),\label{eq:Ueq}
\end{equation}
with the initial condition $U(\Lambda, \Lambda)=1$. The leading-order
RG evolution effectively sums the terms of the form $\alpha_i^n
\log^n(\Lambda/\mu_{\rm EW})$ to all orders. Since some of the
anomalous dimensions are large, we count $\alpha_i
\log(\Lambda/\mu_{\rm EW})\sim {\mathcal O}(1)$. We work to
leading-logarithmic order and thus include all terms that are
${\mathcal O}(1)$. This means that the matching conditions are
calculated to the same order, i.e., are obtained at tree
level. Matching is done at one-loop, if the tree level contribution
vanishes and the one-loop contribution is numerically important, for
details see below.

The first matching arises at the EW scale, $\mu_{\rm EW}\sim m_Z$,
where one integrates out the top quark, Higgs, $W$ and $Z$. For
$\mu<\mu_{\rm EW}$ the propagating degrees of freedom are then the
photon, the gluons, $n_f=5$ quark flavors, and the leptons. The RG
running in the five-flavor theory is given by the anomalous dimension
matrix $\gamma_{[5]}$. It receives QCD and electromagnetic
contributions, so that at one loop order,
\begin{equation}
\gamma_{[n_f]}= \frac{\alpha_s}{4\pi} \gamma_{[n_f],s}^{(0)} + \frac{\alpha}{4\pi}
\gamma_{[n_f],e}^{(0)} +\cdots \, , 
\end{equation}
where $\alpha_s$ and $\alpha$ are the strong and electromagnetic
coupling constants. At $\mu_b\sim m_b$ the bottom quark is integrated
out. The resulting four flavor EFT has as the propagating degrees of
freedom the photon, gluons, leptons, and $n_f=4$ flavors of quarks. It
is valid for $\mu_c<\mu<\mu_b$, where $\mu_c\sim m_c$ is the scale at
which the charm quark and the $\tau$ lepton are integrated out. The
running from $\Lambda$ down to the scale $\mu_{\rm had} \sim 2\,$GeV,
where the hadronic matrix elements are evaluated, can thus formally be
written as
\begin{equation}
\begin{split}
\vec C(\mu_{\rm had})|_{n_f=3} & = U_{[3]}(\mu_\textrm{had}, \mu_{c}) M_{[4\to 3]}(\mu_c) U_{[4]}(\mu_{c}, \mu_b) M_{[5\to 4]}(\mu_b)
\\ & \qquad \times U_{[5]}(\mu_b, \mu_{\rm EW}) M_{[\text{EW}\to 5]}(\mu_{\rm EW})
U(\mu_{\rm EW}, \Lambda) \vec C(\Lambda)\,. \label{eq:evol:sol}
\end{split}
\end{equation}
Here $U_{[n_f]}(\mu, \mu')$ are the evolution operators from $\mu'$ to
$\mu$ in a theory with $n_f$ quark flavors that satisfy an evolution
equation similar to~\eqref{eq:Ueq},
\begin{equation}
\frac{d}{d\ln\mu}U_{[n_f]}(\mu, \mu')=\gamma_{[n_f]}^T U_{[n_f]}(\mu,
\mu')\,. \label{eq:Ueq:nf}
\end{equation}
In the numerics we take $\mu_{\rm had}\sim \mu_c\sim 2$ GeV, and thus
set $U_{[3]}(\mu_\textrm{had}, \mu_{c})=1$.  The $M_{[n_f\to n_f-1]}$
in Eq.~\eqref{eq:evol:sol} are the matching matrices when going from a
theory with $n_f$ quark to a theory with $n_f-1$ quarks, while
$M_{[\text{EW}\to 5]}(\mu_{\rm EW})$ symbolises the matching to the
five-flavor theory at the EW scale. On the left side of
Eq.~\eqref{eq:evol:sol} we have denoted explicitly that the final
Wilson coefficients are in the theory with only three flavors of
quarks, i.e., with just $u$, $d$, and $s$ quarks, along with gluons,
photons and the light leptons.

In the remaining part of this section we present the explicit form of
the anomalous dimension matrix, Eq.~\eqref{eq:ADMnotation}, that
describes the mixing of the operators due to the RG evolution from the
mediator scale $\Lambda$ to $\mu_{\rm EW}$. The subsequent matching
and RG evolutions below the weak scale is given ins
Sec.~\ref{sec:match-ew}. We work in the limit of flavor conservation,
setting the Cabibbo-Kobayashi-Maskawa (CKM) matrix to
unity. Furthermore, in this section we keep only the top, bottom,
charm, and tau Yukawa couplings nonzero.

For the computation of the anomalous dimensions we used two
independent automated setups. In the first, the amplitudes were
generated using \texttt{qgraf}~\cite{Nogueira:1991ex} and the
anomalous dimensions were computed using the computer algebra system
\texttt{form}~\cite{Vermaseren:2000nd}. The second setup relied on
\texttt{Mathematica} packages: the Feynman rules were generated using
\texttt{FeynRules}~\cite{Alloul:2013bka}, the amplitudes with
\texttt{FeynArts}~\cite{Hahn:2000kx}, and the anomalous dimensions
were computed using \texttt{FormCalc}~\cite{Hahn:1998yk}. A large part
of the calculations were also checked using pen and paper.

%%%%%%%%%%%%
%%% DIM5 %%%
%%%%%%%%%%%%

\subsection{Mixing of dimension five operators}
\label{subsec:mix:dim5}
We start by providing the anomalous dimension matrices for mixing of
the CP conserving dimension-five operators
$Q_1^{(5)},\dots,Q_8^{(5)}$, defined in
Eqs.~\eqref{Q12}-\eqref{Q78}. For the column of the dimension-five
Wilson coefficients entering the RG evolution equation
\eqref{eq:RGevol} we use the rescaled Wilson coefficients
\begin{equation}\label{eq:vecC:define}
\vec C^\prime=\Big( \frac{\alpha_1}{2\pi} C_1^{(5)}, \frac{\alpha_2}{2\pi}
C_2^{(5)}, C_3^{(5)}, C_4^{(5)}, \frac{\alpha_1}{2\pi} C_5^{(5)},
\frac{\alpha_2}{2\pi} C_6^{(5)}, C_7^{(5)}, C_8^{(5)}\big)\,. 
\end{equation}
The explicit factors of $\alpha_{1,2}/2\pi=g_{1,2}^2/8\pi^2$ in $\vec
C^\prime$ ensure that the anomalous dimension matrices
$\gamma_i^{(0)}$, still defined by~\eqref{eq:ADMnotation}, do not
depend on coupling constants. The evolution of the primed Wilson
coefficients is given by the analogue of Eq.~\eqref{eq:ADMdef},
namely,
\begin{equation}
\mu\frac{d}{d\mu} \vec C^\prime(\mu) = \gamma^{T} \vec C^\prime(\mu)
\,,
\end{equation}
The corresponding rescaled operators are also denoted by a prime and
read
\begin{align}
Q_{1}^{\prime(5)} &= \frac{1}{g_1} (\bar\chi\sigma^{\mu\nu}\chi)B_{\mu\nu}\,,
&Q_{2}^{\prime(5)} &= \frac{1}{g_2} (\bar\chi\sigma^{\mu\nu}\tilde\tau^a\chi)W_{\mu\nu}^a\,,\label{eq:uv-dipoles}\\
Q_{5}^{\prime(5)} &= \frac{1}{g_1} (\bar\chi\sigma^{\mu\nu}i\gamma_5\chi)B_{\mu\nu}\,,
&Q_{6}^{\prime(5)} &= \frac{1}{g_2}\label{eq:uv-dipoles-g5}\ (\bar\chi\sigma^{\mu\nu}i\gamma_5\tilde\tau^a\chi)W_{\mu\nu}^a\,,
\end{align}
while $Q_{i}^{\prime(5)} \equiv Q_{i}^{(5)}$ for $i=3,4,7,8$.

The anomalous dimension for $Q_1^{\prime(5)},\dots,Q_8^{\prime(5)}$
splits into two blocks, for the CP even operators,
$Q_1^{\prime(5)},\dots,Q_4^{\prime(5)}$, and the CP odd operators,
$Q_5^{\prime(5)},\dots,Q_8^{\prime(5)}$, while there is no mixing
between the two blocks. The QCD anomalous dimensions vanish, since all
fields are color neutral. The remaining one-loop anomalous dimensions
for the $Q_1^{\prime(5)},\dots,Q_4^{\prime(5)}$ block are
\begin{align}
\big[\gamma_1^{(0)}\big]_{Q_{1\dots4}^{\prime(5)} \times Q_{1\cdots 4}^{\prime(5)}} &= 
\begin{pmatrix}
\frac{5}{2}Y_\chi^2-2\beta_1^{(0)}&0&-6Y_\chi&0\\
-4Y_\chi{\cal J}_\chi&\frac{1}{2}Y_\chi^2&0&12Y_\chi\\
0&0&-\frac{3}{2}-\frac{3}{2}Y_\chi^2&0\\
0&0&0&-\frac{3}{2}-\frac{3}{2}Y_\chi^2
\end{pmatrix}\, ,\label{eq:gamma1:dim5}\\[5mm]
%%%
\big[\gamma_2^{(0)}\big]_{Q_{1\cdots 4}^{\prime(5)} \times Q_{1\cdots 4}^{\prime(5)}} &=
\begin{pmatrix}
2{\cal J}_\chi&-4Y_\chi&0&-24\\
0&10{\cal J}_\chi-8-2\beta_2^{(0)}&12{\cal J}_\chi&0\\
0&0&-6{\cal J}_\chi - \frac{9}{2}&0\\
0&0&0&-6{\cal J}_\chi + \frac{3}{2}
\end{pmatrix}\, . \label{eq:gamma2:dim5}
\end{align}
Here ${\cal J}_\chi = I_\chi(I_\chi+1)$, with $d_\chi=2I_\chi+1$ the
dimensionality of the DM electroweak multiplet, and $Y_\chi$ its
hypercharge. The $\beta$ functions for the gauge couplings $g_1$ and
$g_2$ are given by
\begin{equation}
\beta_1^{(0)} = - \frac{41}{6} - \frac{Y_\chi^2}{3} d_\chi N_\chi \, ,
\qquad 
\beta_2^{(0)} = \frac{19}{6} - \frac{4}{9} {\cal J}_\chi d_\chi N_\chi \, , \\ 
\end{equation}
respectively, where $N_\chi$ is the number of DM multiplets in the
representation $I_\chi$ (we will mostly take $N_\chi=1$). The
anomalous dimension matrices for the CP-odd operators
$Q_5^{\prime(5)},\dots,Q_8^{\prime(5)}$ are also given by the same
matrices, $[\gamma_i^{(0)}]_{Q_{5\cdots 8}^{\prime(5)} \times
  Q_{5\cdots 8}^{\prime(5)}} = [\gamma_i^{(0)}]_{Q_{1\cdots
    4}^{\prime(5)} \times Q_{1\cdots 4}^{\prime(5)}}$,
$i=1,2,y,\lambda$, as required by the fact that CP breaking is not
probed by the relevant one-loop diagrams.

\begin{figure}\centering
\includegraphics[scale=0.8]{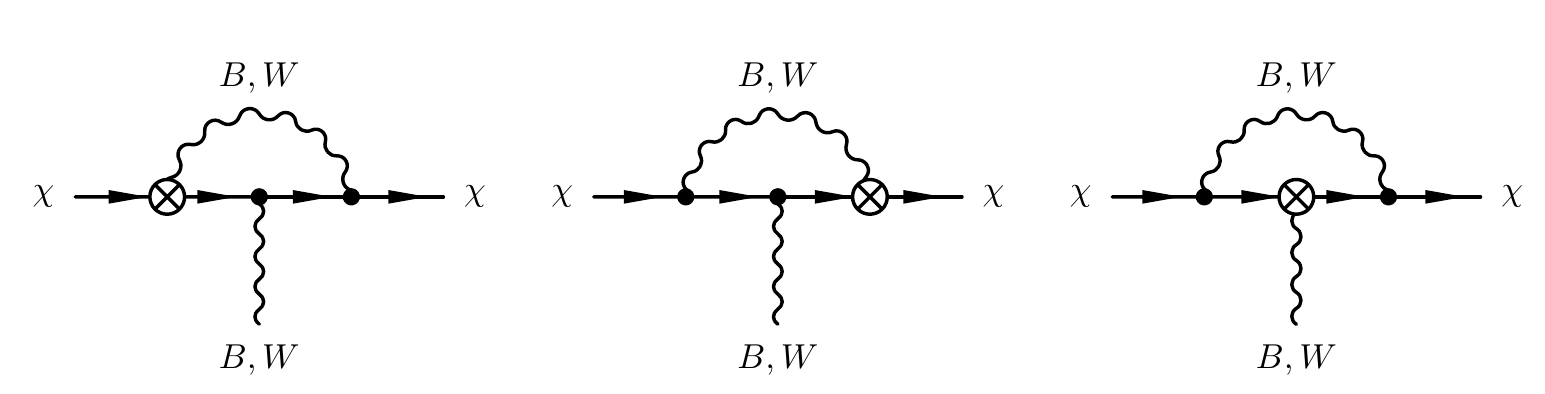}
\caption{Sample diagrams for renormalization of
  $Q_1^{\prime(5)},Q_2^{\prime(5)}$ operators due to $B_\mu, W_\mu^a$
  exchanges at one loop.}\label{fig:D51_diag}
\end{figure}

\begin{figure}\centering
\includegraphics[width=0.24\linewidth]{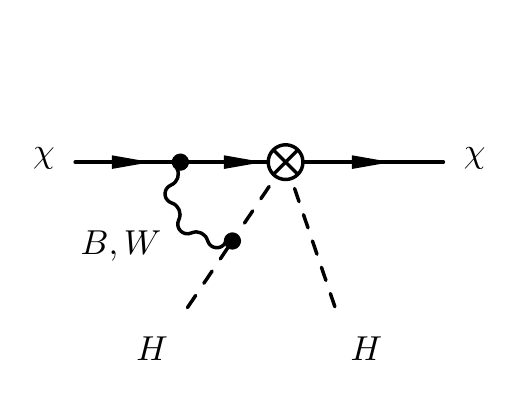}
\includegraphics[width=0.24\linewidth]{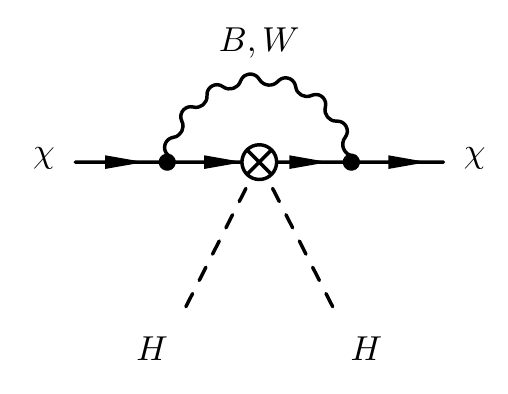}
\includegraphics[width=0.24\linewidth]{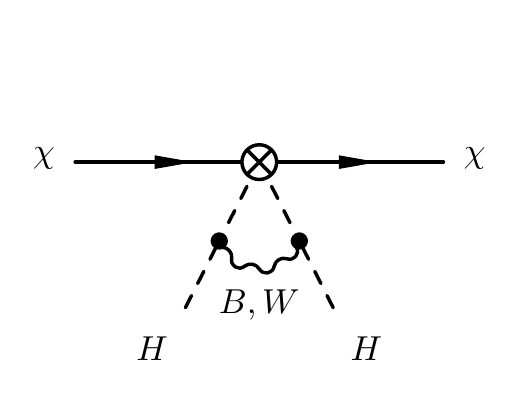}
\includegraphics[width=0.24\linewidth]{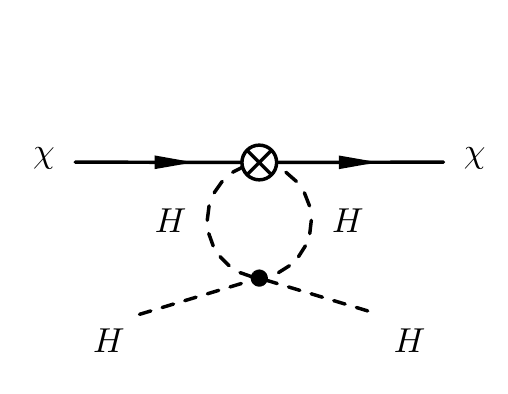}
\caption{The renormalization of the $Q_3^{\prime(5)},Q_4^{\prime(5)}$
  operators at one loop, with only one representative of each class of
  diagrams shown.}
         \label{fig:D53}
\end{figure}

\begin{figure}[t]\centering
	\includegraphics[scale=0.75]{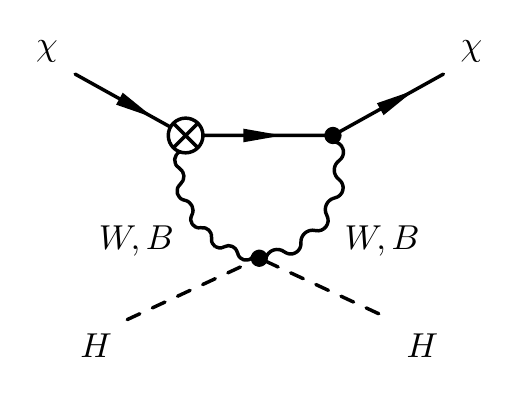}
	\includegraphics[scale=0.75]{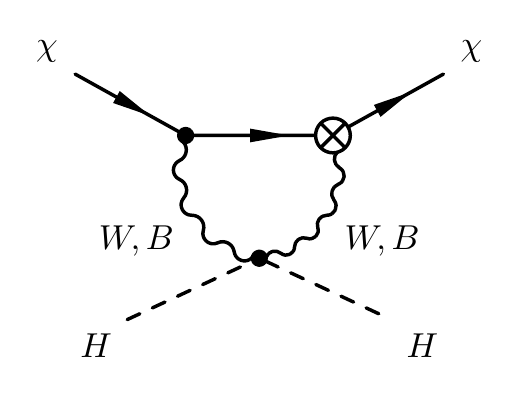}
	\caption{Sample diagrams for mixing of the dipole operators
		$Q_1^{\prime(5)}$ and $Q_2^{\prime(5)}$ into the Higgs
		operators $Q_3^{\prime(5)}$ and $Q_4^{\prime(5)}$.}
	\label{fig:dipole-into-higgs}
\end{figure}

The anomalous dimensions $\gamma_1^{(0)}$ in
Eq.~\eqref{eq:gamma1:dim5} and $\gamma_2^{(0)}$ in
Eq.~\eqref{eq:gamma2:dim5} come from the exchanges of the $B_\mu$ and
$W_\mu^a$ gauge bosons in Figs.~\ref{fig:D51_diag},~\ref{fig:D53},
and~\ref{fig:dipole-into-higgs}, respectively. They are almost
diagonal, with only six nonzero off-diagonal entries.  The
$Q_{1}^{\prime(5)}$ mixing into $Q_{2}^{\prime(5)}$ in
$\gamma_1^{(0)}$ is due to a loop exchange of $B_\mu$ with an emission
of $W_\mu^a$, shown in the left two diagrams in
Fig.~\ref{fig:D51_diag}, while the $Q_{2}^{\prime(5)}$ mixing into
$Q_{1}^{\prime(5)}$ is due to a similar diagram with $B_\mu$ and
$W_\mu^a$ exchanged. The $Q_{1}^{\prime(5)}$ mixing into
$Q_{2}^{\prime(5)}$ in $\gamma_2^{(0)}$ is due to a loop exchange of
$B_\mu$ in the last diagram in Fig.~\ref{fig:D51_diag}. The mixings of
dipole operators, $Q_{1,2}^{\prime(5)}$, into the Higgs current
operators, $Q_{3,4}^{\prime(5)}$, arise from the diagrams in
Fig.~\ref{fig:dipole-into-higgs}. These mixing contributions vanish
for singlet DM ($Y_\chi = {\cal J}_\chi = 0$). This is true also for
the mixing of $Q_{1}^{\prime(5)}$ into $Q_{4}^{\prime(5)}$ in
$\gamma_2^{(0)}$ (recall that the operators $Q_{2}^{\prime(5)}$ and
$Q_{4}^{\prime(5)}$ are absent for singlet DM). The contributions
proportional to Yukawa couplings and the Higgs self coupling to the
anomalous dimension lead only to multiplicative renormalization of
$Q_{3}^{\prime(5)}$ and $Q_{4}^{\prime(5)}$:
\begin{align}
\big[\gamma_{y_t}^{(0)}\big]_{Q_{1\cdots 4}^{\prime(5)} \times Q_{1\cdots
    4}^{\prime(5)}} &= \big[\gamma_{y_b}^{(0)}\big]_{Q_{1\cdots 4}^{\prime(5)}
  \times Q_{1\cdots 4}^{\prime(5)}} =
\big[\gamma_{y_c}^{(0)}\big]_{Q_{1\cdots 4}^{\prime(5)} \times Q_{1\cdots
    4}^{\prime(5)}} = \diag\big(0, 0, 6, 6\big)\,,\\ 
\big[\gamma_{y_\tau}^{(0)}\big]_{Q_{1\cdots 4}^{\prime(5)} \times Q_{1\cdots
    4}^{\prime(5)}} &= \diag\big(0, 0, 2, 2\big)\,,\\ 
\big[\gamma_{\lambda}^{(0)}\big]_{Q_{1\cdots 4}^{\prime(5)} \times Q_{1\cdots
    4}^{\prime(5)}} &= \diag\big(0, 0, 3, 1\big)\,.
\end{align}
They arise from the Higgs wave function renormalization, and in the
case of $\gamma_\lambda^{(0)}$, from the last diagram in
Fig.~\ref{fig:D53}.

After running from $\mu \sim \Lambda$ to $\mu \sim m_Z$, we revert the
rescaling of the Wilson coefficients, i.e.,
\begin{equation}\label{eq:vecC:define:orig}
\vec C=\bigg( \frac{2\pi}{\alpha_1} C_1^{\prime(5)},
\frac{2\pi}{\alpha_2} C_2^{\prime(5)}, C_3^{\prime(5)},
C_4^{\prime(5)}, \frac{2\pi}{\alpha_1} C_5^{\prime(5)},
\frac{2\pi}{\alpha_2} C_6^{\prime(5)}, C_7^{\prime(5)},
C_8^{\prime(5)}\bigg)\,,
\end{equation}
corresponding to our original definition of operators in
Eqs.~\eqref{Q12}-\eqref{Q78}. We use the unprimed Wilson coefficients
for determining the matching conditions in Sec.~\ref{sec:match-ew}.

%%%%%%%%%%%%
%%% DIM6 %%%
%%%%%%%%%%%%

\subsection{Mixing of dimension six operators}
\label{subsec:mix:dim6}
We turn next to the anomalous dimensions for the dimension-six
operators. Counting the three SM fermion generations and keeping only
flavor-diagonal fermion currents, there are 46 operators in total that
couple DM with the SM.  We work in the limit of flavor conservation
which simplifies the structure of the anomalous dimensions.

We split the $46\times 46$ matrix of anomalous dimensions into several
sub-blocks. They correspond to three groups of operators: the
operators with quark currents, $Q_{1,i}^{(6)}, \dots, Q_{8,i}^{(6)}$;
the operators with lepton current, $Q_{9,i}^{(6)}, \dots,
Q_{14,i}^{(6)}$; and the operators with Higgs currents, $Q_{15}^{(6)},
\dots, Q_{18}^{(6)}$ (see
Eqs.~\eqref{eq:dim6:Q15}--\eqref{eq:dim6:Q16Q18} for
definitions). Moreover, we will distinguish between mixing within one
fermion generation, and mixing between different generations.

A technical remark is in order. To project the one-loop matrix
elements onto our operator basis within the context of dimensional
regularization, we have to manipulate Dirac $\gamma$ matrices in
$d\neq 4$ dimensions. Strictly speaking, this requires the extension
of the operator basis by evanescent operators. However, the one-loop
anomalous dimensions are not affected by the choice of the evanescent
operator basis, and we can effectively use four-dimensional Dirac
algebra~\cite{Dugan:1990df}.

We start with the mixing among the operators that are products of DM
and quark currents, $Q_{1,i}^{(6)},\dots, Q_{8,i}^{(6)}$,
Eqs.~\eqref{eq:dim6:Q15}--\eqref{eq:dim6:Q48}, within {\em the same}
quark generation. The corresponding $8\times8$ block of the anomalous
dimension matrix is given by
\begin{equation}
\big[\gamma_1^{(0)} \big]_{Q_{1,i\cdots 8,i}^{(6)} \times Q_{1,i\cdots 8,i}^{(6)}}=
\begin{pmatrix}
 0 & 0 & 0 & 0 & -Y_\chi & 0 & 0 & 0 \\
 0 & \frac{2}{3} d_\chi Y_\chi^2+\frac{2}{9} & \frac{8}{9} & -\frac{4}{9} & 0 & -Y_\chi & 0 & 0 \\
 0 & \frac{4}{9} & \frac{2}{3} d_\chi Y_\chi^2+\frac{16}{9} & -\frac{8}{9} & 0 & 0 & 4 Y_\chi & 0 \\
 0 & -\frac{2}{9} & -\frac{8}{9} & \frac{2}{3} d_\chi Y_\chi^2+\frac{4}{9} & 0 & 0 & 0 & -2 Y_\chi \\
 -Y_\chi & 0 & 0 & 0 & 0 & 0 & 0 & 0 \\
 0 & -Y_\chi & 0 & 0 & 0 & \frac{2}{9} & \frac{8}{9} & -\frac{4}{9} \\
 0 & 0 & 4 Y_\chi & 0 & 0 & \frac{4}{9} & \frac{16}{9} & -\frac{8}{9} \\
 0 & 0 & 0 & -2 Y_\chi & 0 & -\frac{2}{9} & -\frac{8}{9} & \frac{4}{9}
\end{pmatrix}\, ,
\label{eq:gamma1:DMquark}
\end{equation}
for the part of the anomalous dimension matrix proportional to
$g_1^2$, while the part of the anomalous dimension proportional to
$g_2^2$ is
\begin{equation}
\big[\gamma_2^{(0)}\big]_{Q_{1,i\cdots 8,i}^{(6)} \times Q_{1,i\cdots 8,i}^{(6)}}= 
\begin{pmatrix}
 \frac{8}{9} {\mathcal J}_\chi d_\chi-4 & 0 & 0 & 0 & 0 & -3 {\mathcal J}_\chi & 0 & 0 \\
 0 & 0 & 0 & 0 & -12 & 0 & 0 & 0 \\
 0 & 0 & 0 & 0 & 0 & 0 & 0 & 0 \\
 0 & 0 & 0 & 0 & 0 & 0 & 0 & 0 \\
 0 & -3 {\mathcal J}_\chi & 0 & 0 & -4 & 0 & 0 & 0 \\
 -12 & 0 & 0 & 0 & 0 & 0 & 0 & 0 \\
 0 & 0 & 0 & 0 & 0 & 0 & 0 & 0 \\
 0 & 0 & 0 & 0 & 0 & 0 & 0 & 0
\end{pmatrix}\, .\label{eq:gamma2:DMquark}
\end{equation}
Both of the anomalous dimension matrices are diagonal in flavor
indices.  As far as the $U(1)$ gauge interaction is concerned, for
$Y_\chi=0$ the operators $Q^{(6)}_i$, $i=2,3,4,6,7,8$, are partially
conserved currents, and one would naively expect their anomalous
dimensions to vanish.  That this is not the case can be understood as
the result of a non-multiplicative renormalization, allowed for $U(1)$
gauge groups; see Ref.~\cite{Collins:2005nj}. Similar arguments apply
for the QED anomalous dimensions discussed in
Sec.~\ref{sec:RG:below:ew}.

\begin{figure}\centering
\includegraphics[scale=0.8]{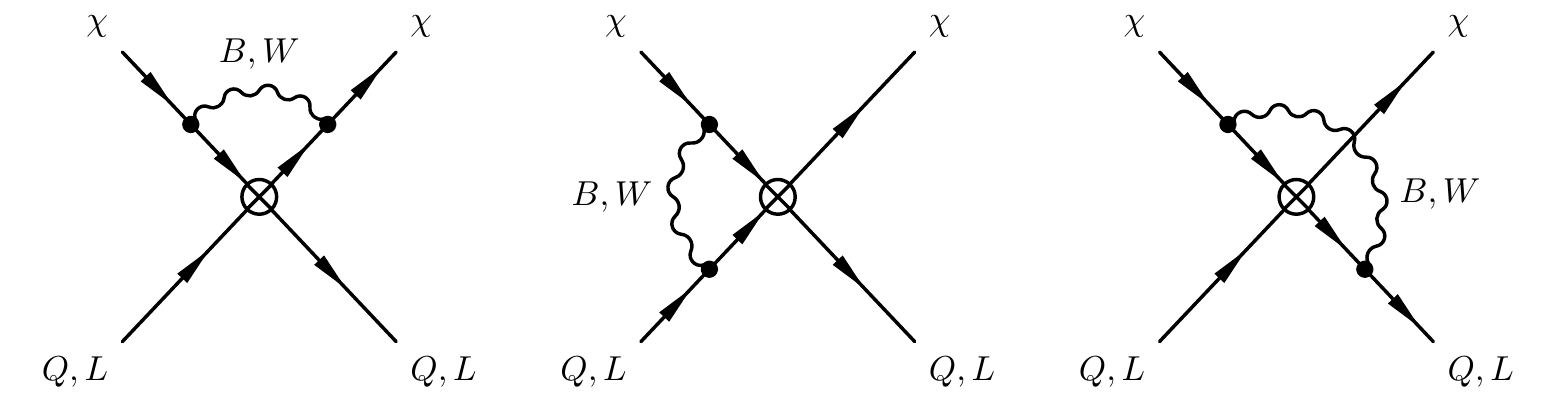}
\caption{Sample diagrams for the renormalization of
  $Q_{1,i}^{(6)},\ldots,Q_{14,i}^{(6)}$ operators due to the exchange
  of $B_\mu,W_\mu^a$ at one loop (right-handed quarks and leptons can
  also be on the external lines).}
\label{fig:D6_4ferm}
\end{figure}

\begin{figure}\centering
\includegraphics[scale=0.8]{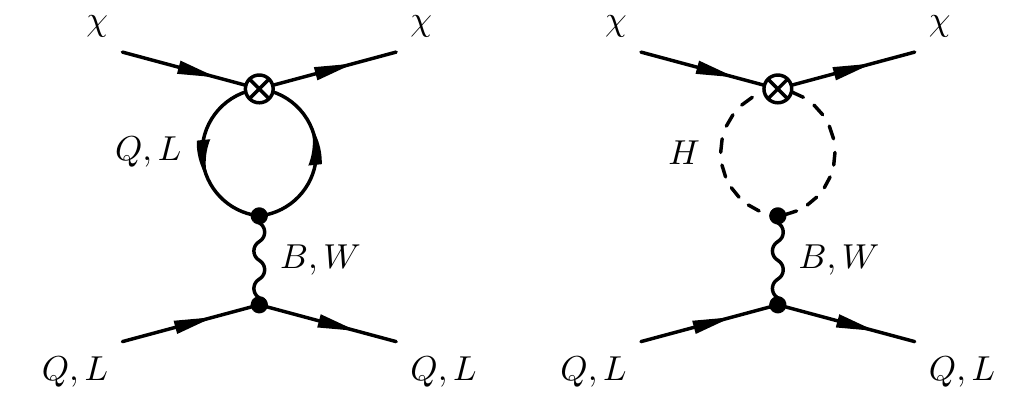}
\includegraphics[scale=0.8]{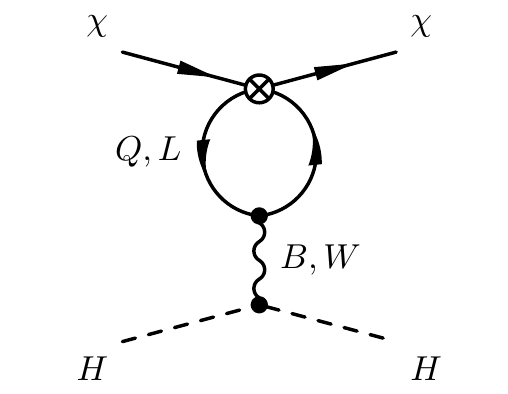}
\caption{The renormalization of the operators
  $Q_{1,i}^{(6)},\ldots,Q_{14,i}^{(6)}$ and
  $Q_{15}^{(6)},\ldots,Q_{18}^{(6)}$ due to the $B_\mu,W_\mu^a$
  penguin insertion (right-handed quarks and leptons can also be on
  the external lines).  }
\label{fig:D6_pengs}
\end{figure}

The Feynman diagrams that lead to nonzero entries in the two matrices
are given in Figs.~\ref{fig:D6_4ferm}~and~\ref{fig:D6_pengs}, with
contributions from gauge boson exchanges between fermion lines, and
penguin diagrams, respectively. We see that an exchange of the
hypercharge boson $B$ between the DM and quark lines, shown in
Fig~\ref{fig:D6_4ferm} (middle and right panel), mixes $Q_{1,i}^{(6)}$
and $Q_{5,i}^{(6)}$, while these operators do not mix with any of the
remaining operators. The same contributions also mix $Q_{2,i}^{(6)}$
and $Q_{6,i}^{(6)}$, $Q_{3,i}^{(6)}$ and $Q_{7,i}^{(6)}$, and
$Q_{4,i}^{(6)}$ and $Q_{8,i}^{(6)}$, respectively. These diagrams are
nonzero only for DM with EW charges. If DM is EW neutral, the
$8\times8$ part of $\gamma_1^{(0)}$ splits into two remaining
$3\times3$ nonzero blocks formed by operators $Q_{2,3,4}^{(6)}$ and
$Q_{6,7,8}^{(6)}$.

In contrast to $\gamma_1^{(0)}$ there are only a few nonzero entries
in $\gamma_2^{(0)}$ in this $8\times 8$ block. The operator
$Q_{1,i}^{(6)}$ gets renormalized through diagrams in
Fig.~\ref{fig:D6_4ferm}, and mixes into $Q_{6,i}^{(6)}$ through the
middle and rightmost diagrams in Fig.~\ref{fig:D6_4ferm}. Equivalent
diagrams mix $Q_{2,i}^{(6)}$ and $Q_{5,i}^{(6)}$. Note that these
contributions to the mixing vanish, if DM is EW neutral, while the
operators $Q_{1,i}^{(6)}$ and $Q_{4,i}^{(6)}$ would be absent.

The penguin insertions, Fig.~\ref{fig:D6_pengs}, also lead to mixing
between operators involving quark currents of different
generations. The corresponding anomalous dimensions are given, for $i
\neq j$, by
\begin{equation}
\big[\gamma_1^{(0)} \big]_{Q_{2,i\dots 4,i}^{(6)} \times Q_{2,j \cdots 4,j}^{(6)}}=
\big[\gamma_1^{(0)} \big]_{Q_{6,i \cdots 8,i}^{(6)} \times Q_{6,j \cdots 8,j}^{(6)}}=
\begin{pmatrix}
\frac{2}{9} & \frac{8}{9} & -\frac{4}{9} \\
\frac{4}{9} & \frac{16}{9} & -\frac{8}{9} \\
-\frac{2}{9} & -\frac{8}{9} & \frac{4}{9}
\end{pmatrix}\, ,
\label{eq:gamma1:DMquark:peng}
\end{equation}
for the part of the anomalous dimension matrix proportional to
$g_1^2$, while the part of the anomalous dimension proportional to
$g_2^2$ has the following non-zero entries for $i\neq j$
\begin{equation}
\big[\gamma_2^{(0)}\big]_{Q_{1,i}^{(6)} Q_{1,j}^{(6)}} = 
    \big[\gamma_2^{(0)}\big]_{Q_{5,i}^{(6)} Q_{5,j}^{(6)}} = 2\,.
\end{equation}
All the other entries vanish.

We turn next to the $6\times6$ block of the anomalous dimension matrix
that describes the mixing of the lepton operators
$Q_{9,i}^{(6)},\dots,Q_{14,i}^{(6)}$,
Eqs.~\eqref{eq:dim6:Q9Q12}-\eqref{eq:dim6:Q11Q14}, among themselves,
giving
\begin{equation}
\big[\gamma_1^{(0)}\big]_{Q_{9,i \cdots 14,i}^{(6)} \times Q_{9,i \cdots 14,i}^{(6)}}= 
\begin{pmatrix}
 0 & 0 & 0 & 3 Y_\chi & 0 & 0 \\
 0 & \frac{2}{3} d_\chi Y_\chi^2+\frac{2}{3} & \frac{4}{3} & 0 & 3 Y_\chi & 0 \\
 0 & \frac{2}{3} & \frac{2}{3} d_\chi Y_\chi^2\frac{4}{3} & 0 & 0 & -6 Y_\chi \\
 3 Y_\chi & 0 & 0 & 0 & 0 & 0 \\
 0 & 3 Y_\chi & 0 & 0 & \frac{2}{3} & \frac{4}{3} \\
 0 & 0 & -6 Y_\chi & 0 & \frac{2}{3} & \frac{4}{3}
\end{pmatrix}\, ,
\end{equation}
and
\begin{equation}
\big[\gamma_2^{(0)}\big]_{Q_{9,i\cdots 14,i}^{(6)} \times Q_{9,i\cdots 14,i}^{(6)}} = 
\begin{pmatrix}
 \frac{8}{9} {\mathcal J}_\chi d_\chi-\frac{16}{3} & 0 & 0 & 0 & -3 {\mathcal J}_\chi & 0 \\
 0 & 0 & 0 & -12 & 0 & 0 \\
 0 & 0 & 0 & 0 & 0 & 0 \\
 0 & -3 {\mathcal J}_\chi & 0 & -\frac{16}{3} & 0 & 0 \\
 -12 & 0 & 0 & 0 & 0 & 0 \\
 0 & 0 & 0 & 0 & 0 & 0
\end{pmatrix}\, .
\end{equation}
The latter two anomalous dimension matrices are straightforward
modifications of the ones for the DM-quark operators in
\eqref{eq:gamma1:DMquark}, \eqref{eq:gamma2:DMquark}, taking into
account different EW charges of the leptons, compared to the quarks,
and the fact that there is no right-handed neutrino in the
SM. (The operators containing the right-handed neutrino could be included, if
  necessary, and would not mix with the operators in our basis.)

Penguin-type insertions lead to mixing between different
generations also for leptons, giving (for $i \neq j$)
\begin{equation}
\big[\gamma_1^{(0)}\big]_{(Q_{10,i}^{(6)},Q_{11,j}^{(6)}) \times (Q_{10,i}^{(6)}, Q_{11,j}^{(6)})}= 
\big[\gamma_1^{(0)}\big]_{(Q_{13,i}^{(6)},Q_{14,j}^{(6)}) \times (Q_{13,i}^{(6)},Q_{14,j}^{(6)})}= 
\begin{pmatrix}
\frac{2}{3} & \frac{4}{3} \\
\frac{2}{3} & \frac{4}{3}
\end{pmatrix}\, ,
\end{equation}
and
\begin{equation}
\big[\gamma_2^{(0)}\big]_{Q_{9,i}^{(6)} Q_{9,j}^{(6)}} =
\big[\gamma_2^{(0)}\big]_{Q_{12,i}^{(6)} Q_{12,j}^{(6)}} = \tfrac{2}{3}\,.
\end{equation}
All the other entries vanish.

A very interesting effect of the one-loop RG running is that the
penguin diagrams mix the operators with quark- and operators with
lepton currents. This is shown in Fig.~\ref{fig:D6_pengs} (left),
where the two quark lines coming from the EFT operator are contracted
into a loop, while the emission of a $B$ converts this into a lepton
current. Conversely, an operator with a leptonic current can be
converted to a DM--quark operator at one-loop. The corresponding
mixing of the quark operators $Q_{1}^{(6)},\dots,Q_{8}^{(6)}$ into the
lepton operators $Q_{9}^{(6)},\dots,Q_{14}^{(6)}$ is given by the
following $8\times 6$ block of $\gamma_1^{(0)}$, now for arbitrary
generation indices $i,j$
\begin{equation}
\big[\gamma_1^{(0)}\big]_{Q_{1,i\cdots 8,i}^{(6)} \times Q_{9,j\cdots 14,j}^{(6)}} =  
\begin{pmatrix}
 0 & 0 & 0 & 0 & 0 & 0 \\
 0 & -\frac{2}{3} & -\frac{4}{3} & 0 & 0 & 0 \\
 0 & -\frac{4}{3} & -\frac{8}{3} & 0 & 0 & 0 \\
 0 & \frac{2}{3}  &  \frac{4}{3} & 0 & 0 & 0 \\
 0 & 0 & 0 & 0 & 0 & 0 \\
 0 & 0 & 0 & 0 & -\frac{2}{3} & -\frac{4}{3} \\
 0 & 0 & 0 & 0 & -\frac{4}{3} & -\frac{8}{3} \\
 0 & 0 & 0 & 0 &  \frac{2}{3}  & \frac{4}{3}
\end{pmatrix}\, . \label{eq:gamma1:quarktolepton}
\end{equation}
The corresponding block of the $\gamma_2^{(0)}$ matrix has only two
nonzero entries,
\begin{equation}
\big[\gamma_2^{(0)}\big]_{Q_{1,i}^{(6)} Q_{9,j}^{(6)}} =\big[\gamma_2^{(0)}\big]_{Q_{5,i}^{(6)} Q_{12,j}^{(6)}} =2\,,
\end{equation}
while the remaining entries in this $8\times 6$ block of $\gamma_2^{(0)}$ are zero. 

The mixing of the lepton operators, $Q_{9}^{(6)},\dots, Q_{14}^{(6)}$,
into the quark operators, $Q_{1}^{(6)},\dots,Q_{8}^{(6)}$, is given for
arbitrary generation indices $i,j$ by the following $6\times 8$ block
of the $\gamma_1^{(0)}$ anomalous matrix
\begin{equation}
\big[\gamma_1^{(0)}\big]_{Q_{9,i\cdots14,i}^{(6)} \times Q_{1,j\cdots 8,j}^{(6)}} = 
\begin{pmatrix}
 0 & 0 & 0 & 0 & 0 & 0 & 0 & 0 \\
 0 & -\frac{2}{9} & -\frac{8}{9} & \frac{4}{9} & 0 & 0 & 0 & 0 \\
 0 & -\frac{2}{9} & -\frac{8}{9} & \frac{4}{9} & 0 & 0 & 0 & 0 \\
 0 & 0 & 0 & 0 & 0 & 0 & 0 & 0 \\
 0 & 0 & 0 & 0 & 0 & -\frac{2}{9} & -\frac{8}{9} & \frac{4}{9} \\
 0 & 0 & 0 & 0 & 0 & -\frac{2}{9} & -\frac{8}{9} & \frac{4}{9}
\end{pmatrix}\, .
\end{equation}
The corresponding $6\times 8$ block of the $\gamma_2^{(0)}$ anomalous
matrix has only two nonzero entries,
\begin{equation}
\big[\gamma_2^{(0)}\big]_{Q_{9,i}^{(6)} Q_{1,j}^{(6)}} =\big[\gamma_2^{(0)}\big]_{Q_{12,i}^{(6)} Q_{5,j}^{(6)}} =\frac23\,.
\label{eq:gamma2:leptontoquark}
\end{equation}

The mixing of DM--quark and DM--lepton operators,
Eqs.~\eqref{eq:gamma1:quarktolepton}--\eqref{eq:gamma2:leptontoquark},
has important phenomenological consequences. One implication is that,
in any theory where one introduces DM--quark interactions, one-loop
mixing will generate DM--lepton interactions. The converse is also
true: a theory of purely ``leptophilic'' DM is impossible. An
interaction between DM and leptons will lead to an interaction between
DM and quarks via one-loop mixing. Note that the mixing is nonzero
irrespective of whether or not DM carries any electroweak
charge. Penguin insertions will also generate DM-quark and DM-lepton
interactions, when initially only the pure DM operators
(Eq.~\eqref{eq:op:DM-DM}) are present; see App.~\ref{app:DM}.

\begin{figure}\centering
\includegraphics[scale=0.8]{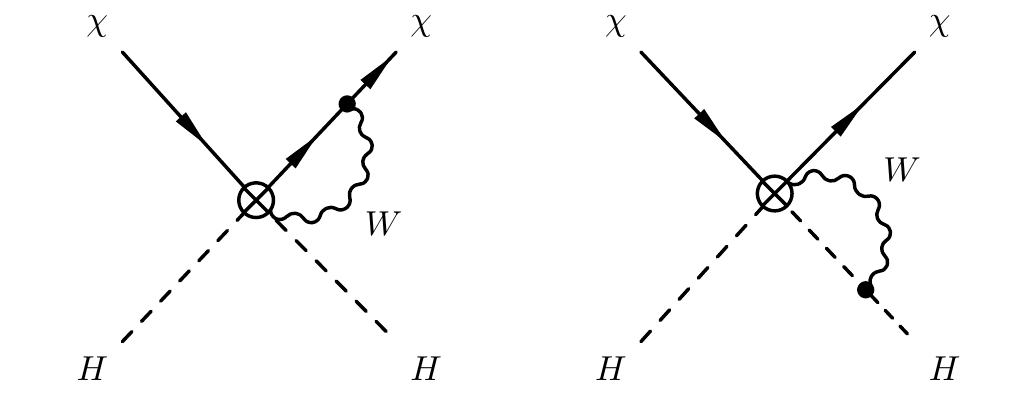}
\caption{Additional diagrams for the renormalization of
  the $Q_{15}^{(6)},\ldots, Q_{18}^{(6)}$ operators at one loop. These diagrams are in
  addition to the set of diagrams in Fig.~\ref{fig:D53} which, at
  dimension six, include an additional derivative w.r.t. to the
  dimension-five operators.}
\label{fig:D6_XXHH}
\end{figure}

Finally, we move to the mixing of dimension-six operators with Higgs
currents, $Q_{15}^{(6)},\ldots,Q_{18}^{(6)}$,
Eqs.~\eqref{eq:dim6:Q15Q17}--\eqref{eq:dim6:Q16Q18}. We start with the
$4\times 4$ blocks of the anomalous dimension matrices that give the
mixing of these operators among themselves,\begin{align}
\big[\gamma_1^{(0)}\big]_{Q_{15\cdots18}^{(6)}\times Q_{15\cdots18}^{(6)}} &= \diag\big(0, \tfrac{1}{3}+\tfrac{2}{3} d_\chi Y_\chi^2, 0, \tfrac{1}{3}\big)\,,\\
\big[\gamma_2^{(0)}\big]_{Q_{15\cdots 18}^{(6)}\times Q_{15\cdots
		18}^{(6)}} &=\diag\big(\tfrac{8}{9} {\cal J}_\chi d_\chi - \tfrac{17}{3},0,-\tfrac{17}{3},0\big)\,.
\end{align}
The relevant diagrams are shown in
Figs.~\ref{fig:D53}~and~\ref{fig:D6_XXHH}. The renormalization induced
by these contributions is multiplicative and does not lead to mixing
of the DM-Higgs operators.

In addition there is mixing of the operators with quark and lepton
currents into the Higgs-current operators and vice versa (see
Fig.~\ref{fig:D6_pengs}). The resulting mixing of the DM--quark
operators, $Q_{1,i}^{(6)}, \ldots, Q_{8,i}^{(6)}$, and the lepton
operators $Q_{9,i}^{(6)},\dots,Q_{14,i}^{(6)}$ into the DM--Higgs
operators, $Q_{15}^{(6)}, \ldots, Q_{18}^{(6)}$, are given by the
following $8\times 4$ and $6\times 4$ blocks in the $\gamma_1^{(0)}$
anomalous dimension matrix ($i=1,2,3$), respectively, 
\begin{equation}\label{eq:mix:g1:Q1Q8-Q15Q18}
\big[\gamma_1^{(0)}\big]_{Q_{1,i\dots8,i}^{(6)}\times Q_{15\cdots 18}^{(6)}} = 
\begin{pmatrix}
0 & 0 & 0 & 0 \\
0 & \frac{2}{3} & 0 & 0 \\
0 & \frac{4}{3} & 0 & 0 \\
0 & -\frac{2}{3} & 0 & 0 \\
0 & 0 & 0 & 0 \\
0 & 0 & 0 & \frac{2}{3} \\
0 & 0 & 0 & \frac{4}{3} \\
0 & 0 & 0 & -\frac{2}{3}
\end{pmatrix}\,,
\qquad
\big[\gamma_1^{(0)}\big]_{Q_{9,i\cdots 14,i}^{(6)}\times Q_{15\cdots 18}^{(6)}} = 
\begin{pmatrix}
0 & 0 & 0 & 0 \\
0 & -\frac{2}{3} & 0 & 0 \\
0 & -\frac{2}{3} & 0 & 0 \\
0 & 0 & 0 & 0 \\
0 & 0 & 0 & -\frac{2}{3} \\
0 & 0 & 0 & -\frac{2}{3}
\end{pmatrix}\,,
\end{equation}
and by the corresponding $8\times 4$ and $6\times 4$ blocks in the $\gamma_2^{(0)}$
anomalous dimension matrix, which, however, only have two nonzero
entries each, 
\begin{equation}
\big[\gamma_2^{(0)}\big]_{Q_{1,i}^{(6)} Q_{15}^{(6)}} =\big[\gamma_2^{(0)}\big]_{Q_{5,i}^{(6)} Q_{17}^{(6)}} =2\,,
\qquad
\big[\gamma_2^{(0)}\big]_{Q_{9,i}^{(6)} Q_{15}^{(6)}} =\big[\gamma_2^{(0)}\big]_{Q_{12,i}^{(6)} Q_{17}^{(6)}} =\frac{2}{3}\,.
\label{eq:gamma2:quarktohiggs}
\end{equation}

The mixing of the DM--Higgs operators,
$Q_{15}^{(6)},\dots,Q_{18}^{(6)}$, into the DM--quark operators,
$Q_{1}^{(6)},\dots,Q_{8}^{(6)}$, and into the DM--lepton operators,
$Q_{9}^{(6)},\dots,Q_{14}^{(6)}$, is given by
\begin{equation}
\big[\gamma_1^{(0)} \big]_{Q_{15\cdots 18}^{(6)}\times Q_{1,i\cdots 8,i}^{(6)}}= 
\begin{pmatrix}
 0 & 0 & 0 & 0 & 0 & 0 & 0 & 0 \\
 0 & \frac{1}{9} & \frac{4}{9} & -\frac{2}{9} & 0 & 0 & 0 & 0 \\
 0 & 0 & 0 & 0 & 0 & 0 & 0 & 0 \\
 0 & 0 & 0 & 0 & 0 & \frac{1}{9} & \frac{4}{9} & -\frac{2}{9}
\end{pmatrix}\, ,
\end{equation}
and 
\begin{equation}
\big[\gamma_1^{(0)}\big]_{Q_{15\cdots 18}^{(6)}\times Q_{9,i\cdots 14,i}^{(6)}} = 
\begin{pmatrix}
 0 & 0 & 0 & 0 & 0 & 0 \\
 0 & -\frac{1}{3} & -\frac{2}{3} & 0 & 0 & 0 \\
 0 & 0 & 0 & 0 & 0 & 0 \\
 0 & 0 & 0 & 0 & -\frac{1}{3} & -\frac{2}{3}
\end{pmatrix}\, ,
\end{equation}
respectively, for the corresponding blocks of $\gamma_1^{(0)}$, while
the nonzero $\gamma_2^{(0)}$ entries are given by 
\begin{equation}
\big[\gamma_2^{(0)}\big]_{Q_{15}^{(6)} Q_{1,i}^{(6)}}
=\big[\gamma_2^{(0)}\big]_{Q_{17}^{(6)} Q_{5,i}^{(6)}}
=\big[\gamma_2^{(0)}\big]_{Q_{15}^{(6)} Q_{9,i}^{(6)}}
=\big[\gamma_2^{(0)}\big]_{Q_{17}^{(6)} Q_{12,i}^{(6)}} =\frac{1}{3}\,. 
\label{eq:gamma2:higgstoleptonquark}
\end{equation}
Note that both the mixing of DM--quark and DM--lepton operators into
the DM--Higgs ones and vice versa is present even if the DM does not
carry any electroweak charge.

\begin{figure}\centering
	\includegraphics[scale=0.70]{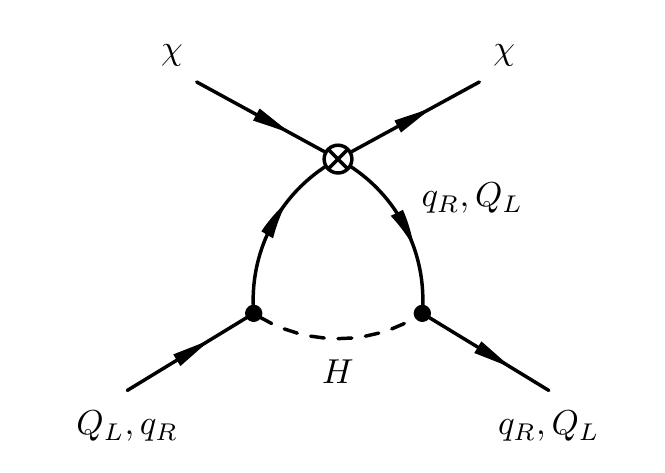}
	\includegraphics[scale=0.70]{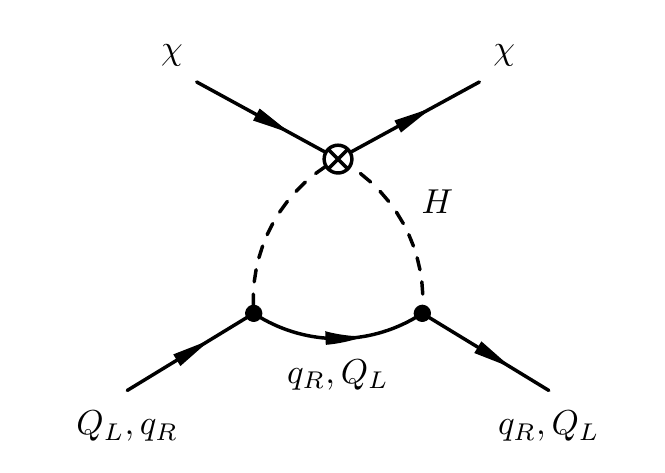}
	\includegraphics[scale=0.70]{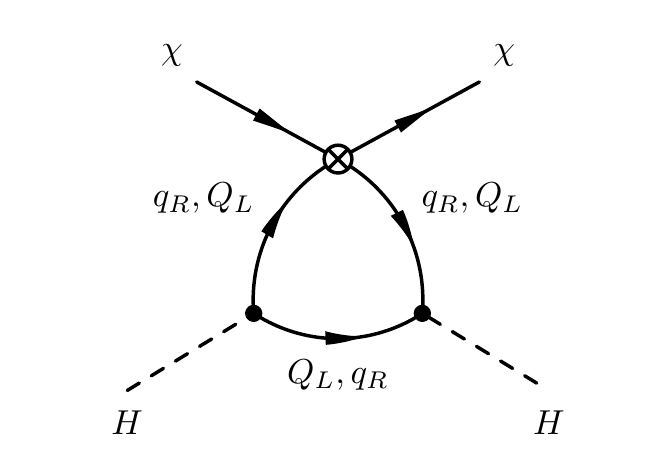}
	\caption{Mixing of the quark operators
          proportional to the quark yukawa coupling due to the insertion of
          DM-quark and DM-Higgs operators. We consider only the top, bottom, charm Yukawas here and so $q\in\{t,b,c\}$. The same diagrams with $Q_L\to L_L$ and $q_R\to\tau_R$ mix DM-lepton and DM-Higgs operators. These diagrams only contribute to off-diagonal mixing.
}
	\label{fig:mixing-propto-yt}
\end{figure}

For the third-generation DM--quark operators, $Q_{1,3}^{(6)},\dots
Q_{8,3}^{(6)}$, there is also the renormalization due to the Yukawa
interaction with the Higgs (we neglect all the Yukawa interactions
except with the third fermion generation and the charm Yukawa), giving
\begin{equation}\label{eq:mix:yt:Q1Q8-Q1Q8}
\big[\gamma_{y_c}^{(0)}\big]_{Q_{1,2\cdots 8,2}^{(6)}\times Q_{1,2\cdots 8,2}^{(6)}}  =
\big[\gamma_{y_t}^{(0)}\big]_{Q_{1,3\cdots 8,3}^{(6)}\times Q_{1,3\cdots 8,3}^{(6)}}  =
\begin{pmatrix}
1 & 0 & 0 & 0 & 0 & 0 & 0 & 0  \\
 0 & 1 & -2 & 0 & 0 & 0 & 0 & 0 \\
 0 & -1 & 2 & 0 & 0 & 0 & 0 & 0 \\
 0 & 0 & 0 & 0 & 0 & 0 & 0 & 0  \\
 0 & 0 & 0 & 0 & 1 & 0 & 0 & 0  \\
 0 & 0 & 0 & 0 & 0 & 1 & -2 & 0  \\
 0 & 0 & 0 & 0 & 0 & -1 & 2 & 0  \\
 0 & 0 & 0 & 0 & 0 & 0 & 0 & 0 
\end{pmatrix},
\end{equation}
and
\begin{equation}\label{eq:mix:yb:Q1Q8-Q1Q8}
\big[\gamma_{y_b}^{(0)}\big]_{Q_{1,3\cdots 8,3}^{(6)}\times Q_{1,3\cdots 8,3}^{(6)}}  =
\begin{pmatrix}
1& 0& 0& 0& 0& 0& 0& 0\\
 0&1& 0& - 2& 0& 0& 0& 0\\
 0& 0& 0& 0& 0& 0& 0& 0\\
 0& - 1& 0&2& 0& 0& 0& 0\\
 0& 0& 0& 0&1& 0& 0& 0\\
 0& 0& 0& 0& 0&1& 0& - 2\\
 0& 0& 0& 0& 0& 0& 0& 0\\
 0& 0& 0& 0& 0& - 1& 0&2
\end{pmatrix}.
\end{equation}
The off-diagonal entries in Eq.~(\ref{eq:mix:yt:Q1Q8-Q1Q8}) are
generated by the left-most diagram in Fig.~\ref{fig:mixing-propto-yt}
while the diagonal entries result from the field renormalization constants.
The Yukawa interactions also lead to mixing of the
DM--third-generation quark operators into the DM-Higgs operators,
$Q_{15}^{(6)},\dots,Q_{18}^{(6)}$, 
\begin{equation}\label{eq:yt-quark-higgs}
\big[\gamma_{y_c}^{(0)} \big]_{Q_{1,2\cdots 8,2}^{(6)}\times Q_{15\cdots 18}^{(6)}}= 
\big[\gamma_{y_t}^{(0)} \big]_{Q_{1,3\cdots 8,3}^{(6)}\times Q_{15\cdots 18}^{(6)}}= 
\begin{pmatrix}
 -6 & 0 & 0 & 0 \\
 0 & 6 & 0 & 0 \\
 0 & -6 & 0 & 0 \\
 0 & 0 & 0 & 0 \\
 0 & 0 & -6 & 0 \\
 0 & 0 & 0 & 6 \\
 0 & 0 & 0 & -6 \\
 0 & 0 & 0 & 0 
\end{pmatrix},
\end{equation}
and
\begin{equation}\label{eq:yb-quark-higgs}
\big[\gamma_{y_b}^{(0)} \big]_{Q_{1,3\cdots 8,3}^{(6)}\times Q_{15\cdots 18}^{(6)}}= 
\begin{pmatrix}
 - 6& 0& 0& 0\\
 0& - 6& 0& 0\\
 0& 0& 0& 0\\
 0&6& 0& 0\\
 0& 0& - 6& 0\\
 0& 0& 0& - 6\\
 0& 0& 0& 0\\
 0& 0& 0&6
\end{pmatrix},
\end{equation}
as well as to the mixing of the DM-Higgs operators into the
DM--third-generation quark operators, 
\begin{equation}
\big[\gamma_{y_c}^{(0)}\big]_{Q_{15\cdots 18}^{(6)}\times Q_{1,2\cdots 8,2}^{(6)}} = 
\big[\gamma_{y_t}^{(0)}\big]_{Q_{15\cdots 18}^{(6)}\times Q_{1,3\cdots 8,3}^{(6)}} = 
\begin{pmatrix}
 -1 & 0 & 0 & 0 & 0 & 0 & 0 & 0  \\
 0 & 1 & -2 & 0 & 0 & 0 & 0 & 0 \\
 0 & 0 & 0 & 0 & -1 & 0 & 0 & 0  \\
 0 & 0 & 0 & 0 & 0 & 1 & -2 & 0 
\end{pmatrix},
\end{equation}
\begin{equation}
\big[\gamma_{y_b}^{(0)}\big]_{Q_{15\cdots 18}^{(6)}\times Q_{1,3\cdots 8,3}^{(6)}} = 
\begin{pmatrix}
 - 1& 0& 0& 0& 0& 0& 0& 0\\
 0& - 1& 0&2& 0& 0& 0& 0\\
 0& 0& 0& 0& - 1& 0& 0& 0\\
 0& 0& 0& 0& 0& - 1& 0&2
\end{pmatrix}.
\end{equation}

The Yukawa interactions also renormalize the Higgs operators
themselves due to the renormalization of the Higgs fields, giving
\begin{equation}
\big[\gamma_{y_t}^{(0)}\big]_{Q_{15\cdots 18}^{(6)}\times Q_{15\cdots
    18}^{(6)}} = \big[\gamma_{y_b}^{(0)}\big]_{Q_{15\cdots
    18}^{(6)}\times Q_{15\cdots 18}^{(6)}} =
\big[\gamma_{y_c}^{(0)}\big]_{Q_{15\cdots 18}^{(6)}\times Q_{15\cdots
    18}^{(6)}} = \diag (6,6,6,6)\,.
\end{equation}

Finally, we also present the anomalous dimensions due to the tau
Yukawa coupling, leading to mixing among the four-fermion operators,
\begin{equation}\label{eq:mix:ytau:Q9Q14-Q9Q14}
\big[\gamma_{y_\tau}^{(0)}\big]_{Q_{9,3\cdots 14,3}^{(6)}\times Q_{9,3\cdots 14,3}^{(6)}}  =
\begin{pmatrix}
1& 0& 0& 0& 0& 0\\
 0&1& - 2& 0& 0& 0\\
 0& - 1&2& 0& 0& 0\\
 0& 0& 0&1& 0& 0\\
 0& 0& 0& 0&1& - 2\\
 0& 0& 0& 0& - 1&2
\end{pmatrix},
\end{equation}
mixing of four-fermion into Higgs operators,
\begin{equation}\label{eq:ytau-lepton-higgs}
\big[\gamma_{y_\tau}^{(0)} \big]_{Q_{9,3\cdots 14,3}^{(6)}\times Q_{15\cdots 18}^{(6)}}= 
\begin{pmatrix}
 - 2& 0& 0& 0\\
 0& - 2& 0& 0\\
 0&2& 0& 0\\
 0& 0& - 2& 0\\
 0& 0& 0& - 2\\
 0& 0& 0&2
\end{pmatrix},
\end{equation}
Higgs operators into four-fermion operators,
\begin{equation}
\big[\gamma_{y_\tau}^{(0)}\big]_{Q_{15\cdots 18}^{(6)}\times Q_{9,3\cdots 14,3}^{(6)}} = 
\begin{pmatrix}
 - 1& 0& 0& 0& 0& 0\\
 0& - 1&2& 0& 0& 0\\
 0& 0& 0& - 1& 0& 0\\
 0& 0& 0& 0& - 1&2
\end{pmatrix},
\end{equation}
and diagonal self mixing of the Higgs operators due to Higgs field
renormalization,
\begin{equation}
\big[\gamma_{y_\tau}^{(0)}\big]_{Q_{15\cdots 18}^{(6)}\times
  Q_{15\cdots 18}^{(6)}} = \diag (2,2,2,2)\,.
\end{equation}

The results given in this section are sufficient if one works to
leading-log accuracy without resummation of the logarithms. However,
the set of operators $Q_{1,i}^{(6)},\dots,Q_{18}^{(6)}$,
Eqs.~\eqref{eq:dim6:Q15}-\eqref{eq:dim6:Q16Q18}, does not close under
renormalization, unless the DM self-interaction operators,
Eq.~\eqref{eq:op:DM-DM}, and the SM EFT operators are included. We
provide the anomalous dimension that give the mixing with the SM
effective operators in App.~\ref{app:SMEFT}, and with the DM
self-interaction operators in App.~\ref{app:DM}.

%%%%%%%%%%%%%%%%%%%%%%%%%%%%%%%%%%%
\section{Matching to EFT below the weak scale}
\label{sec:match-ew}
%%%%%%%%%%%%%%%%%%%%%%%%%%%%%%%%%%%%

The running from the mediator scale, $\mu\sim \Lambda$, down to the
weak scale, $\mu\sim m_Z$, is described by the evolution operator
$U(\mu_{\rm EW}, \Lambda)$ in Eq.~\eqref{eq:evol:sol}. The relevant
anomalous dimension matrix~$\gamma$, appearing in Eq.~\eqref{eq:Ueq},
was presented in Sections~\ref{subsec:mix:dim5}
and~\ref{subsec:mix:dim6}. The next step is to calculate the matching
onto a five-flavor theory at $\mu\sim v_{\rm EW}$ by integrating out
the top quark, the Higgs and $W,Z$ gauge bosons. This gives the matrix
$M_{\text{EW}\to (5)}$ in the evolution
equation~\eqref{eq:evol:sol}. Since we are interested in the elastic
direct detection scattering we can, below the electroweak scale,
ignore all the charged components of the $\chi$ multiplet. From now on
$\chi$ will thus denote {\em only the neutral component} of the DM
electroweak multiplet.

After the matching at $\mu\sim m_Z$ we arrive at the $n_f=5$ effective
Lagrangian which we organize in terms of the dimensionality of the
operators,
\begin{equation}\label{eq:Lchi:EW}
{\cal L}_\chi|_{n_f=5}={\cal L}_\chi^{(4)}|_{n_f=5}+{\cal
  L}_\chi^{(5)}|_{n_f=5}+{\cal L}_\chi^{(6)}|_{n_f=5}+{\cal
  L}_\chi^{(7)}|_{n_f=5}+\cdots. 
\end{equation} 
In the matching we keep all the numerically leading terms. The leading
contributions from dimension-five $(\propto 1/\Lambda$) and
dimension-six operators ($\propto 1/\Lambda^2$),
Eq.~\eqref{eq:Lchi:Lambda}, generically arise already at tree level,
with the exception of phenomenologically important one-loop matchings
onto the dimension-seven gluonic operators. In these matching
calculations we allow for DM to carry arbitrary $SU(2)\times U(1)$
gauge quantum numbers.  In addition, there are contributions from
renormalizable interactions. We include these in our numerical
examples in Sec.~\ref{sec:RG-run}, taking $Y_\chi=0$, so that there is
no tree-level $Z$ coupling to DM. The first nonzero contributions from
gauge interactions are then due to the one- and two-loop electroweak
threshold corrections, shown in Fig.~\ref{fig:DD-Higgs_loop}, for
which we use the results of Ref.~\cite{Hisano:2011cs}.

We consider two discrete options for the DM mass: i) light DM, $m_\chi
\ll m_Z$, and ii) DM with the EW scale mass, $m_\chi \sim {\mathcal
  O}( m_Z)$. The case of heavy DM, $m_\chi \gg m_Z$, is relegated to
future work (dimension-four interactions are discussed in
Refs.~\cite{Hill:2011be, Hill:2014yka}).  In
Section~\ref{subsec:lightDM} we perform the matching for light DM. In
this case the time component and the spatial components of the DM
current are of the same size at the matching scale. The situation is
different for weak scale DM. For $\mu\lesssim m_\chi\sim {\mathcal O}(
m_Z)$ DM becomes non-relativistic, and thus the time component is
parametrically larger than the spatial ones.  In the matching we
therefore need to simultaneously perform an expansion in $1/m_\chi$,
which is done in Section~\ref{subsec:EW:DM}.

Before proceeding we remark that both the DM mass, $m_\chi$, and the
DM field, $\chi$, get shifted by the Higgs vacuum expectation value
due to the contributions from the $Q_{3,4}^{(5)}$ operators,
Eq.~\eqref{Q34}, and from the $Q_{7,8}^{(5)}$ operators,
Eq.~\eqref{Q78}. The dimension-four part of the effective
Lagrangian~\eqref{eq:Lchi:EW} in terms of the shifted fields, $\chi'$,
is
\begin{equation}\label{eq:redef-mass-field}
{\cal L}_\chi^{(4)}|_{n_f=5}=i \bar \chi' \slashed \partial \chi'
-m_\chi' \bar \chi' \chi'\,.
\end{equation}
The redefinition of the $\chi$ field is a simple chiral rotation,
$\chi'= \exp\big({\frac{i}{2} \gamma_5 \phi}\big)\chi$, with (see also
Ref.~\cite{Fedderke:2014wda})
\begin{equation}\label{eq:chiral-rotation-angle}
	\tan \phi =
        \Big(C_7^{(5)}+\tfrac{Y_\chi}{4}C_8^{(5)}\Big)\Big/\Big[ 2\pi\alpha_2 
        m_\chi\Lambda /(c_w^2 m_Z^2)-\left({C_3^{(5)}+\tfrac{Y_\chi}{4}C_4^{(5)}}\right)
        \Big]\,,
\end{equation}
while the new mass term is
\begin{equation}\label{eq:field:redef:light}
m_\chi'=m_\chi\cos\phi + \frac{c_w^2 m_Z^2}{2\pi\alpha_2\Lambda}
\bigg[\bigg(C_7^{(5)}+\frac{Y_\chi}{4} C_8^{(5)}\bigg) \sin \phi -
  \bigg(C_3^{(5)}+\frac{Y_\chi}{4} C_4^{(5)}\bigg) \cos \phi \bigg]
\,.
\end{equation}
The field redefinition also changes the operators $Q_1^{(5)}, \dots,
Q_8^{(5)}$ in Eqs.~\eqref{Q12}-\eqref{Q78} and the corresponding
Wilson coefficients, $C_i^{(5)}{}'=C_i^{(5)} \cos \phi + C_{i+4}^{(5)}
\sin \phi \,$, $C_{i+4}^{(5)}{}' = C_{i+4}^{(5)} \cos \phi -
C_{i}^{(5)} \sin \phi \,,$ for $i=1,\dots, 4$, while there is no
change in the dimension-six Wilson coefficients.  In the case $m_\chi
\sim \mathcal{O}(m_Z)$ we expand in $m_Z/\Lambda$ which gives
\begin{equation}
\begin{split}
C_i^{(5)}{}'&=C_i^{(5)}+\frac{c_w^2m_Z^2}{2\pi\alpha_2\Lambda m_\chi}
\left(C_7^{(5)}+\tfrac{Y_\chi}{4}C_8^{(5)}\right)C_{i+4}^{(5)},\\
C_{i+4}^{(5)}{}'&=C_{i+4}^{(5)}-\frac{c_w^2m_Z^2}{2\pi\alpha_2\Lambda m_\chi}
\left(C_7^{(5)}+\tfrac{Y_\chi}{4}C_8^{(5)}\right)C_{i}^{(5)}.
\end{split}
\end{equation}
From now on we will assume that the above field and mass redefinitions
have been performed and drop the primes on the Wilson coefficients,
the DM fields, and the DM mass.

\subsection{Light dark matter}
\label{subsec:lightDM}

In the case of light DM, $m_\chi\ll m_Z$, we can use relativistic DM
fields to construct the effective theory below the weak scale. The
effective Lagrangians containing operators of dimensionality $d$ in
Eq. \eqref{eq:Lchi:EW} are given by
\begin{equation}\label{eq:lightDM:Lnf5}
{\cal L}_\chi^{(d)}|_{n_f=5}=\sum_{a}
\hat \C_{a}^{(d)}|_{n_f=5} {\cal Q}_a^{(d)},
\end{equation}
where we introduced the dimensionful Wilson coefficients $\hat
\C_{a}^{(d)}|_{n_f=5}$ in order to simplify the notation. They are
suppressed by inverse powers of the NP scale $\Lambda$ and/or the top,
$W$, $Z$ and Higgs masses.  The DM mass, $m_\chi$, can be set to zero
in the matching except when calculating the electroweak threshold
corrections from the gauge interactions, where one needs to expand to
first order in $m_\chi$.

The electroweak EFT Lagrangian~\eqref{eq:Lchi:Lambda} with operators
up to dimension six matches onto the ``five-flavor'' EFT in the broken
electroweak phase, Eq.~\eqref{eq:Lchi:EW}. This gives rise to
operators up to dimension seven, if one keeps only the leading
contributions. We first give the basis of the operators $\Q_a^{(d)}$
in the five-flavor EFT, required for the matching, and then present
their respective Wilson coefficients $\hat \C_a^{(d)}$.

At dimension five there are only two operators,
\begin{equation}
\label{eq:dim5:nf5:Q1Q2:light}
{\cal Q}_{1}^{(5)} = \frac{e}{8 \pi^2} (\bar \chi \sigma^{\mu\nu}\chi)
 F_{\mu\nu} \,, \qquad {\cal Q}_2^{(5)} = \frac{e }{8 \pi^2} (\bar
\chi \sigma^{\mu\nu} i\gamma_5 \chi) F_{\mu\nu} \,,
\end{equation}
where $F_{\mu\nu}$ is the electromagnetic field strength tensor. The
operator $\Q_1^{(5)}$ is CP even, while $\Q_2^{(5)}$ is CP odd. The
dimension-six operators are
\begin{align}
{\cal Q}_{1,f}^{(6)} & = (\bar \chi \gamma_\mu \chi) (\bar f \gamma^\mu f)\,,
 &{\cal Q}_{2,f}^{(6)} &= (\bar \chi\gamma_\mu\gamma_5 \chi)(\bar f \gamma^\mu f)\,, \label{eq:dim6EW:Q1Q2:light}
  \\ 
{\cal Q}_{3,f}^{(6)} & = (\bar \chi \gamma_\mu \chi)(\bar f \gamma^\mu \gamma_5 f)\,,
  & {\cal Q}_{4,f}^{(6)}& = (\bar
\chi\gamma_\mu\gamma_5 \chi)(\bar f \gamma^\mu \gamma_5 f)\,.\label{eq:dim6EW:Q3Q4:light}
\end{align}
Here $f$ denotes any quark, $f=u,d,s,c,b$, or charged lepton flavor,
$f=e,\mu, \tau$. We find it convenient to express the operators in
terms of (axial-)vector and (pseudo-)scalar currents, which have
definite non-relativistic limits. Operators with neutrinos are not
needed for our purposes as they do not run below the EW scale.

In the effective Lagrangian Eq.~\eqref{eq:Lchi:EW} we need to include
a subset of dimension-seven operators. These are generated from
dimension-five and -six operators in the effective
Lagrangian~\eqref{eq:Lchi:Lambda} when integrating out the Higgs and
the $Z$ boson at $\mu_{\rm EW}\sim m_Z$. They are thus suppressed by
${\mathcal O}(1/\Lambda^2 m_{h,Z})$ or ${\mathcal O}(1/\Lambda
m_{h,Z}^2)$, instead of ${\mathcal O}(1/\Lambda^3)$, and can lead to
contributions in direct detection comparable to those of the
dimension-six operators,
Eqs.~\eqref{eq:dim6EW:Q1Q2:light}-\eqref{eq:dim6EW:Q3Q4:light}.

The relevant dimension-seven operators involving the DM and gluon
fields are given by
\begin{align}
{\cal Q}_1^{(7)} & = \frac{\alpha_s}{12\pi} (\bar \chi \chi)
 G^{a\mu\nu}G_{\mu\nu}^a\,, 
 & {\cal Q}_2^{(7)} &= \frac{\alpha_s}{12\pi} (\bar \chi i\gamma_5 \chi) G^{a\mu\nu}G_{\mu\nu}^a\,,\label{eq:dim7:Q1Q2:light}
 \\
{\cal Q}_3^{(7)} & = \frac{\alpha_s}{8\pi} (\bar \chi \chi) G^{a\mu\nu}\widetilde
 G_{\mu\nu}^a\,, 
& {\cal Q}_4^{(7)}& = \frac{\alpha_s}{8\pi}
(\bar \chi i \gamma_5 \chi) G^{a\mu\nu}\widetilde G_{\mu\nu}^a \,, \label{eq:dim7:Q3Q4:light}
\end{align}
where $\widetilde G_{\mu\nu} =
\frac{1}{2}\varepsilon_{\mu\nu\rho\sigma} G^{\rho\sigma}$ and
$a=1,\dots,8$ are the color indices. The strong coupling constant
$\alpha_s$ is defined in the five-flavor scheme. The normalization
reflects the fact that these operators are typically generated at
one-loop level. Note that ${\cal Q}_2^{(7)}$ and ${\cal Q}_3^{(7)} $
are CP odd.

There are also four scalar operators
\begin{align}
{\cal Q}_{5,f}^{(7)} & = m_f (\bar \chi \chi)( \bar f f)\,, 
&{\cal
  Q}_{6,f}^{(7)} &= m_f (\bar \chi i \gamma_5 \chi)( \bar f f)\,,\label{eq:dim7EW:Q5Q6:light}
  \\
{\cal Q}_{7,f}^{(7)} & = m_f (\bar \chi \chi) (\bar f i \gamma_5 f)\,, 
&{\cal Q}_{8,f}^{(7)} & = m_f (\bar \chi i  \gamma_5 \chi)(\bar f i \gamma_5
f)\,, \label{eq:dim7EW:Q7Q8:light} 
\end{align}
with $f$ denoting any quark $(f=u,d,s,c,b)$ or charged lepton flavor
$(f=e,\mu, \tau)$. The definitions of ${\cal Q}_{5,f}^{(7)},\dots,
{\cal Q}_{8,f}^{(7)} $ include an explicit power of the corresponding
quark or lepton mass. This reflects the leading contributions to their
Wilson coefficients, see below.

In the remainder of the subsection we give the results of the matching
at $\mu_{\rm EW} \sim m_Z$.  We start with the dimension-five
operators where the contributions come from $W$ and $B$ dipole
operators above $m_Z$ after rotating the EW gauge eigenstates into the
mass eigenstates after EWSB:\footnote{Note that in
  Eqs.~\eqref{eq:match:lightDM:C1:light}
  and~\eqref{eq:match:lightDM:C2:light} we use the original definition
  of operators, Eqs.~\eqref{Q12} and~\eqref{Q56}.}
\begin{align}
\label{eq:match:lightDM:C1:light}
\hat \C_{1}^{(5)}|_{n_f=5} & = \frac{1}{\Lambda} \bigg( C_{1}^{(5)} +
\frac{Y_\chi}{2} C_{2}^{(5)} \bigg)+\ldots\,,
\\
\label{eq:match:lightDM:C2:light}
\hat \C_{2}^{(5)}|_{n_f=5} &=  \frac{1}{\Lambda} \bigg( 
C_{5}^{(5)} + \frac{Y_\chi}{2} C_{6}^{(5)} \bigg)\,.  
\end{align}
Equation~\eqref{eq:match:lightDM:C1:light} also receives a one-loop
contribution from dimension-four gauge interactions, denoted by the
ellipsis, proportional to the hypercharge of the DM multiplet. We omit
this contribution here since a non-zero hypercharge leads to a
tree-level $Z$ exchange with nuclei which is excluded by direct
detection experiments.

\begin{figure}[t]\centering
\includegraphics[scale=0.8]{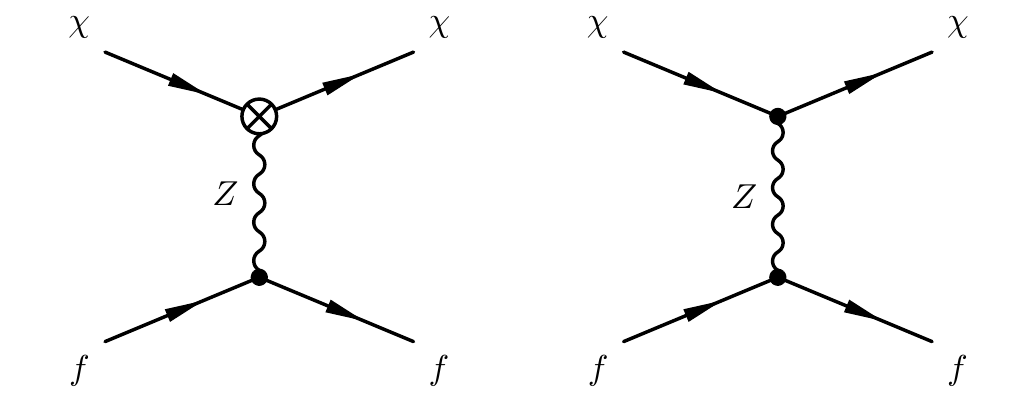}
\includegraphics[scale=0.8]{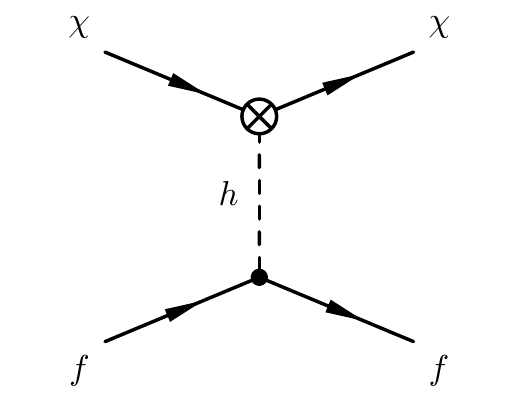}
\caption{Matching contributions to the effective operators at $\mu
  \sim m_Z$. The left diagram shows the contribution of the
  high-energy operators, the middle one the contribution from the
  dimension-four gauge interactions onto the dimension-six operators,
  respectively; the right diagram shows the contribution to the
  matching onto the dimension-seven operators.}
\label{fig:d6-matching}
\end{figure}

For the dimension-six operators we start with the operators with
external quark legs. The contributions from dimension-six UV operators
with external quark legs are
\begin{align}
\hat \C_{1,u_i (d_i)}^{(6)}|_{n_f=5} &= \frac{1}{\Lambda^2}\left[\mp \frac{Y_\chi }{8} C_{1,i}^{(6)} + \frac{C_{2,i}^{(6)}}{2}  +
\frac{C_{3(4),i}^{(6)}}{2} \pm \frac{3 - 8 (4) s_w^2}{6}
	\left(\frac{Y_\chi}{4} C_{15}^{(6)} +C_{16}^{(6)}
          \right)\right] \,,\label{eq:match-6-up-1}
\\
\hat \C_{2,u_i(d_i)}^{(6)}|_{n_f=5} &=\frac{1}{\Lambda^2}\left[ \mp \frac{Y_\chi }{8} C_{5,i}^{(6)} + \frac{C_{6,i}^{(6)}}{2}  +
\frac{C_{7(8),i}^{(6)}}{2}  \pm \frac{3 - 8(4) s_w^2}{6}
		\left(\frac{Y_\chi}{4} C_{17}^{(6)} +C_{18}^{(6)} \right)\right]
			\,,\label{eq:match-6-up-2}
\\
\begin{split}
\hat \C_{3,u_i(d_i)}^{(6)}|_{n_f=5} &=  \frac{1}{\Lambda^2}\left[\pm\frac{Y_\chi }{8} C_{1,i}^{(6)} - \frac{C_{2,i}^{(6)}}{2}  +
\frac{C_{3(4),i}^{(6)}}{2}  \mp \frac{1}{2}
		\left(\frac{Y_\chi }{4}C_{15}^{(6)} + C_{16}^{(6)}\right)\right] \,,
\end{split}
\\
\hat \C_{4,u_i(d_i)}^{(6)}|_{n_f=5} &=  \frac{1}{\Lambda^2}\left[\pm\frac{Y_\chi }{8} C_{5,i}^{(6)} - \frac{C_{6,i}^{(6)}}{2}  +
\frac{C_{7(8),i}^{(6)}}{2}  \mp \frac{1}{2}
		\left(\frac{Y_\chi }{4}C_{17}^{(6)} +
                C_{18}^{(6)}\right)\right] + \ldots\,,\label{eq:match-6-up-4}
\end{align}
where $i$ is a generation index ($u_1 \equiv u$, $u_2 \equiv c$ and
$d_1 \equiv d$, $d_2 \equiv s$, $d_3 \equiv b$) and the upper(lower)
signs apply for up(down) quarks. For each of the Wilson coefficients
the last $1/\Lambda^2$-suppressed term is due to $Z$ exchange, shown
in Fig.~\ref{fig:d6-matching} (left). For a DM multiplet with nonzero
hypercharge $Y_\chi$, $Z$ exchange due to the renormalizable gauge
coupling~\eqref{eq:gauge-int}, see Fig.~\ref{fig:d6-matching} (right),
gives the additional contributions $\hat \C_{1,u_i
  (d_i)}^{(6)}|_{n_f=5} = \pm \frac{\pi\alpha_2}{6c_w^2 m_Z^2} (3 - 8
(4) s_w^2) Y_\chi$ and $\hat \C_{3,u_i(d_i)}^{(6)}|_{n_f=5} = \mp
\frac{\pi\alpha_2}{2c_w^2 m_Z^2} Y_\chi$. $\hat
\C_{4,u_i(d_i)}^{(6)}|_{n_f=5}$ receives a contribution, denoted by
the ellipsis, from gauge interactions at one-loop (see
Fig.~\ref{fig:DD-Higgs_loop}) that does not vanish for
$Y_\chi=0$. This requires a two-loop matching calculation with
$m_\chi$ kept parametrically small, which is beyond the scope of
present paper. In the numerical evaluations we thus use the results
from Ref.~\cite{Hisano:2011cs}, that were obtained assuming that
$m_\chi$ is not much smaller than $m_Z$.

Similarly we find for the dimension-six operators with leptons
($\ell_1 \equiv e$, $\ell_2 \equiv \mu$, $\ell_3 \equiv \tau$) on the
external legs
\begin{align}
\hat \C_{1,\ell_i}^{(6)}|_{n_f=5} & =  \frac{1}{\Lambda^2}\left[\frac{Y_\chi }{8} C_{9,i}^{(6)} + \frac{C_{10,i}^{(6)}}{2}  +
\frac{C_{11,i}^{(6)}}{2} 
- \frac{1 - 4 s_w^2}{2} \left( \frac{Y_\chi}{4}C_{15}^{(6)} +
C_{16}^{(6)} \right)\right] \,,\label{eq:match-6-lep-1}
\\
\hat\C_{2,\ell_i}^{(6)}|_{n_f=5} &=\frac{1}{\Lambda^2}\left[  \frac{Y_\chi }{8} C_{12,i}^{(6)} + \frac{C_{13,i}^{(6)}}{2}  +
\frac{C_{14,i}^{(6)}}{2}  - \frac{1 - 4 s_w^2}{2} \left( \frac{Y_\chi}{4}C_{17}^{(6)} +
C_{18}^{(6)} \right)\right]\,,
\\
\hat \C_{3,\ell_i}^{(6)}|_{n_f=5} &= \frac{1}{\Lambda^2}\left[-\frac{Y_\chi }{8} C_{9,i}^{(6)} - \frac{C_{10,i}^{(6)}}{2}  +
\frac{C_{11,i}^{(6)}}{2}  + \frac{1}{2} \left( \frac{Y_\chi}{4}C_{15}^{(6)} +
C_{16}^{(6)} \right)\right] \,,
\\
\hat \C_{4,\ell_i}^{(6)}|_{n_f=5} &= \frac{1}{\Lambda^2}\left[-\frac{Y_\chi }{8} C_{12,i}^{(6)} - \frac{C_{13,i}^{(6)}}{2}  +
\frac{C_{14,i}^{(6)}}{2}  +
	\frac{1}{2} \left( \frac{Y_\chi}{4}C_{17}^{(6)} + C_{18}^{(6)}
        \right)\right] + \ldots\,.\label{eq:match-6-lep-4} 
\end{align}
As before, $Z$-boson exchange due to the renormalizable gauge
coupling~\eqref{eq:gauge-int} leads to the additional contributions
$\hat \C_{1,\ell_i}^{(6)}|_{n_f=5} = - \frac{\pi\alpha_2}{2c_w^2
  m_Z^2} (1 - 4 s_w^2) Y_\chi$ and $\hat \C_{3,\ell_i}^{(6)}|_{n_f=5}
= \frac{\pi\alpha_2}{2c_w^2 m_Z^2} Y_\chi$. Also for leptons, $\hat
\C_{4,\ell_i}^{(6)}|_{n_f=5}$ receives a one-loop contribution from
gauge interaction that does not vanish for $Y_\chi=0$, see
Ref.~\cite{Hisano:2011cs}.

\begin{figure}[t]\centering
	\includegraphics[scale=0.8]{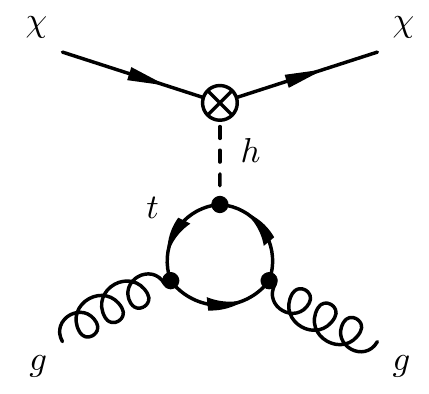}
	\caption{Matching contributions to dimension-seven effective
          operators involving gluons for $\mu<m_Z$ that arise
          from integrating out the top quark.}
	\label{fig:d7-gg}
\end{figure}

The dimension-seven operators receive contributions from both the
renormalizable electroweak interactions of the DM multiplet as well as
from the higher dimension operators. For the gluonic operators
$\Q_{1,2}^{(7)}$ the higher dimension UV operators give a contribution
after integrating out the top quark at one loop, see
Fig.~\ref{fig:d7-gg}. We then have
\begin{equation}\label{eq:match-dim5UV-dim7}
\hat \C_{1(2)}^{(7)}|_{n_f=5} = \frac{1}{\Lambda m_h^2}
\bigg(C_{3(7)}^{(5)} + \frac{Y_\chi}{4} C_{4(8)}^{(5)}\bigg) + \ldots
\,,
\end{equation}
and $\hat \C_{3(4)}^{(7)}|_{n_f=5}=0$.  Note that the loop factor is
already included in the definition of the operators $\Q_i^{(7)}$.  The
explicit top-quark mass dependence drops out because we expand to
leading (quadratic) order in the small external momenta. This
  limit is equivalent to the limit of heavy top mass in on-shell Higgs
  decays to two photons or gluons, where the non-decoupling of chiral
  fermions is a familiar result. The ellipsis denotes the two-loop
contributions from renormalizable electroweak interactions, see
Ref.~\cite{Hisano:2011cs}. 

For scalar operators we have
\begin{align}
\label{eq:match-5-7}
\hat \C_{5,f}^{(7)}|_{n_f=5} &= -\frac{1}{\Lambda m_h^2} \bigg(C_3^{(5)} +
\frac{Y_\chi}{4} C_4^{(5)}\bigg) + \ldots \,,
\\
  \label{eq:match-6-7}
\hat \C_{6,f}^{(7)}|_{n_f=5} &= -\frac{1}{\Lambda m_h^2}
\bigg(C_7^{(5)} + \frac{Y_\chi}{4}C_8^{(5)}\bigg)\,,
\end{align}
and $\hat \C_{7,f}^{(7)}|_{n_f=5}=\hat \C_{8,f}^{(7)}|_{n_f=5}=0$.
The right diagram in Fig.~\ref{fig:d6-matching} shows the tree-level
contributions from higher dimension operators. Here, the ellipsis
denotes the one-loop ``Higgs penguin'' contribution from gauge
interactions (see Fig.~\ref{fig:DD-Higgs_loop} and
Ref.~\cite{Hisano:2011cs}).

\subsection{Electroweak scale dark matter}
\label{subsec:EW:DM}
The case that the DM mass is comparable to the electroweak scale,
$m_\chi\sim {\mathcal O}(m_Z)$, needs to be treated separately. In
this case we integrate out at the electroweak scale, in addition to
the top quark, the Higgs and the $W, Z$ bosons, also the high-momentum
fluctuations of the DM field. In this way we arrive at the Heavy Dark
Matter Effective Theory (HDMET). The HDMET is an effective theory the
describes the scattering of a heavy DM, where the momenta exchanges
are much smaller than the DM mass, $q\ll m_\chi$. The HDMET uses the
HQET (Heavy Quark Effective Theory) formalism~\cite{Grinstein:1990mj,
  Eichten:1989zv, Georgi:1990um} applied to DM direct detection
physics~\cite{Hill:2013hoa}. The result is an effective theory where
terms are organized as an expansion in $1/m_\chi$. In practice the
HDMET description is not necessary if one performs electroweak
matching only at tree level. However, some of the one-loop matching
corrections are important phenomenologically and need to be kept,
requiring the use of HDMET.

The construction of the requisite terms in HDMET has been presented in
Ref.~\cite{Bishara:2016hek}. Here, we just collect the main results
and refer the interested reader to the literature for details.

The HDMET Lagrangian is given by%
%%%%
%%%%footnote
%%%%
\footnote{For very heavy DM, $m_\chi\gg m_Z$, the DM mass is
  integrated out before the weak gauge bosons \cite{Hill:2013hoa,
    Hill:2014yka, Hill:2014yxa, Berlin:2015njh}, giving
\begin{equation}\label{eq:HDMET:above EW}
{\cal L}_{\rm HDMET}= \bar \chi_v (i v \cdot D)
\chi_v+\frac{1}{2m_\chi}\bar \chi_v (i D_\perp)^2 \chi_v+ \frac{g_2
  c_W}{4 m_\chi} \bar \chi_v \sigma_{\mu\nu} \, \tilde \tau \ncdot
W^{\mu \nu}\chi_v + \frac{g_1 c_B}{4 m_\chi}\bar \chi _v
\sigma_{\mu\nu} B^{\mu\nu} \chi_v+\cdots,
\end{equation}
where at tree level $c_W=c_B=1$, and the ellipsis denotes terms of
higher order in $1/m_\chi$, as well as the $1/\Lambda$ suppressed
interactions. The covariant derivative contains the $W^a_\mu$ and
$B_\mu$ gauge fields, so that in the infinite mass limit the DM
multiplet, $\chi_v$, acts as a static source of the electroweak gauge
fields.}
%%%
%%% end of footnote
%%%
\begin{equation} 
\label{{eq:HDMET}}
{\cal L}_{\rm HDMET}= \bar \chi_v (i
v \cdot \partial) \chi_v
+\frac{1}{2m_\chi}\bar \chi_v (i \partial_\perp)^2 \chi_v+\ldots+\sum_d {\cal L}_{\chi_v}^{(d)}|_{n_f=5}.
\end{equation}
Here, $\chi_v$ denotes only the neutral component of the DM
electroweak multiplet, i.e., only the DM state. The first term is the
LO HDMET Lagrangian and describes an infinitely heavy DM particle, and
contains no explicit dependence on $m_\chi$. The ${\mathcal
  O}(1/m_\chi)$ term is fixed by reparametrization invariance
\cite{Luke:1992cs}, with ellipsis denoting terms of higher order in
the $1/m_\chi$ expansion.
 
The effective Lagrangians ${\cal L}_{\chi_v}^{(d)}|_{n_f=5}$ comprise
the interactions of DM with the SM. They are expanded in powers of
$1/m_\chi, 1/\Lambda$ and $1/m_Z$, mirroring the case of light
DM in Eq.~\eqref{eq:lightDM:Lnf5}. The only difference is that we now
denote explicitly at which order in $1/m_\chi$ the operators enter,
\begin{equation}\label{eq:ewDM:Lnf5}
{\cal L}_{\chi_v}^{(d)}|_{n_f=5}=\sum_{a,m}
\hat \C_{a}^{(d,m)}|_{n_f=5} {\cal Q}_a^{(d,m)}\,,
\end{equation}
such that $\hat \C_{a}^{(d,m)}|_{n_f=5} \propto (\Lambda, m_Z)^{4+m-d}
m_\chi^{-m}$, where $(\Lambda, m_Z)^{4+m-d}$ symbolizes a product of
powers of $\Lambda$ and $m_Z$ with total power $4+m-d$. The double
superscripts on $\hat \C_{a}^{(d,m)}$ and $ {\cal Q}_a^{(d,m)}$ thus
signal that they are defined in the HDMET, while a single superscript
on $\hat \C_{a}^{(d)}$ or on $ {\cal Q}_a^{(d)}$ means that we are
considering light DM.

The difference $d-m$ gives the {\em ``mediator dimensionality''}. This
is the dimension of the relativistic operator ${\cal Q}_a^{(d-m)}$
that gives the HDMET operator ${\cal Q}_a^{(d,m)}$ upon expanding the
DM currents to order $1/m_\chi^m$ (see Ref.~\cite{Bishara:2016hek,
  Bishara:2017pfq} for the explicit expressions).\footnote{Note that
  the $\Lambda\gg m_{Z}\sim m_{\chi}$ limit reduces the set of HDMET
  operators that are generated. For instance, at dimension seven the
  operator $(\bar \chi_v \chi_v) G^{a\mu\nu}G^{a}_{\mu\nu}$ arises in
  the matching, but not the operator $(\bar \chi_v \chi_v) v^\mu v^\nu
  G^{a}_{\mu\rho}G^{a\rho}_{\nu}$. The latter would arise from the
  dimension-nine UV operator $(\bar \chi \partial^\mu \partial^\nu
  \chi) G^{a}_{\mu\rho}G^{a\rho}_{\nu}$ and is thus
  $m_\chi^2/\Lambda^2$ suppressed. In contrast, for $m_\chi \sim
  \Lambda$ the two operators are of the same size, and thus both arise
  in the matching to HDMET at scale $\mu\sim \Lambda$ (see, e.g., the
  discussion of twist-two operators in Ref.~\cite{Hisano:2015bma}).}
We group the operators in terms of their mediator dimensionality,
$d-m$. The operators that arise at LO in $1/m_\chi$, i.e., for which
the Wilson coefficients start at order ${\mathcal O}(1/m_\chi^0)$, are
the HDMET counterparts of the operators in
Eqs.~\eqref{eq:dim5:nf5:Q1Q2:light}-\eqref{eq:dim7EW:Q7Q8:light}. The
two dimension-five operators in Eq.~\eqref{eq:dim5:nf5:Q1Q2:light} get
replaced by the HDMET operators
\begin{equation}
\label{eq:dim5:HDM:nf5:Q1Q2}
{\cal Q}_{1}^{(5,0)} = \frac{e}{4 \pi^2} \epsilon_{\mu\nu\alpha\beta} (\bar \chi_v S_\chi^{\alpha} v^\beta \chi_v)
 F^{\mu\nu}\,,\qquad
{\cal Q}_2^{(5,0)} = \frac{e }{2 \pi^2} (\bar
\chi_v S_\chi^{\mu}v^{\nu} \chi_v) F_{\mu\nu}\,.
\end{equation}
We also need the following two subleading operators
\begin{equation}
\label{eq:dim5:HDM:nf5:Q1NLO}
{\cal Q}_{1}^{(6,1)} = \frac{ie}{8\pi^2} \big( \bar\chi_v v^{\mu} \sigma_\perp^{\nu\rho} \lrpartial_{\rho} \chi_v \big)
 F_{\mu\nu}\,,\qquad
{\cal Q}_{2}^{(6,1)} = -\frac{e}{8\pi^2} \big( v^{\mu} \partial^{\nu} \bar\chi_v \chi_v\big)
F_{\mu\nu}\,, 
\end{equation}
since the presence of the photon pole in the interaction of the
magnetic dipole with the nuclear current requires that we go to the
second order in the expansion of the DM tensor current. We defined
$\sigma_\perp^{\mu\nu}=i [\gamma_\perp^\mu, \gamma_\perp^\nu]/2$,
$\gamma_\perp^\mu=\gamma^\mu -v^\mu \slashed v$, $\bar \chi_v
\lrpartial^\mu \chi_v=\bar \chi_v (\partial^\mu \chi_v)-
(\partial^\mu\bar \chi_v) \chi_v$, and $S^\mu=\gamma_\perp^\mu
\gamma_5/2 $ is the spin operator, while $v^\mu = (1,\vec 0\,)$ is the
velocity label of the nonrelativistic DM field (cf.
Ref.~\cite{Bishara:2016hek}).

At tree-level we have
\begin{align}
\hat {\cal C}_{1}^{(5,0)} |_{n_f=5}&\overset{\rm tree}{=} \hat \C_1^{(5)}|_{n_f=5}+\cdots\,,
& \hat \C_2^{(5,0)}|_{n_f=5}\overset{\rm tree}{=} &\hat \C_2^{(5)}|_{n_f=5}\,,
\\
\label{eq:tree:dim5:identity}
\hat {\cal C}_{1}^{(6,1)}|_{n_f=5}&\overset{\rm
  tree}{=}\frac{1}{m_\chi}\hat {\cal C}_{1}^{(5,0)}|_{n_f=5}\,, &\hat
     {\cal C}_{2}^{(6,1)}|_{n_f=5}\overset{\rm tree}{=}&
     \frac{1}{m_\chi}\hat {\cal C}_{1}^{(5,0)}|_{n_f=5}\,, 
\end{align}
where the equalities get corrections at loop level. Again, $\hat {\cal
  C}_{1}^{(5,0)} |_{n_f=5}$ receives a photon penguin contribution
proportional to $Y_\chi$, denoted by the ellipsis and omitted in the
following. The Wilson coefficients for the dipole operator in the case
of light DM, $\hat \C_{1,2}^{(5)}$, are given in
Eqs.~\eqref{eq:match:lightDM:C1:light},~\eqref{eq:match:lightDM:C2:light}.

The dimension-six operators of LO in $1/m_\chi$ are
\begin{align}
{\cal Q}_{1,f}^{(6,0)} & = (\bar \chi_v  \chi_v) (\bar f \slashed v f)\,,
& {\cal Q}_{2,f}^{(6,0)} &= 2 (\bar \chi_vS_{\chi,\mu} \chi_v)(\bar f \gamma^\mu f)\,,
\label{eq:dim6EW:Q1Q2:HDM}
  \\ 
{\cal Q}_{3,f}^{(6,0)} & = (\bar \chi_v \chi_v)(\bar f \slashed v \gamma_5 f)\,,
\qquad 
& {\cal Q}_{4,f}^{(6,0)}& = 2 (\bar
\chi_v S_{\chi,\mu} \chi_v)(\bar f \gamma^\mu \gamma_5 f)\,.
 \label{eq:dim6EW:Q3Q4:HDM}
\end{align}
In addition, we need the following $d-m=6$ operators that are
$1/m_\chi$ suppressed 
\begin{align}
{\cal Q}_{1,f}^{(7,1)} & = \frac{1}{2}(\bar \chi_v i \lrpartial_{\perp}^\mu \chi_v) (\bar f \gamma_\mu f)\,,
& {\cal Q}_{2,f}^{(7,1)} &= -i (\bar \chi_vS_\chi \cdot \lrpartial \chi_v)(\bar f \slashed v f)\,,
\label{eq:dim6EW:Q1Q2:HDM1}
  \\ 
{\cal Q}_{3,f}^{(7,1)} & = \frac{1}{2}(\bar \chi_v i \lrpartial_{\perp}^\mu \chi_v) (\bar f \gamma_\mu \gamma_5 f)\,,
\qquad 
& {\cal Q}_{4,f}^{(7,1)}& = -i (\bar
\chi_v S_\chi \cdot \lrpartial \chi_v)(\bar f \slashed v \gamma_5 f)\,,
 \label{eq:dim6EW:Q3Q4:HDM1}
 \\
 {\cal Q}_{5,f}^{(7,1)} & = \frac{1}{2}\partial_\nu (\bar \chi_v \sigma_\perp^{\mu\nu} \chi_v) (\bar f \gamma_\mu f)\,,
& {\cal Q}_{6,f}^{(7,1)} & = \frac{1}{2}\partial_\nu (\bar \chi_v \sigma_\perp^{\mu\nu} \chi_v) (\bar f \gamma_\mu \gamma_ 5f)\,,
\label{eq:dim6EW:Q5Q6:HDM1}
\end{align}
where our convention is that the derivatives act only within the
brackets or on the nearest bracket. The ${\cal Q}_{1,f}^{(7,1)}, {\cal
    Q}_{2,f}^{(7,1)}$ operators do not enter the phenomenological
  analysis, but we keep them for completeness and transparency of
  notation. For the matching conditions we
have
\begin{equation}
\label{eq:rel:C:dim6}
\hat {\cal C}_{i,f}^{(6,0)}|_{n_f=5}=m_\chi \hat {\cal
  C}_{i,f}^{(7,1)}|_{n_f=5}\overset{\rm tree}{=}\hat {\cal C}_{i,f}^{(6)}|_{n_f=5}\,,
\quad i=1,\ldots,4\,; 
\end{equation}
and in addition
\begin{equation}
 {\cal C}_{5,f}^{(7,1)}|_{n_f=5}\overset{\rm tree}{=}\frac{1}{m_\chi}\hat {\cal C}_{1,f}^{(6)}|_{n_f=5}\,, 
 \qquad
  {\cal C}_{6,f}^{(7,1)}|_{n_f=5}\overset{\rm tree}{=}\frac{1}{m_\chi}\hat {\cal C}_{3,f}^{(6)}|_{n_f=5}\,. 
\end{equation}
Note that the equalities denoted by ``tree'' are only valid for
tree-level matching, while the remaining relations are valid to all
orders due to reparametrization invariance. The light DM Wilson
coefficients $C_{i,f}^{(6)}$ are given in
Eqs.~\eqref{eq:match-6-up-1}-\eqref{eq:match-6-lep-4} .

The relevant dimension-seven operators in
Eqs.~\eqref{eq:dim7:Q1Q2:light}-\eqref{eq:dim7EW:Q7Q8:light} involve
scalar and pseudoscalar DM currents. The HDMET scalar current operator
starts at ${\mathcal O}(1/m_\chi^0)$, while pseudoscalar current
starts at ${\mathcal O}(1/m_\chi)$, see
Ref.~\cite{Bishara:2016hek}. We thus define the following $d-m=7$
HDMET operators

\begin{align}
{\cal Q}_1^{(7,0)} & = \frac{\alpha_s}{12\pi} (\bar \chi_v \chi_v)
 G^{a\mu\nu}G_{\mu\nu}^a\,, 
 & {\cal Q}_2^{(8,1)} &= \frac{\alpha_s}{12\pi} \partial_\mu \big(\bar \chi_v S_\chi^\mu \chi_v\big) G^{a\mu\nu}G_{\mu\nu}^a\,,\label{eq:dim7:Q1Q2:HDM}
 \\
{\cal Q}_3^{(7,0)} & = \frac{\alpha_s}{8\pi} (\bar \chi_v \chi_v) G^{a\mu\nu}\widetilde
 G_{\mu\nu}^a\,, 
& {\cal Q}_4^{(8,1)}& = \frac{\alpha_s}{8\pi}
\partial_\mu \big(\bar \chi_v S_\chi^\mu \chi_v\big) G^{a\mu\nu}\widetilde G_{\mu\nu}^a \,, \label{eq:dim7:Q3Q4:HDM}
\\
{\cal Q}_{5,f}^{(7,0)} & = m_f (\bar \chi_v \chi_v)( \bar f f)\,, 
&{\cal
  Q}_{6,f}^{(8,1)} &= m_f \partial_\mu \big(\bar \chi_v S_\chi^\mu \chi_v\big) ( \bar f f)\,,\label{eq:dim7EW:Q5Q6:HDM}
  \\
{\cal Q}_{7,f}^{(7,0)} & = m_f (\bar \chi_v \chi_v) (\bar f i \gamma_5 f)\,, 
&{\cal Q}_{8,f}^{(8,1)} & = - m_f \partial_\mu \big(\bar \chi_v S_\chi^\mu \chi_v\big)(\bar f i \gamma_5
f)\,. \label{eq:dim7EW:Q7Q8:HDM} 
\end{align}
The top-quark loop contributions to the gluonic operators,
Eq.~\eqref{eq:dim7:Q1Q2:HDM} and~\eqref{eq:dim7:Q3Q4:HDM}, are the
same as in Eq.~\eqref{eq:match-dim5UV-dim7}, so that
\begin{equation}
\label{eq:match-dim5UV-dim7:HDM}
\hat \C_1^{(7,0)}|_{n_f=5} = \hat \C_1^{(7)}|_{n_f=5} \,, \qquad \hat \C_2^{(8,1)}|_{n_f=5} =\hat \C_2^{(7,0)}|_{n_f=5}\,.
\end{equation}
The Wilson coefficients $\hat \C_{3}^{(7,0)}|_{n_f=5}$ and $\hat
\C_4^{(8,1)}|_{n_f=5}$ vanish.

The Wilson coefficients for the scalar operators are
\begin{equation}
  \label{eq:match-5-7-HDMET}
\hat \C_{5,f}^{(7,0)}|_{n_f=5} = \hat \C_{5,f}^{(7)}|_{n_f=5}\,,
\qquad
\hat \C_{6,f}^{(8,1)}|_{n_f=5} = \frac{1}{m_\chi} \hat \C_{6,f}^{(7)}|_{n_f=5}\,,
\end{equation}
while $\hat \C_{7,f}^{(7,0)}|_{n_f=5}=\hat
\C_{8,f}^{(8,1)}|_{n_f=5}=0$. The dimension-five UV operators
$Q_{3,4}^{(5)}$ in Eq.~\eqref{Q34} and $Q_{7,8}^{(5)}$ in
Eq.~\eqref{Q78} contribute through a Higgs exchange at tree level, see
Fig.~\ref{fig:d6-matching} (right panel), and give the same matching
conditions as in the case of light DM, Eqs.~\eqref{eq:match-5-7}
and~\eqref{eq:match-6-7}. Note that within this subsection, the full
(unexpanded) results of Ref.~\cite{Hisano:2011cs} should be used.

The following twist-two operators are needed for the two-loop
electroweak matching contributions (the numbering is chosen such that
we avoid inconsistencies with the numbering in
Ref.~\cite{Brod:2017bsw}):
\begin{align}\label{eq:twist2:op}
  {\cal Q}_{23,q}^{(7,0)} & = \tfrac{1}{2} (\bar \chi_v \chi_v)
  \big[\bar q \big(\slashed{v} \, i \!\! \stackrel{\leftrightarrow}{D} \!\! \cdot v
    - \tfrac{1}{4} i \!\! \stackrel{\leftrightarrow}{\slashed{D}}\big) q\big]\,, \\
  {\cal Q}_{25}^{(7,0)} & = (\bar \chi_v \chi_v)
  \big[\tfrac{1}{4} G_{\alpha\beta}^a G^{a,\alpha\beta}
    - v_\mu v_\nu G^{a,\mu\lambda} G\indices{^{a,\nu}_{\lambda}}\big]\,.
\end{align}
The first operator, ${\cal Q}_{23,q}^{(7,0)}$, receives a
non-vanishing matching contribution at the electroweak scale. It can
be extracted from Ref.~\cite{Hisano:2011cs} if in their results one
takes the leading HDMET limit of the DM bilinears $\bar \chi i
\partial^\mu \chi \to m_\chi \, \bar \chi_v v^\mu \chi_v$ and $\bar
\chi \gamma^\mu \chi \to \bar \chi_v v^\mu \chi_v$. It is then given
by ${\cal C}_{23,q}^{(7,0)} = g_q^{(1)} + g_q^{(2)}$, with the loop
functions given in Ref.~\cite{Hisano:2011cs}. The operator ${\cal
  Q}_{25}^{(7,0)}$ does not receive an initial condition at the weak
scale, but is generated by QCD RG evolution below the weak scale, to
be discussed in the following section.

\subsection{RG running below the electroweak scale}\label{sec:RG:below:ew}

The matching at $\mu\sim \mu_{\rm EW}$ is followed by the QCD and QED
RG running from $\mu_{\rm EW}$ to $\mu_c\sim\mu_{\rm had}\sim 2$
GeV. The five-flavor theory below $\mu_{\rm EW}$ is matched onto the
four-flavor theory at the bottom quark threshold, $\mu_b$, and then
onto the three-flavor theory at the charm quark threshold, $\mu_c$;
see Eq.~\eqref{eq:evol:sol}. There is no running in the three flavor
basis because of our choice of scales, $\mu_c=\mu_{\rm had}$. This RG
evolution was discussed in detail in Ref.~\cite{Hill:2014yxa}. For
completeness and convenience we convert the results of
Ref.~\cite{Hill:2014yxa} to our notation. (See also
Ref.~\cite{Bishara:2017nnn} for a computer implementation of the RG
evolution, as well as Ref.~\cite{DEramo:2016gos} for the case of
vector mediators.)

\begin{figure}[t]\centering
	\includegraphics[scale=0.8]{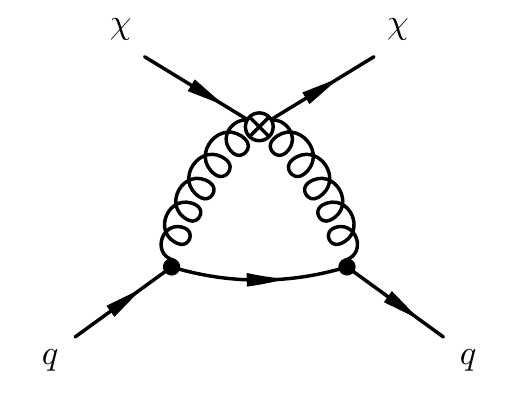}
	\includegraphics[scale=0.8]{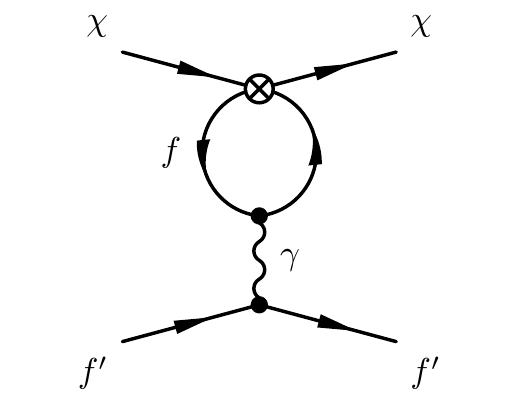}
	\caption{\emph{Left panel:} the mixing of the gluonic
          operators into operators with scalar and pseudoscalar quark
          currents. \emph{Right panel:} the mixing of dimension-six
          four-fermion operators into each other via the photon
          penguin insertion.}
	\label{fig:d6-photon-peng}
\end{figure}

\paragraph*{QCD running.}
Since the vector currents are conserved, $\hat C_{1,q}^{(6)}$ and
$\hat C_{3,q}^{(6)}$ in Eqs.~\eqref{eq:dim6EW:Q1Q2:light}
and~\eqref{eq:dim6EW:Q3Q4:light} do not run. Moreover, the axial
currents have vanishing anomalous dimensions at
$\mathcal{O}(\alpha_s)$ and so the Wilson coefficients $\hat
C_{2,q}^{(6)}$ and $\hat C_{4,q}^{(6)}$ in
Eqs.~\eqref{eq:dim6EW:Q1Q2:light} and~\eqref{eq:dim6EW:Q3Q4:light} do
not run at one-loop order. At dimension seven, the only non-zero
effect is the mixing of the gluonic operators,
Eqs.~\eqref{eq:dim7:Q1Q2:light}-\eqref{eq:dim7:Q3Q4:light}, into the
scalar operators Eq.~\eqref{eq:dim7EW:Q5Q6:light} -- see left panel in
Fig.~\ref{fig:d6-photon-peng} -- with anomalous dimension
(cf. Ref.~\cite{Hill:2014yxa})
\begin{equation}
\big[\gamma_s^{(1)}\big]_{\mathcal{Q}_{1\cdots 4}^{(7)} \times \mathcal{Q}_{5,q\cdots 8,q}^{(7)}} =  
8\,\diag (C_F,C_F,-1,-1)\,,
\end{equation}
where $q$ runs over active quark flavors and $C_F = 4/3$.
This anomalous dimension arises at $\mathcal{O}(\alpha_s^2)$
since the $GG$ operators are defined with an additional factor of
$\alpha_s$ to reflect the fact that they are loop generated, and thus,
\begin{equation}
\big[\gamma_s^{(0)}\big]_{\mathcal{Q}_{1\cdots 4}^{(7)} \times \mathcal{Q}_{5,q\cdots 8,q}^{(7)}} = 0\,.
\end{equation}
The mixing of the two operators ${\cal Q}_{23,q}^{(7,0)}$ and ${\cal
  Q}_{25}^{(7,0)}$ in Eq.~\eqref{eq:twist2:op} is given by the
anomalous dimension matrix
\begin{equation}
\big[\gamma_s^{(0)}\big]_{\mathcal{Q}_{23,q}^{(7,0)} \mathcal{Q}_{25}^{(7,0)} \times \mathcal{Q}_{23,q}^{(7,0)} \mathcal{Q}_{25}^{(7,0)}} =
\begin{pmatrix}
\frac{64}{9} & -\frac{4}{3} \\
-\frac{64}{9} & \frac{4}{3} N_f \\
\end{pmatrix}
\,.
\end{equation}

\paragraph*{QED running.} In general, the QED contribution to the RG
evolution can be neglected due to the smallness of the electromagnetic
coupling constant. The one exception are the off-diagonal mixings of
the operators $\Q_{1,f}^{(6)}$ for different fermion flavors $f$ (and
similarly for $\Q_{2,f}^{(6)}$), induced by the photon penguin
diagrams, see Fig.~\ref{fig:d6-photon-peng}. These mixings lead to
nonzero scattering on nuclei even if DM couples only to leptons at
leading order~\cite{Kopp:2009et, DEramo:2017zqw}.  The conservation of
parity forbids the mixing of $\Q_{1,f}^{(6)}$ into $\Q_{2,f}^{(6)}$
and vice versa.  The required anomalous dimensions can be found in
Ref.~\cite{D'Eramo:2014aba}, and are
\begin{equation}
\big[\gamma_e^{(0)}\big]_{\mathcal{Q}_{1,f}^{(6)}, \mathcal{Q}_{1,f'}^{(6)}} =
\big[\gamma_e^{(0)}\big]_{\mathcal{Q}_{2,f}^{(6)}, \mathcal{Q}_{2;f'}^{(6)}} =
\frac{8}{3}\,Q_f\,Q_{f'}\,N_c^f\,, 
\end{equation}
where $Q_f$ is the electric charges of the SM fermion
$f$, while $N_c^f=1 (3)$, if $f$ is a lepton
(quark).  In analogy with Eq.~(\ref{eq:ADMnotation}), we use the
notation
\begin{equation}
\gamma_e = \frac{\alpha}{4\pi}\gamma_e^{(0)} + \ldots\,,
\end{equation}
where the ellipsis denotes higher orders.

Finite corrections arise at each heavy flavor threshold. Beside the
usual threshold corrections to $\alpha_s$ (see, e.g.,
Ref.~\cite{Chetyrkin:1997un}), there are also finite threshold
corrections for the operators in
Eqs.~\eqref{eq:dim7:Q1Q2:light}-\eqref{eq:dim7:Q3Q4:light}, where at
$\mu=\mu_b$,
\begin{equation}
\begin{split}\label{eq:threshold:mb}
  \hat \C_{1(2)}^{(7)}|_{n_f=4} (\mu_b) &= \hat \C_{1(2)}^{(7)}|_{n_f=5} (\mu_b) - 
  \hat \C_{5,b(6,b)}^{(7)}|_{n_f=5} (\mu_b) \,, \\
  \hat \C_{3(4)}^{(7)}|_{n_f=4} (\mu_b) &= \hat \C_{3(4)}^{(7)}|_{n_f=5} (\mu_b) +
  \hat \C_{7,b(8,b)}^{(7)}|_{n_f=5} (\mu_b) \,, 
\end{split}
\end{equation}
while at $\mu=\mu_c$,
\begin{equation}
\begin{split}\label{eq:threshold:mc}
  \hat \C_{1(2)}^{(7)}|_{n_f=3} (\mu_c) &= \hat \C_{1(2)}^{(7)}|_{n_f=4} (\mu_c) - 
  \hat \C_{5,c(6,c)}^{(7)}|_{n_f=4} (\mu_c) \,, \\
  \hat \C_{3(4)}^{(7)}|_{n_f=3} (\mu_c) &= \hat \C_{3(4)}^{(7)}|_{n_f=4} (\mu_c) +
  \hat \C_{7,c(8,c)}^{(7)}|_{n_f=4} (\mu_c) \,,
\end{split}
\end{equation}
such that the effects of the heavy quarks appear, at low energies, as
additional contributions to the gluonic
operators, Eqs.~\eqref{eq:dim7:Q1Q2:light}-\eqref{eq:dim7:Q3Q4:light}. All
the other Wilson coefficients cross the thresholds continuously, $\hat
\C_i^{(d)}|_{n_f-1} = \hat \C_i^{(d)}|_{n_f}$.

\subsection{DM interactions with nucleons and nuclei}

The final step in the RG evolution is the matching at $\mu\sim
\mu_{\rm had}$ onto an effective theory describing interactions of DM
with nonrelativistic protons and neutrons. The momenta exchanged in
direct detection experiments are $q\lesssim 200$ MeV, with a typical
value of $20-60$ MeV, which is well below the chiral symmetry breaking
scale $4 \pi f_\pi\sim m_N$. One can thus use chiral perturbation
theory (ChPT) to organize different contributions in terms of an
expansion in $(q/4\pi f_\pi)^n$, see Refs.~\cite{Cirigliano:2012pq,
  Hoferichter:2015ipa, Menendez:2012tm, Klos:2013rwa, Baudis:2013bba,
  Vietze:2014vsa, Bishara:2016hek, Bishara:2017pfq}.  The
leading-order contributions come from the interactions involving a DM
field and a single nucleon inside the nucleus (these can still be
coherently summed over all the neutrons and protons in the nucleus).

The effective Lagrangian for DM scattering on nonrelativistic nucleons
(see Refs.~\cite{Anand:2013yka, Fitzpatrick:2012ib,
  Fitzpatrick:2012ix, Bishara:2017pfq}),
\begin{equation}\label{eq:LNR}
{\cal L}_{\rm NR}=\sum_{i,N} c_i^N(q^2) \op_i^N\,,
\end{equation}
contains 14 operators with up to two derivatives which are needed to
describe the chirally leading interactions. The momentum-independent
nonrelativistic operators are
\begin{align}
\label{eq:O1pO4p}
{\mathcal O}_1^N&= \mathbbm{1}_\chi \mathbbm{1}_N\,,
&{\mathcal O}_4^N&= \vec S_\chi \cdot \vec S_N \,,
\end{align}
while the relevant subset of momentum-dependent operators consists of
\begin{align}
\label{eq:O5pO6p}
{\mathcal O}_5^N&= \vec S_\chi \cdot \Big(\vec v_\perp \times \frac{i\vec q}{m_N} \Big) \, \mathbbm{1}_N \,,
&{\mathcal O}_6^N&= \Big(\vec S_\chi \cdot \frac{\vec q}{m_N}\Big) \, \Big(\vec S_N \cdot \frac{\vec q}{m_N}\Big)\,,
\\
\label{eq:O7pO8p}
{\mathcal O}_7^N&= \mathbbm{1}_\chi \, \big( \vec S_N \cdot \vec v_\perp \big)\,,
&{\mathcal O}_8^N&= \big( \vec S_\chi \cdot \vec v_\perp \big) \, \mathbbm{1}_N\,,
\\
\label{eq:O9pO11p}
{\mathcal O}_9^N&= \vec S_\chi \cdot \Big(\frac{i\vec q}{m_N} \times \vec S_N \Big)\,,
&{\mathcal O}_{11}^N&= - \Big(\vec S_\chi \cdot \frac{i\vec q}{m_N} \Big) \, \mathbbm{1}_N \,,
\end{align}
with $N=p,n$. We use the conventions of \cite{Bishara:2017pfq,
  Bishara:2017nnn}, so that
\begin{equation}
\label{eq:kinematics:qvperp}
\vec q = \vec k_2-\vec k_1=\vec p_1 -\vec p_2\,, \qquad \vec v_\perp=
\frac{\vec p_1+\vec p_2}{2{m_\chi}} - \frac{\vec k_1+\vec
  k_2}{2{m_N}}\,,
\end{equation}
where $\vec p_{1(2)}$ and $\vec k_{1(2)}$ are the incoming (outgoing)
nucleon and DM three-momenta, respectively.

The coefficients of the two momentum-independent operators
\eqref{eq:O1pO4p} are, schematically,
\begin{align}
\label{eq:NR-coeff-scaling:c1p}
c_{1}^{N} & \sim \frac{C_{1,\dots,4,f;15,16}^{(6)}}{\Lambda^2} + \frac{2}{27}\frac{ m_N}{\Lambda
  m_h^2}C_{3,4}^{(5)} + \frac{\sigma_q}{\Lambda m_h^2}C_{3,4}^{(5)} + \frac{\alpha}{\Lambda m_\chi}C_{1,2}^{(5)} \,, 
  \\
  \label{eq:NR-coeff-scaling:c4p}
c_{4}^{N}& \sim \frac{C_{5,\dots,8,f;17,18}^{(6)}}{\Lambda^2} + \frac{\alpha}{\Lambda m_N}C_{1,2}^{(5)}\,.
\end{align}
At leading chiral order one also has the contributions from the
operators with two derivatives, $\op_{5,6}^N$, whose coefficients are
\begin{align}
\label{eq:NR-coeff-scaling:c5p}
c_{5}^{N} & \sim \delta_{N,p} \frac{\alpha m_N}{\Lambda q^2}C_{1,2}^{(5)}\,,
\qquad
&c_{6}^{N} & \sim 
\frac{m_N^2}{m_\pi^2} \frac{C_{5,\dots,8,f;17,18}^{(6)}}{\Lambda^2}+ \frac{\alpha m_N}{\Lambda q^2}C_{1,2}^{(5)}\,. 
\end{align}
The sums in
Eqs.~\eqref{eq:NR-coeff-scaling:c1p}-\eqref{eq:NR-coeff-scaling:c5p}
are to be understood in the scaling sense, i.e., we only indicate a
rough order of magnitude for the contribution of each of the UV Wilson
coefficients, $C_a^{(d)}$.  Above we equated the weak scale with
$\mu_{\rm EW}\sim m_h\sim m_Z$. The complete expressions can be
obtained, for instance, from Refs.~\cite{Bishara:2017nnn,
  Bishara:2017pfq}, using the matching results given in
Sections~\ref{subsec:lightDM} and~\ref{subsec:EW:DM}. The additional
contributions arising from the twist-two operators are collected in
App.~\ref{sec:cNR}.

The $\op_1^N$ operator receives contributions from the
vector$\times$vector parts of the operators $Q_{1,f}^{(6)},\ldots
Q_{4,f}^{(6)}$, Eqs.~\eqref{eq:dim6:Q15}-\eqref{eq:dim6:Q48}, and from
tree-level $Z$ exchange due to the $Q_{15,16}^{(6)}$ operators,
Eqs.~\eqref{eq:dim6:Q15Q17} and~\eqref{eq:dim6:Q16Q18}. The analogous
operators with an axial-vector DM current, $Q_{5,f}^{(6)},\ldots
Q_{8,f}^{(6)}$ and $Q_{17,18}^{(6)}$, lead to spin--spin coupling in
the nonrelativistic limit, and contribute to both $\op_4^N$ and
$\op_6^N$. The two contributions are parametrically of the same order,
since the coefficient $c_6^N$ is enhanced by the pion pole, which
compensates the ${\mathcal O}(q^2)$ suppression of $\op_6^N$ for
$q^2/m_\pi^2\sim {\mathcal O}(1)$ (numerically, the compensation is
still only partial for electroweak scale DM~\cite{Bishara:2017pfq}).
The dipole operators $Q_{1,2}^{(5)}$ give contributions to all four
nonrelativistic operators, while the scalar operators $Q_{3,4}^{(5)}$
give leading contributions only to $Q_1^N$, through tree-level Higgs
exchange. The parameters $\sigma_q$ in
Eq.~\eqref{eq:NR-coeff-scaling:c1p} are related to the matrix elements
of $\bar q q$ quark scalar currents and are of order ${\cal
  O}(20-40)\,$MeV.

The coefficients of the single-derivative operators,
Eqs.~\eqref{eq:O7pO8p} and~\eqref{eq:O9pO11p}, are schematically
\begin{align}
\label{eq:NR-coeff-scaling:c3p1}
c_{7}^{N} & \sim \frac{C_{1,\dots,4,f}^{(6)}}{\Lambda^2}\,, 
\\
\label{eq:NR-coeff-scaling:c4p1}
c_{8}^{N} & \sim \frac{C_{5,\dots,8,f}^{(6)}}{\Lambda^2}\,, 
\\
\label{eq:NR-coeff-scaling:c5p1}
c_{9}^{N} & \sim \frac{C_{5,\dots,8,f}^{(6)}}{\Lambda^2} + \frac{m_N}{m_\chi}\frac{C_{1,\dots,4,f}^{(6)}}{\Lambda^2}\,,
\\
\label{eq:NR-coeff-scaling:c2p1}
c_{11}^{N} & \sim 
\frac{2}{27} \frac{m_N^2}{\Lambda m_h^2 m_\chi}C_{7,8}^{(5)} + \frac{\sigma_q m_N}{\Lambda m_h^2 m_\chi}C_{7,8}^{(5)} + \frac{\alpha m_N}{\Lambda\,q^2} \delta_{N,p} C_{5,6}^{(5)}\,. 
\end{align}
The coefficients $c_{7}^{N}$ and $c_{8}^{N}$ arise from
vector$\times$axial and axial$\times$vector parts of the operators
$Q_{1,f}^{(6)},\ldots Q_{4,f}^{(6)}$ and $Q_{5,f}^{(6)},\ldots
Q_{8,f}^{(6)}$, respectively, while all of these contribute to
$c_{9}^{N}$. Since these operators are momentum (velocity) suppressed,
they will give subleading contributions to the scattering rates,
unless the leading contributions (to $c_{1}^N$ from the
vector$\times$vector parts, and to $c_{4,6}^N$ from axial$\times$axial
parts) cancel. In the next section we will discuss in more detail how
realistic this is. The operators $\op_{7}^{N}, \ldots, \op_{9}^{N}$,
also receive contributions from $Q_{15}^{(6)},\ldots Q_{ 18}^{(6)}$
due to $Z$ exchange, where no such cancellation can occur. These
contributions are thus always subleading and were neglected in
Eqs.~\eqref{eq:NR-coeff-scaling:c3p1}-\eqref{eq:NR-coeff-scaling:c5p1}. The
operators $Q_{7,8}^{(5)}$ lead to $q^2$-suppressed contributions to
the scattering rate in the nonrelativistic limit, while
$Q_{5,6}^{(5)}$ induce an electric dipole moment for DM, giving a
$1/q^2$-enhanced direct detection scattering rate~\cite{Banks:2010eh}.

If the EFT above the weak scale is extended to mass dimension seven,
then also the nonrelativistic operators, ${\mathcal O}_{10}^N= -
\mathbbm{1}_\chi \, \big(\vec S_N \cdot {i\vec q}/{m_N} \big)$,
${\mathcal O}_{12}^N= \vec S_\chi \cdot \big( \vec S_N \times \vec
v_\perp \big)$, ${\mathcal O}_{14}^N = -\big(\vec S_\chi \cdot i\vec
q/m_N \big) \, \big(\vec S_N\cdot \vec v_\perp \big)$ become
phenomenologically important~\cite{Brod:2017bsw}. They arise from
dimension-seven operators with tensor DM currents and from
interactions of DM with the $G\tilde G$ current. The scaling estimates
for the corresponding coefficients are $c_{10}^{N} \sim {\mathcal
  O}\big({m_N}/{\Lambda^3}\big)$, $c_{12}^{N} \sim {\mathcal
  O}\big({m_q}/{\Lambda^3}\big)$, $c_{14}^{N} \sim {\mathcal
  O}\big({m_N}/{\Lambda^3}\big)$, setting the dimensionless Wilson
coefficients to unity.

Having obtained the coefficients in the effective Lagrangian for DM
scattering on nonrelativistic nucleons, Eq.~\eqref{eq:LNR}, the final
step is to calculate the DM--nucleus scattering cross section
\cite{Anand:2013yka, Fitzpatrick:2012ib, Fitzpatrick:2012ix},
\begin{equation}
\label{eq:dsigmadER}
\begin{split}
\frac{d\sigma}{d E_R} = \frac{2 m_A}{(2 j_A+1)v^2}\sum_{\tau,\tau'}
\bigg[ & R_M^{\tau\tau'}W_M^{\tau\tau'} +
  R_{\Sigma''}^{\tau\tau'}W_{\Sigma''}^{\tau\tau'} +
  R_{\Sigma'}^{\tau\tau'}W_{\Sigma'}^{\tau\tau'} \\ & + \frac{\vec
    q^{\,\,2}}{m_N^2}\Big(R_{\Delta}^{\tau\tau'}W_{\Delta}^{\tau\tau'}+R_{\Delta
    \Sigma'}^{\tau\tau'}W_{\Delta\Sigma''}^{\tau\tau'}\Big)\bigg]\,. 
\end{split}
\end{equation}
Here, $E_R$ is the recoil energy of the nucleus, $m_A$ the mass of the
nucleus, $j_A$ its spin, and $v$ the initial DM velocity in the lab
frame. The kinematic factors contain the $c_i^N$ coefficients,
\begin{align}
R_M^{\tau\tau'}&=c_1^\tau c_1^{\tau'}+\frac{1}{4}\Big(\frac{\vec q^{\,\,2}}{m_N^2}\vec v_T^{\perp 2} c_5^\tau c_5^{\tau'} +\vec v_T^{\perp 2}c_8^\tau c_8^{\tau'}+\frac{\vec q^{\,\,2}}{m_N^2} c_{11}^\tau c_{11}^{\tau'}\Big)\,,
\\
R_{\Sigma''}^{\tau\tau'}&=\frac{1}{16}\Big(c_4^{\tau}c_4^{\tau'}+\frac{\vec q^{\,\,2}}{m_N^2}\big(c_4^\tau c_6^{\tau'}+c_6^\tau c_4^{\tau'}\big) + \frac{\vec q^{\,\,4}}{m_N^4}c_6^\tau c_6^{\tau'}\Big)\,,
\\
R_{\Sigma'}^{\tau\tau'}&=\frac{1}{8}\vec v_T^{\,\,2} c_7^\tau c_7^{\tau'}+\frac{1}{16}\Big(c_4^\tau c_4^{\tau'}+\frac{\vec q^{\,\,2}}{m_N^2}c_9^\tau c_9^{\tau'}\Big)\,,
\\
R_{\Delta}^{\tau\tau'}&=\frac{1}{4}\Big(\frac{\vec q^{\,\,2}}{m_N^2}c_5^\tau c_5^{\tau'}+c_8^\tau c_8^{\tau'}\Big)\,,
\\
R_{\Delta\Sigma'}^{\tau\tau'}&=\frac{1}{4}\Big(c_5^\tau c_4^{\tau'}-c_8^\tau c_9^{\tau'}\Big)\,,
\end{align}
where $\vec v_T^\perp$ is defined as in
Eq.~\eqref{eq:kinematics:qvperp}, but with the nucleus replacing the
nucleon. The sum in \eqref{eq:dsigmadER} is over isospin,
$\tau,\tau'=0,1$, so that
\begin{equation}
c_i^{0(1)}=\frac{1}{2}\big(c_i^p\pm c_i^n\big)\,.
\end{equation}

The nuclear response functions depend on $|\vec q\,|$ and have the
approximate scaling (see, e.g., Fig. 2 in \cite{Kang:2018odb})
\begin{equation}\label{eq:scaling}
W_M\sim {\mathcal O}(A^2)\,, \qquad W_{\Sigma'}, W_{\Sigma''},
W_{\Delta}, W_{\Delta\Sigma'}\sim{\mathcal O}(10^{-2})-{\mathcal O}(1)\,.
\end{equation}
The $W_{\Sigma'}, W_{\Sigma''}, W_{\Delta}$, and $W_{\Delta\Sigma'}$
response functions depend strongly on the detailed properties of
nuclei, for instance, whether or not they have an un-paired nucleon in
the outer shell. Here $W_{\Sigma',\Sigma''}$ measure the spin content
of the nucleus, $W_{\Delta}$ the average angular momentum in the
nucleus, and $W_{\Delta\Sigma'}$ the interference of the two. Their
sizes can thus differ drastically between different isotopes of the
same element.
  
The $W_M$ response function encodes the coherent scattering
enhancement, ${\mathcal O}(A^2)$, where $A$ is the atomic mass
number. This is achieved in the long-wavelength limit, $q\to 0$, where
DM scatters coherently on the whole nucleus, for instance, due to the
$\op_1^N$ contact interaction. The coherent scattering due to
$\op_5^N$ is ${\mathcal O}(q^2 v_T^2)$ suppressed. However, since its
coefficient is $1/q^2$ enhanced, the corresponding contribution is of
leading order~\cite{Bishara:2017pfq}. The contributions due to
$\op_{8,11}^N$, though coherently enhanced, are at the same time
velocity suppressed.

%% file: EffectOfRG.tex
%!TEX root = paper.tex

\section{The Effects of RG running}
\label{sec:RG-run}

The impact of the mixing of electroweak operators on the scattering
cross section depends on two factors: {\it i)} the structure of the
anomalous dimension and thus the sizes of the induced Wilson
coefficients in Eqs.~\eqref{Q12}-\eqref{eq:dim6:Q16Q18}, and {\it ii)}
on the sizes of the nuclear response functions,
Eq.~\eqref{eq:scaling}, for each of the operators involved in the
mixing. In Section~\ref{sec:eff:norun} we first give the scalings of
the scattering cross sections {\em without} mixing effects, for
several benchmark choices of UV Wilson coefficients. In
Section~\ref{sec:eff:run} we then include the mixing and perform the
actual numerical analysis using the full expressions for the
DM-nucleus scattering cross sections derived in the sections above.

\subsection{Low energy phenomenology ignoring RG running}\label{sec:eff:norun}
We first estimate the size of the DM--nucleus scattering cross section
induced by each of the UV operators, Eqs.~\eqref{Q12}-\eqref{Q78} and
\eqref{eq:dim6:Q15}-\eqref{eq:dim6:Q16Q18}, neglecting the RG
running. In the estimates we use the scaling for nonrelativistic
Wilson coefficients in
Eqs.~\eqref{eq:NR-coeff-scaling:c1p}-\eqref{eq:NR-coeff-scaling:c2p1},
and the rough scalings of nuclear response functions in
Eq.~\eqref{eq:scaling}, but setting for simplicity $W_{\Sigma'},
W_{\Sigma''}, W_{\Delta}, W_{\Delta\Sigma'}\sim{\mathcal O}(1)$, with
the knowledge that the sizes of the latter contributions have high
variations between different target materials.

\addtocontents{toc}{\SkipTocEntry}
\subsubsection{Magnetic or electric dipole operators}

The DM magnetic dipole operators, $Q_{1,2}^{(5)}\sim (\bar \chi
\sigma_{\mu\nu}\chi)\{B^{\mu\nu}, W^{\mu\nu}\}$, Eq.~\eqref{Q12},
induce both spin-dependent and spin-independent interactions. These
give parametrically similar contributions to the DM--nucleus
scattering cross section. Schematically,
\begin{equation}
\begin{split} \label{eq:dipole:op:matchings}
\frac{d\sigma}{dE_R}&\sim \left( c_{1}^{2}+ \vec v_T^{\perp 2}\frac{\vec q^{\,\,2}}{m_N^2} c_{5}^{2}\right) W_M +\left\{c_{4}, \frac{\vec q^{\,\,2}}{m_N^2} c_{6}\right\}^2 W_{\Sigma',\Sigma''} +\left\{\frac{\vec q^{\,\,2}}{m_N^2} c_5, c_4\right\}^2\, W_{\Delta,\Delta\Sigma'}
\\
&\sim
\bigg(\frac{\alpha C_{1,2}^{(5)}}{\Lambda}\bigg)^2\bigg[\Big(\frac{1}{m_\chi^2}+\frac{\vec v_T^{\perp^2}}{\vec q^{\,\,2}}
\Big)A^2+\frac{1}{m_N^2}+\frac{1}{m_N^2}\bigg]\,,
\end{split}
\end{equation}
where we shortened the notation, $c_i^N\to c_i$, and dropped common
factors.  The scaling estimates for each of the three terms are given
in the second line. The spin-independent scattering has two
contributions, both ${\mathcal O}(A^2)$ coherently enhanced: the
contribution from $\op_1^N$ is suppressed by ${\mathcal
  O}(1/m_\chi^2)$, while the contribution from $\op_5^p$ contains a
photon pole, leading to a net suppression of ${\mathcal O}(\vec
v_T^\perp{}^2/\vec q^{\,\,2})$. Using $|\vec v_T^\perp|\sim 10^{-3}$,
$|\vec q\,|\sim 0.1\,m_N$, the two contributions are comparable for
$m_\chi\sim {\mathcal O}(100{\rm~GeV})$.  The two spin-dependent terms
carry a much smaller mass suppression of ${\mathcal O}(1/m_N^2)$, but
no coherent enhancement. Which term dominates then depends on the
details of the nuclear structure for the nuclei in the
target~\cite{Bishara:2017pfq}.

The DM electric dipole operators, $Q_{5,6}^{(5)}\sim (\bar \chi
\sigma_{\mu \nu}\gamma_5\chi) \{B^{\mu\nu}, W^{\mu\nu}\}$,
Eq.~\eqref{Q56}, match onto the nuclear operator $\op_{11}^{p} \sim i
\vec q \cdot \vec S_\chi$. This leads to coherently enhanced
scattering independent of the nuclear spin, with the $1/|\vec q\,|^2$
pole only partially cancelled,
\begin{equation}
\frac{d \sigma}{dE_R} \sim \frac{\vec q^{\,\,2}}{m_N^2}\,
(c_{11}^p)^{2} \, W_M\sim \bigg(\frac{\alpha}{\Lambda}\frac{1}{|\vec
  q\,|} C_{5,6}^{(5)} \bigg)^2 A^2\,.
\end{equation}
Compared to the magnetic dipole operators, the bounds on the NP scale
$\Lambda$ for electric dipole interactions of DM are thus more
stringent by a factor of order $m_\chi/|\vec q\,|$.

\addtocontents{toc}{\SkipTocEntry}
\subsubsection{Operators with DM scalar currents} 
The operators $Q_{3,4}^{(5)}\sim (\bar \chi \chi) (H^\dagger H)$,
Eq.~\eqref{Q34}, generate DM interactions with a scalar quark current
once the Higgs is integrated out at $\mu\sim m_Z$. Integrating out the
top, bottom, and charm quarks at the respective thresholds generates
an effective coupling of DM to gluons. At $\mu\sim \mu_{\rm had}$ DM
thus couples to both the gluonic and light-quark scalar currents. Both
of these match onto the nuclear operator $\op_1^N \sim
\mathbbm{1}_\chi\,\mathbbm{1}_N$, giving a coherently enhanced
spin-independent cross section,
\begin{equation}\label{eq:xsec:Higgs}
\frac{d\sigma}{dE_R} \sim c_{1}^2\, W_M\sim \Big(\frac{2}{27}
\frac{m_N}{\Lambda m_h^2}C_{3,4}^{(5)}\Big)^2  A^2\,, 
\end{equation}
where in the last term we kept the numerically important factor $2/27$.

The operators with pseudoscalar DM current, $Q_{7,8}^{(5)}\sim (\bar
\chi \gamma_5\chi) (H^\dagger H)$, Eq.~\eqref{Q78}, follow a similar
series of matchings. The only significant difference arises in the
nonrelativistic limit, where the DM pseudoscalar current gives an
${\mathcal O}(q)$ suppressed operator, $\op_{11}^N\sim \vec i q\cdot
\vec S_\chi$. The resulting DM--nucleus scattering cross section is
still coherently enhanced, but suppressed by ${\mathcal O}(\vec
q^{\,\,2}/m_\chi^2)$ compared to~\eqref{eq:xsec:Higgs},
\begin{equation}
\frac{d\sigma}{dE_R} \sim \frac{\vec q^{\,\,2}}{m_N^2} c_{11}^2\,
W_M\sim \frac{\vec q^{\,\,2}}{m_\chi^2}
\Big(\frac{2}{27}\frac{m_N}{\Lambda m_h^2}C_{7,8}^{(5)}\Big)^2
A^2\,.
\end{equation}

\addtocontents{toc}{\SkipTocEntry}
\subsubsection{Operators with DM vector current and with quark vector or axial-vector currents} 
We focus next on the operators $Q_{1,i}^{(6)},\dots, Q_{4,i}^{(6)}\sim
(\bar \chi \gamma^\mu \chi) \{\bar q_{L} \gamma_\mu q_{L}, q_{R}
\gamma_\mu q_{R}\}$, Eqs.~\eqref{eq:dim6:Q15}-\eqref{eq:dim6:Q48}.
Barring cancellations, the leading contribution is due to the
vector$\times$vector part of the operators, $(\bar \chi \gamma_\mu
\chi)(\bar q\gamma^\mu q)$. For couplings to the first generation
quarks this leads to coherently enhanced spin-independent scattering,
\begin{equation}\label{eq:spin-indep:xsec:vectors:1}
\frac{d\sigma}{dE_R} \sim c_{1}^2\, W_M\sim
\Big(\frac{1}{\Lambda^2} C_{1q,\dots,4q}^{(6)}\Big)^2 A^2 \qquad
\text{($1^{\rm st}$ generation quarks)}\,.
\end{equation}

The estimate is different, if DM only couples to quarks of the second
or third generation. For these the nuclear matrix element of the
vector current vanishes, and the leading contribution comes from
closing the quarks in a loop, exchanging a photon with the up- or
down-quark vector currents. This also results in a spin-independent
scattering, with a cross section
\begin{equation}\label{eq:spin-indep:xsec:vectors:23}
\frac{d\sigma}{dE_R} \sim \Big(\frac{\alpha}{4\pi}\frac{1}{\Lambda^2} C_{1q,\dots,4q}^{(6)}\Big)^2
A^2 \qquad \text{($2^{\rm nd}$ and $3^{\rm rd}$ generation quarks)}\,.
\end{equation}
In addition there are subleading contributions from matching onto
higher dimension operators with gluons, as well as spin-dependent,
velocity-suppressed scattering from the axial currents.

The situation is qualitatively different if the UV physics is such
that at $\mu\sim \Lambda$ it projects the $Q_{1,i}^{(6)},\dots,
Q_{4,i}^{(6)}$ operators only on the vector$\times$axial-vector
structure. For instance, if the Wilson coefficients obey
$C_{1,i}^{(6)}(\Lambda) = 0$, while $C_{2,i}^{(6)}(\Lambda) =
-C_{3,i}^{(6)}(\Lambda) = -C_{4,i}^{(6)}(\Lambda)$, then, neglecting
RG effects, only operators of the form $(\bar \chi \gamma_\mu
\chi)(\bar q \gamma^\mu \gamma_5 q)$ are generated. If the operators
involve light quarks, this gives a spin-dependent cross section that
scales as (for $q=u,d,s$)
\begin{equation}
\label{eq:scaling:V_chixA_q}
\frac{d\sigma}{dE_R}\sim \Big(\vec v_T^{\perp2} c_{7}^2 + \frac{\vec
  q^{\,\,2}}{m_N^2} c_{9}^{2} \Big) W_{\Sigma'}
   \sim \bigg(\vec v_T^{\,\,2} + \frac{\vec
  q^{\,\,2}}{m_\chi^2}\bigg) 
\bigg(\frac{\C_{1i,\dots,4i}^{(6)}}{\Lambda^2}\bigg)^2, \qquad \text{(axial vector)}.
\end{equation}
The two contributions are comparable for $|\vec q\,|\sim 0.1 m_N$ and
$m_\chi\sim {\mathcal O}(100{\rm~GeV})$. If the
vector$\times$axial-vector operators involve only the heavy quarks,
$q=c,b,t$, the scattering cross section is further severely suppressed
by the small contributions of the heavy quarks to the nucleon spin
(see Sec.~\ref{sec:noRG:AA} below and Ref.~\cite{Brod:2018ust} for a
more detailed discussion).

Note that the spin-dependent scattering in
Eq.~\eqref{eq:scaling:V_chixA_q} is suppressed by $\vec
v_T^{\,\,2}\sim {\vec q^{\,\,2}}/{m_\chi^2}\sim 10^{-6}$. There is no
such suppression for the spin-independent cross section,
Eq. \eqref{eq:spin-indep:xsec:vectors:1}, which is, in addition,
enhanced by the coherence factor $A^2$. This means that the Wilson
coefficients contributing to the quark vector currents at the scale
$\mu\gtrsim m_Z$ need to cancel to the level $\sim |\vec q\,|/(m_\chi
A)\sim {\mathcal O}(10^{-6})$ if the spin-dependent scattering is to
be the dominant DM-nucleus interaction. Perfect cancellation at all
scales is impossible to arrange, since the contributions come from
operators in different representations of the SM gauge group,
$\big(\bar \chi \gamma_\mu \chi \big)\big(\bar Q_L\gamma^\mu Q_L)$,
$\big(\bar \chi \gamma_\mu \chi \big) \big(\bar u_R\gamma^\mu u_R)$,
$\big(\bar \chi \gamma_\mu \chi \big) \big(\bar d_R\gamma^\mu
d_R)$. Even if one engineers the Wilson coefficients of these
operators such that the vector currents are zero at one scale, a small
amount of running will make them nonzero at a different scale. The
required cancellation is numerically of three-loop order, so that even
the radiative corrections may need to be canceled by fine tuning in
order for the spin-dependent scattering to be the leading effect.

\addtocontents{toc}{\SkipTocEntry}
\subsubsection{Operators with DM axial-vector and with quark vector or axial-vector currents}\label{sec:noRG:AA}
A qualitatively different situation is encountered for the operators
that involve DM axial-vector currents, $Q_{5,i}^{(6)},\dots,
Q_{8,i}^{(6)}\sim (\bar \chi \gamma_\mu\gamma_5 \chi) \{\bar q_{L}
\gamma^\mu q_{L}, \bar q_{R} \gamma^\mu q_{R}\}$,
Eqs.~\eqref{eq:dim6:Q15}-\eqref{eq:dim6:Q48}. In this case the $(\bar
\chi \gamma_\mu\gamma_5 \chi)(\bar q \gamma^\mu \gamma_5 q)$ operators
lead to spin-dependent scattering, while the $(\bar \chi
\gamma_\mu\gamma_5 \chi)(\bar q \gamma^\mu q)$ operators lead to
coherently enhanced, but momentum-suppressed scattering. We discuss
each of the two limiting cases separately.

If the operator $(\bar \chi \gamma_\mu \gamma_5 \chi)(\bar q
\gamma^\mu \gamma_5 q)$ involves light quarks, $q=u,d,s$, this results
in a spin-dependent cross section (not displaying explicitly the
suppression for strange quark due to its small axial charge, $\Delta
s=-0.031(5)$~\cite{Bishara:2017pfq, QCDSF:2011aa, Engelhardt:2012gd,
  Bhattacharya:2015gma, Alexandrou:2017hac}),
\begin{equation}\label{eq:xsec:spin-dependent:axial-axial:light}
\frac{d\sigma}{dE_R} \sim \Big\{c_{4}, \frac{\vec q^{\,\,2}}{m_N^2}
c_{6}\Big\}^2 W_{\Sigma',\Sigma''} \sim
\Big(\frac{C_{5i,\dots,8i}^{(6)}}{\Lambda^2}\Big)^2 \qquad \text{(light quarks)}\,.
\end{equation}
If the operator $(\bar \chi \gamma_\mu \gamma_5 \chi)(\bar q
\gamma^\mu \gamma_5 q)$ involves only the heavy quarks, $q=t,b,c$, the
scattering cross section is generally very small. The axial charges of
charm and bottom quarks are tiny and poorly determined.
Ref.~\cite{Polyakov:1998rb} obtained $\Delta c\approx - 5 \cdot
10^{-4}$, $\Delta b\approx - 5 \cdot 10^{-5}$, with probably at least
a factor of two uncertainty on these estimates. Despite this, for
heavy quark axial--axial interactions the heavy quark axial charges
still dominate the cross section over the contributions from mixing
induced couplings to light quarks, discussed in the next section (see
also Ref.~\cite{Brod:2018ust}).

We focus next on the limiting case where at $\mu\sim m_Z$ only the
axial-vector$\times$vector operators, $(\bar \chi \gamma_\mu \gamma_5
\chi)(\bar q \gamma^\mu q)$, are generated. For $q=u,d,s$ these match
on two nonrelativistic operators with one derivative, $\op_{8}^{N}$,
$\op_{9}^{N}$.  Both lead to momentum suppressed incoherent
scattering, with $\op_8^N$ giving rise, in addition, to
spin-independent scattering that is coherently enhanced, but velocity
suppressed,
\begin{equation}
\label{eq:xsec:spin-dependent:axial-vector:light}
\frac{d\sigma}{dE_R} \sim
\vec v_T^{\perp 2} c_{8}^2 W_{M} + \frac{\vec q^{\,\,2}}{m_N^2}\Big\{c_{8}, c_{9}\Big\}^2 W_{\Delta,\Delta\Sigma',\Sigma'} \sim
 \Big\{ \vec v_T^{\perp2} A^2, \frac{\vec q^{\,\,2}}{m_N^2}\Big\}
\bigg(\frac{C_{5i,\dots,8i}^{(6)}}{\Lambda^2}\bigg)^2 \quad
\text{(light quarks)}\,.
\end{equation}
The two contributions are of parametrically similar size for heavy
nuclei, $A\sim {\mathcal O}(100)$, in which case $|\vec v_T| A \sim
|\vec q|/m_N$. Which of the two contributions dominates then depends
on the details of the nuclear structure for the particular isotope.

For $(\bar \chi \gamma_\mu \gamma_5 \chi)(\bar q \gamma^\mu q)$ with
$q=t,b,c$, the leading contribution comes from closing the heavy quark
loop, exchanging a photon with the up- or down-quark vector
current. The cross section is suppressed with respect to
Eq.~\eqref{eq:xsec:spin-dependent:axial-vector:light} by an additional
factor of $(\alpha/4\pi)^2$,
\begin{equation}\label{eq:xsec:spin-dependent:axial-vector:heavy}
\frac{d \sigma}{d E_R} \sim \Big(\frac{\alpha}{4\pi}\Big)^2 \Big\{ \vec v_T^{\perp2} A^2, \frac{\vec q^{\,\,2}}{m_N^2}\Big\}
\bigg(\frac{C_{5i,\dots,8i}^{(6)}}{\Lambda^2}\bigg)^2 \qquad \text{(heavy quarks)}\,.
\end{equation}
There is also a contribution from matching onto higher dimension
operators with gluons, which is expected to be at most of similar
size.

In general the sum of $Q_{5,i}^{(6)},\dots, Q_{8,i}^{(6)}$ operators
matches onto both $(\bar \chi \gamma_\mu \gamma_5 \chi)(\bar q
\gamma^\mu \gamma_5 q)$ and $(\bar \chi \gamma_\mu \gamma_5 \chi)(\bar
q \gamma^\mu q)$ operators at $\mu=\mu_{\rm EW}$, giving a cross
section that is a sum of
Eqs.~\eqref{eq:xsec:spin-dependent:axial-axial:light}
and~\eqref{eq:xsec:spin-dependent:axial-vector:light}. The
spin-dependent scattering in
Eq.~\eqref{eq:xsec:spin-dependent:axial-axial:light} is parametrically
the largest. Since the parametric enhancement is not large, however,
this expectation does depend on the target material, and
spin-independent scattering could be equally important.

\addtocontents{toc}{\SkipTocEntry}
\subsubsection{Operators with Higgs vector currents} 
The operators $Q_{15,16}^{(6)}\sim(\bar \chi \gamma^\mu
\chi)(H^\dagger D_\mu H)$ and $Q_{17,18}^{(6)}\sim (\bar \chi
\gamma^\mu \gamma_5 \chi)(H^\dagger D_\mu H)$,
Eqs.~\eqref{eq:dim6:Q15Q17} and~\eqref{eq:dim6:Q16Q18}, give rise to a
DM--DM--$Z$ boson vertices after the Higgs obtains its vacuum
expectation value. Integrating out the $Z$ at $\mu\sim m_Z$ leads to a
coupling of DM to vector and axial-vector quark currents.  The
relative strength of the two is fixed by the $Z$ couplings to the
left- and right-handed quarks. This is different from the case of the
operators $Q_{1,i}^{(6)},\dots, Q_{8,i}^{(6)}$ that we discussed
before, where a more general structure of DM couplings to quarks was
allowed.

For the operators $Q_{15,16}^{(6)}\sim (\bar \chi \gamma^\mu
\chi)(H^\dagger D_\mu H)$, the dominant contribution comes from a
quark vector current, giving a coherently enhanced, spin-independent
scattering cross section
\begin{equation}\label{eq:xsec:Z:vector}
\frac{d\sigma}{dE_R} \propto c_{1}^2 W_M\sim  \Big(\frac{1}{\Lambda^2} C_{15, 16}^{(6)}\Big)^2  A^2\,.
\end{equation}
The $Z$-boson exchange at $\mu\sim m_Z$ also generates the $\big(\bar
\chi \gamma_\mu \chi\big) \big(\bar q\gamma^\mu \gamma_5 q\big)$
operator. This leads to momentum-suppressed, spin-dependent scattering
that is always subleading.

On the other hand, for the operators with DM axial-vector currents,
$Q_{17,18}^{(6)}\sim (\bar \chi \gamma^\mu \gamma_5 \chi)(H^\dagger
D_\mu H)$, one needs to keep both the spin-dependent and
spin-independent scattering contributions. The induced
axial-vector$\times$vector and axial-vector$\times$axial-vector
interactions lead to a cross section that scales as the sum of
Eqs.~\eqref{eq:xsec:spin-dependent:axial-axial:light}
and~\eqref{eq:xsec:spin-dependent:axial-vector:light},
\begin{equation}\label{eq:xsec:Z:axial}
\begin{split}
\frac{d\sigma}{dE_R} &\sim\Big\{c_{4}, \frac{\vec q^{\,\,2}}{m_N^2} c_{6}\Big\}^2 W_{\Sigma',\Sigma''}+
\vec v_T^{\perp 2} c_{8}^2 W_{M} + \frac{\vec q^{\,\,2}}{m_N^2}\Big\{c_{8}, c_{9}\Big\}^2 W_{\Delta,\Delta\Sigma',\Sigma'} 
\\
&\sim
 \Big( 1+\vec v_T^{\perp2} A^2+ \frac{\vec q^{\,\,2}}{m_N^2}\Big)
\bigg(\frac{C_{17,18}^{(6)}}{\Lambda^2}\bigg)^2\,.
\end{split}
\end{equation}
From scaling considerations, spin-dependent scattering is expected to
be dominant in nuclei with an unpaired nucleon that is not in an
$s$-shell. But even then the spin-independent scattering contributions
may need to be included, depending on the nucleus. An example is
dicsussed in Section~\ref{sec:AA3}.

\subsection{Inclusion of RG running}
\label{sec:eff:run}

The modifications due to RG running can significantly impact the cross
section predictions. We will show several examples where the RG
running effects are particularly large. While the sizes and patterns of
the induced corrections does depend on the electroweak charges of DM,
the effects themselves are not ``optional''. They are due to SM
particles in the loops, and are thus always present.

Consider, for instance, $SU(2)_L$-singlet DM, where all mixing
proportional to $g_2$ vanishes, as can be seen by inspecting
$\gamma_2^{(0)}$. Another example is DM that is hypercharge neutral,
$Y_\chi=0$, for which all the mixings due to $B_\mu$ exchanges with
the DM line vanish. However, in both cases there is still mixing due
to the running of the non-conserved SM currents. For instance, for DM
that is a complete SM singlet the main mixing is induced by the
top-quark Yukawa interaction. This case has been discussed in detail
in the literature~\cite{Crivellin:2014qxa, D'Eramo:2014aba,
  DEramo:2016gos} (see also Ref.~\cite{Brod:2018ust} for the
discussion of weak-mixing effects below the weak scale).

Here, we will use our general results from Sec.~\ref{sec:RGrunning}
and apply them to the simplest nontrivial example of DM with
electroweak charges -- a Dirac fermion multiplet that is hypercharge
neutral, $Y_\chi=0$, and an electroweak triplet, $I_\chi=1$.  The
choice $Y_\chi=0$ is imposed on us by the phenomelogical requirement
that DM (the neutral component of the multiplet) should not couple to
the $Z$ boson at tree level, in order to avoid a too large direct detection
scattering scattering cross section.

We will illustrate the effects of RG running for several different
choices of non-renormalizable DM interactions, taking $I_\chi=1$ as an
example.  The scattering rates then receive two types of
contributions. First, there are contributions from higher dimension
operators. These vanish in the limit $\Lambda\to\infty$. However, for
$I_\chi \ne0$ there are also contributions from renormalizable
electroweak interactions that are independent of $\Lambda$. The
leading contributions of this type are due to the small ``Higgs
penguin'', the one-loop and two-loop contributions shown in
Fig.~\ref{fig:DD-Higgs_loop}. They lead to coherently enhanced
scattering of parametric size (for $I_\chi=1$, using the right diagram
in Fig.~\ref{fig:DD-Higgs_loop})
\begin{equation}\label{eq:Higgs:penguin:scaling}
\left.  \frac{d\sigma}{dE_R}\right|_{\text{``Higgs~penguin''}} \propto A^2 \bigg( \frac{\alpha_2}{4\pi}
  \bigg)^4 \frac{g_s^4}{M_W^4} \sim \frac{10^{-19} A^2}{\text{GeV}^4} \,,
\end{equation}
where $A^2 = {\cal O}(10^4)$ for scattering on Xenon, and $A^2 = {\cal
  O}(10^2)$ for scattering on Fluorine. If the target nucleus has
non-zero spin, the $W$ box shown in Fig.~\ref{fig:DD-Higgs_loop} gives
an additional contribution scaling as
\begin{equation}\label{eq:W:box:scaling}
\left.  \frac{d\sigma}{dE_R}\right|_{\text{``$W$ box''}} \propto \bigg( \frac{\alpha_2}{4\pi}
  \bigg)^4 \frac{1}{M_W^4} \sim \frac{10^{-16}}{\text{GeV}^4} \,.
\end{equation}
These scalings omit a proportionality factor that depends on the DM
mass and velocity, the recoil energy, and the detailed structure of
the nucleus. For a typical scattering event with $m_\chi=100\,$GeV and
$E_R=20\,$keV, this factor is roughly of the order of $10^8\,$GeV. In
our numerical evaluations we use the exact results from
Ref.~\cite{Hisano:2011cs} (for heavy DM see also \cite{Hill:2011be,
  Hill:2014yka, Chen:2018uqz}). The ratio of scaling estimates in
Eqs.~\eqref{eq:Higgs:penguin:scaling} and~\eqref{eq:W:box:scaling}
agrees with the ratio of full results within an order of
magnitude. Note that for DM that is a complete electroweak singlet the
gauge contribution is absent.

In Figs.~\ref{fig:A_j1_3rd} to \ref{fig:VA_j3} we show numerical
examples for DM scattering rates in two fictitious, yet realistic
detectors. For a Xenon target we integrate the differential rates over
$E_R\in [5\,{\rm keV},40\,{\rm keV}]$, and for Fluorine over
$E_R\in[3.3\,{\rm keV},200\,{\rm keV}]$. We average over the natural
abundances of the xenon isotopes and assume a standard
Maxwell-Boltzmann velocity distribution with mean velocity
240\,km/s. For nuclear response functions we use the predictions of
Ref.~\cite{Fitzpatrick:2012ix, Anand:2013yka}, while for nuclear form
factors we use the inputs collected in Ref.~\cite{Bishara:2017pfq}. In
the figures the DM mass varies in the range $m_\chi \in [30\,{\rm
    GeV}, 1\,{\rm TeV}]$. While in the lower part of the range the
shown benchmarks are likely excluded by LEP constraints and LHC
searches, we keep them for illustration purposes.

In the three examples that we show below  the effective interactions involve
axial-vector quark currents. The reason for this choice is easy to understand. In the case where we have DM current coupling to  either only LH or only RH quarks, 
the vector-vector part of the interaction always dominates, and the RG effects are subdominant. In the case where we have DM axial-vector current coupling to either  only LH or only RH quarks, the
mixing effects are larger and are ${\mathcal O}(1)$.  We instead show the cases where the RG running induces the largest corrections, i.e., the case of DM interacting with axial-vector quark currents. 

In all our examples DM couples to both up- and down-type
quarks. Using the triplet operators $Q_{1,i}^{(6)}$ and
$Q_{5,i}^{(6)}$ it is possible to construct interactions of DM with
only up- or down-quark currents separately. This would, however,
require a nonzero DM hypercharge, $Y_\chi\neq 0$, which is phenomenologically not viable
for Dirac fermion DM.

\addtocontents{toc}{\SkipTocEntry}
\subsubsection{Operators with DM axial-vector current and 3rd generation quark axial-vector current}\label{sec:AA3}

For the first example we assume that at the high scale, $\Lambda$, the only nonzero
Wilson coefficients are
\begin{equation}
C_{6,3}^{(6)}(\Lambda) = -C_{7,3}^{(6)}(\Lambda) = -C_{8,3}^{(6)}(\Lambda) \,.
\end{equation}
That is, we assume that DM couples to the SM through renormalizable
weak interactions $(I_\chi=1, Y_\chi=0)$ and, in addition, through the dimension 6 effective
operator
\begin{equation}
\label{eq:axial-axial:top}
-Q_{6,3}^{(6)} + Q_{7,3}^{(6)} + Q_{8,3}^{(6)} = (\bar \chi
\gamma_\mu\gamma_5 \chi) (\bar t \gamma^\mu \gamma_5 t + \bar b
\gamma^\mu \gamma_5 b)\,.
\end{equation}
At tree-level this operator has vanishingly small nuclear matrix
element, see Sec.~\ref{sec:eff:norun}. Appreciable DM--nucleus
scattering is generated only once we close the heavy quark loop. The
RG running captures the logarithmically enhanced part of this
contribution. Starting with
\begin{equation}
- C_{6,3}^{(6)}(\Lambda) = C_{7,3}^{(6)}(\Lambda) =
C_{7,3}^{(6)}(\Lambda) = 1\,,
\end{equation}
the RG running from $\Lambda$ to $\mu\sim M_W$ generates the Wilson
coefficients (keeping only the linear logarithmic term)
\begin{align}
\label{eq:C13:6}
C_{1,3}^{(6)}(m_W) & = - \frac{g_2^2}{16\pi^2} \big[\gamma_2^{(0)}
  \big]_{Q_{6,3}^{(6)}, Q_{1,3}^{(6)}} \log\frac{m_W}{\Lambda} =
  \frac{g_2^2}{16\pi^2} 12 \log\frac{m_W}{\Lambda}\,,
  \\
   \begin{split}
   \label{eq:C18:6}
C_{18}^{(6)}(m_W) & = \frac{y_t^2}{16\pi^2} \Big( -
\big[\gamma_{y_t}^{(0)} \big]_{Q_{6,3}^{(6)}, Q_{18}^{(6)}} +
\big[\gamma_{y_t}^{(0)} \big]_{Q_{7,3}^{(6)}, Q_{18}^{(6)}} \Big)
\log\frac{m_W}{\Lambda} \\ & = - \frac{y_t^2}{16\pi^2} 4N_c
\log\frac{m_W}{\Lambda}\,.
\end{split}
\end{align}
In deriving Eqs.~\eqref{eq:C13:6} and~\eqref{eq:C18:6} we took into
account the cancelations of contributions that arise due to the actual
values of the anomalous dimensions. For instance, the mixing via
penguin insertions generically results in DM coupling to the first two
generations of quarks by generating the operators
$Q_{6,i}^{(6)},\ldots, Q_{8,i}^{(6)}$. However, for the initial
conditions in Eq.~\eqref{eq:axial-axial:top} this mixing vanishes at
leading-logarithmic order. There are also mixings into higher
dimension operators coupling DM to photons or gluons instead of quark
currents. These involve at least two gauge field strengths and an
additional derivative, so that the scattering contributions are
further power suppressed. The leading contributions to the scattering
rates therefore come from the mixing induced operators $Q_{1,3}^{(6)}$
and $Q_{18}^{(6)}$. The mixing into $Q_{18}^{(6)}$ is due to the
top-quark Yukawa interaction of the Higgs. It is present whether or
not DM is part of an electroweak multiplet, i.e., even if DM is an
electroweak singlet.

\begin{figure}
\begin{center}
\includegraphics[width=0.49\linewidth]{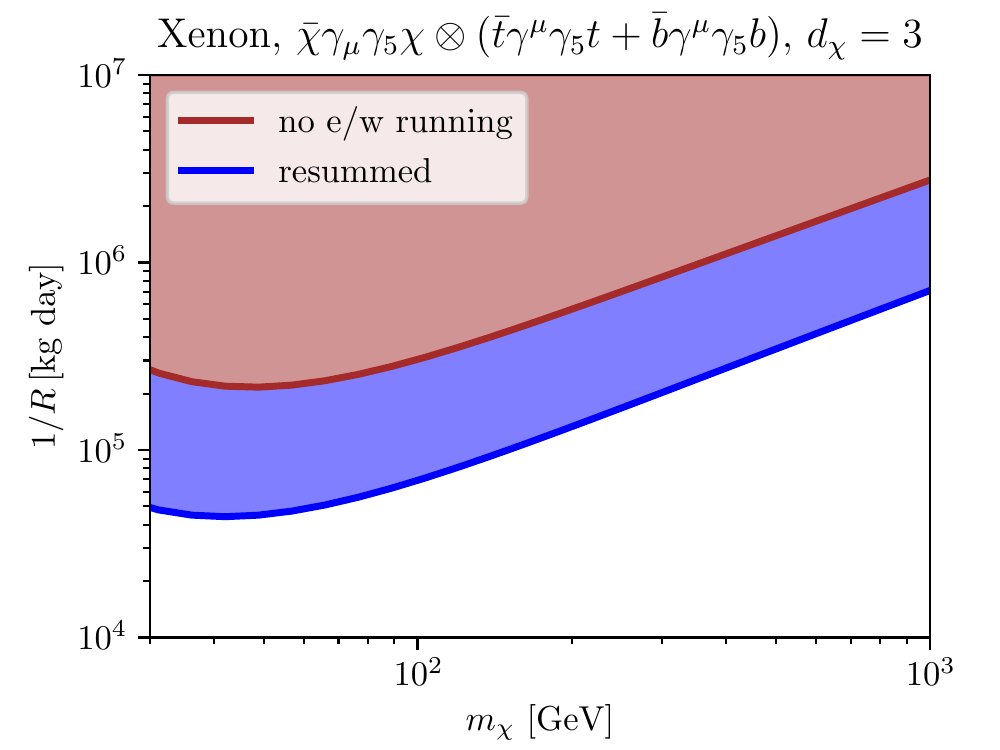}~~~~~
\includegraphics[width=0.49\linewidth]{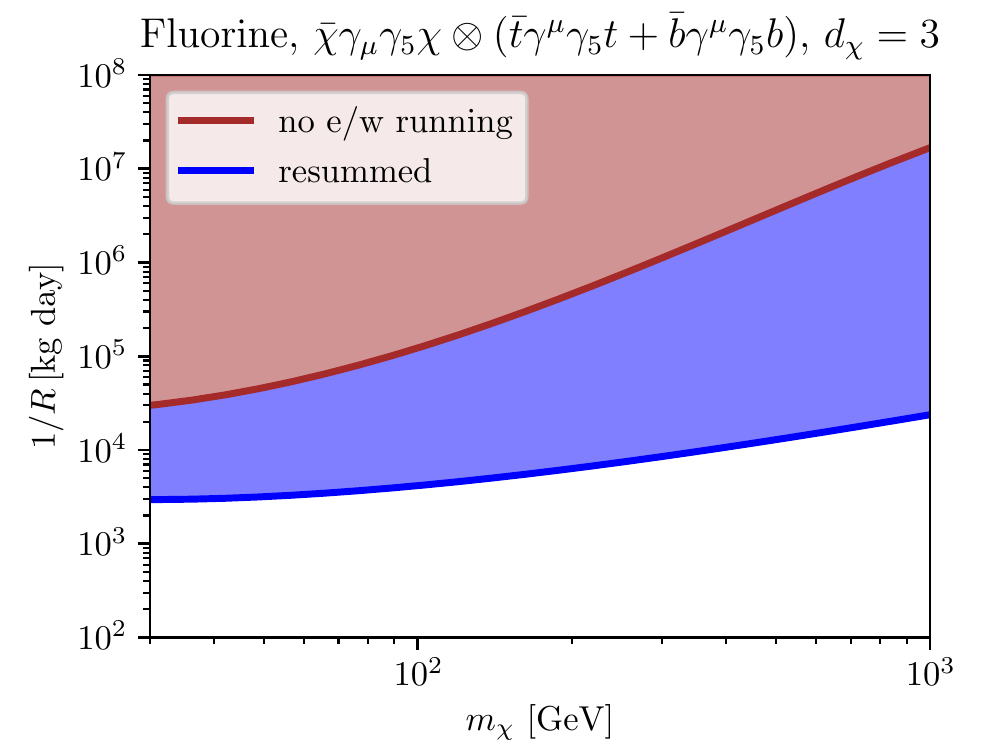}
\end{center}
\caption{The inverse scattering rates on Xenon (left) and Fluorine
  (right) for $I_\chi=1, Y_\chi=0$ Dirac fermion DM with additional
  dimension-six interactions coupling a DM axial-vector current to the
  SM axial-vector third generation current, setting
  $\Lambda=1\,$TeV. The red lines show the predicted rates without
  mixing; the blue lines after RG resummation. The plots extend in $m_\chi$ below the electroweak scale only for illustration purposes -- realistic models require extended sets of non-trivial electroweak multiplets that can modify the running (cf. Sec. \ref{sec:UV}).}
\label{fig:A_j1_3rd}
\end{figure}

In Fig.~\ref{fig:A_j1_3rd} we compare the predicted rates for
scattering on Xenon and Fluorine, obtained with (blue lines) and
without (red lines) RG evolution. In the case of no RG evolution the
scattering is almost entirely due to the contribution of
renormalizable weak interactions. For $\Lambda=1$ TeV the RG induced
contributions from dimension-six operators dominate over the
renormalizable ones, in the case of Fluorine by up to two orders of
magnitude.

The sizes of the different contributions can be qualitatively
understood from their parametric scalings, given for the gauge
contributions in Eqs.~\eqref{eq:Higgs:penguin:scaling}
and~\eqref{eq:W:box:scaling}. The mixing induced $Q_{18}^{(6)}$ leads
to a cross section that scales roughly as
\begin{equation}\label{eq:scaling:C18}
  \frac{d\sigma}{dE_R} \propto \bigg[ 1 + \frac{q^2}{m_N^2} +
    (v_T^\perp)^2 A^2 \bigg] \bigg( \frac{\alpha_t}{4\pi} \bigg)^2
  \bigg( 12 \log \frac{M_W}{\Lambda} \bigg)^2 \frac{1}{\Lambda^4} \sim
  \frac{10^{-14} + 10^{-18} + 10^{-20} A^2}{\text{GeV}^4} \,,
\end{equation}
while the cross section induced by the mixing into
$Q_{1,3}^{(6)}$, scales roughly as
\begin{equation}\label{eq:scaling:C613}
  \frac{d\sigma}{dE_R} \propto A^2 \bigg( \frac{\alpha_2}{4\pi} \bigg)^2
  \bigg( 12 \log \frac{M_W}{\Lambda} \bigg)^2 \bigg(
  \frac{\alpha}{4\pi} \bigg)^2 \bigg( \frac{16}{9} \log \frac{m_b}{M_W} \bigg)^2
  \frac{1}{\Lambda^4} \sim \frac{10^{-19} A^2}{\text{GeV}^4} \,.
\end{equation}
In the numerical estimates we assumed a typical momentum transfer of
$q = {\cal O}(10)\,$MeV and set $\Lambda=1$ TeV. The first two terms
in the square bracket in Eq.~\eqref{eq:scaling:C18} are due to
spin-dependent scattering, with the parametric and numerical estimates
shown for Fluorine and ${}^{129}$Xe, while they are much smaller for
the other main Xenon isotopes. The last term in
Eq.~\eqref{eq:scaling:C18} is due to spin-independent scattering. The
spin-dependent terms give the dominant contribution to the scattering
rates on Fluorine.

The scattering contribution in \eqref{eq:scaling:C613} involves QED
mixing, converting the third generation quark current to the first
generation one, see Sec.~\ref{sec:RG:below:ew}. This contribution is
relevant only for scattering on Xenon, where it is, for
$\Lambda=1$TeV, comparable to the gauge contribution as well as to the
spin-dependent scattering in Eq.~\eqref{eq:scaling:C18}. Indeed, the
left panel of Fig.~\ref{fig:A_j1_3rd} shows that the contributions are
of the same size.

As already mentioned, the scattering on Fluorine is dominated by the
$Q_{18}^{(6)}$-induced contributions, Eq.~\eqref{eq:scaling:C18},
where the leading term comes from spin-dependent
scattering. Inspection of the $\Sigma'$, $\Sigma''$ response
functions, Ref.~\cite{Fitzpatrick:2012ix}, shows that spin-dependent
scattering on Fluorine is about ten times larger than for
${}^{129}$Xe, while the other Xenon isotopes give negligible
contributions. In Fig.~\ref{fig:A_j1_3rd} we weighted the
contributions according to the natural abundance of Xenon isotopes,
giving an additional roughly five-fold suppression of the
spin-dependent rate for Xenon. Consequently, the effect of the RG
evolution is large only for scattering on fluorine (right panel of
Fig.~\ref{fig:A_j1_3rd}).

\addtocontents{toc}{\SkipTocEntry}
\subsubsection{Vector -- axial-vector (first generation)}

\begin{figure}
\begin{center}
\includegraphics[width=0.49\linewidth]{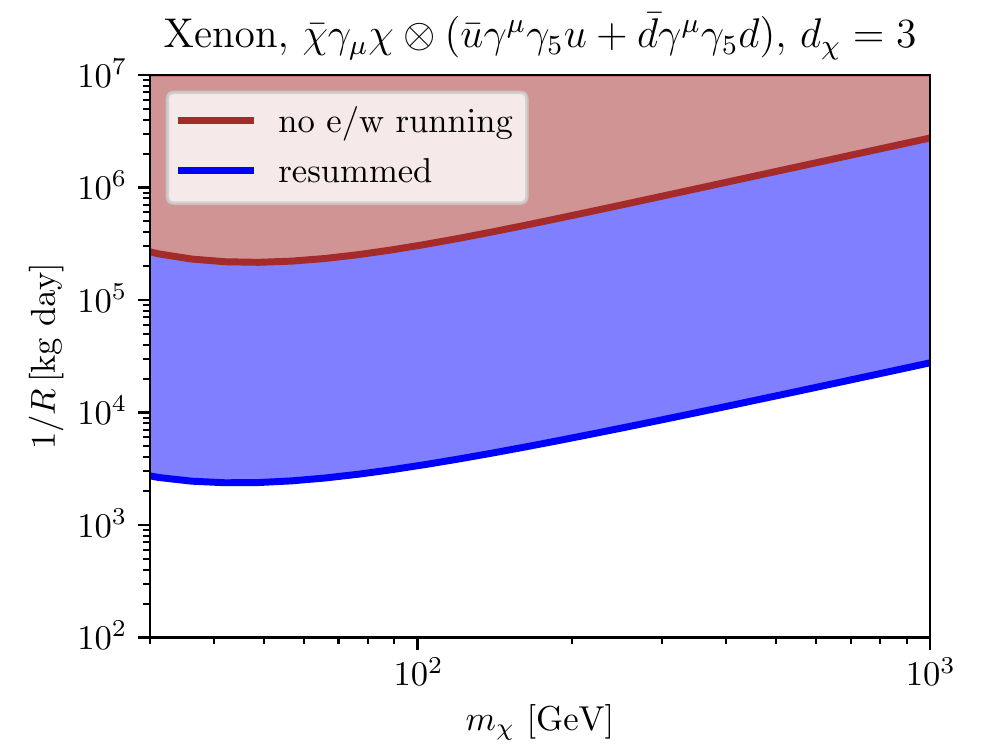}~~~~~
\includegraphics[width=0.49\linewidth]{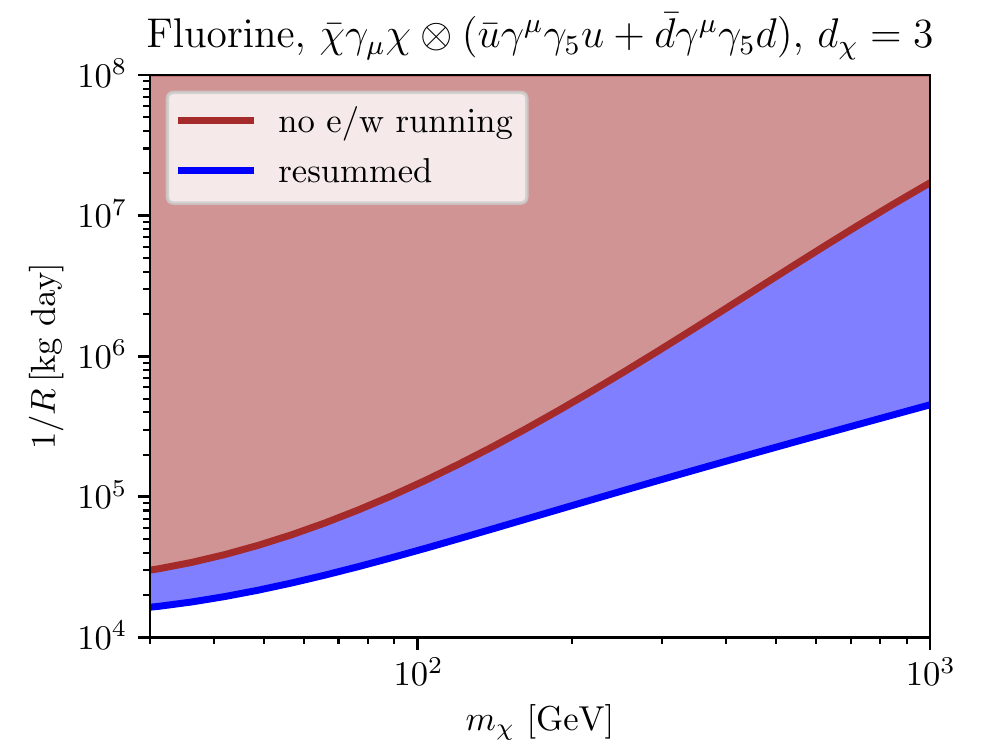}
\end{center}
\caption{ Same as Fig.~\ref{fig:A_j1_3rd}, but for $I_\chi=1,
  Y_\chi=0$ Dirac fermion DM with additional dimension-six
  interactions coupling a DM vector current to the SM axial-vector
  first generation current, setting $\Lambda=1\,$TeV.}
         \label{fig:VA_j1}
\end{figure}

Next, we assume that at $\Lambda=1$ TeV the only nonzero Wilson
coefficients are
\begin{equation}\label{eq:VA:1st:alignment}
-C_{2,1}^{(6)}(\Lambda) = C_{3,1}^{(6)}(\Lambda) =
C_{4,1}^{(6)}(\Lambda) \,,
\end{equation}
so that the non-renormalizable DM interactions are due to the operator
\begin{equation}
- Q_{2,1}^{(6)} + Q_{3,1}^{(6)} + Q_{4,1}^{(6)} = (\bar \chi
\gamma_\mu \chi) (\bar u \gamma^\mu \gamma_5 u + \bar d \gamma^\mu
\gamma_5 d).
\end{equation}
This leads to spin-dependent scattering rate that scales roughly as
\begin{equation}
  \frac{d\sigma}{dE_R} \propto (v_T^\perp)^2 \frac{1}{\Lambda^4}
  \sim \frac{10^{-18}}{\text{GeV}^4} \,,
\end{equation}
see Section~\ref{sec:RG-run}. The Higgs penguin contribution,
Eq.~\eqref{eq:Higgs:penguin:scaling}, dominates over this rate by
orders of magnitude. The dominant contribution, however, is mixing
induced. The Wilson coefficient $C_{2,1}^{(6)}(\Lambda)$ gets modified
by the two-step mixing in the RG evolution to (we neglect numerically
subleading contributions)
\begin{equation}
\begin{split}
C_{2,1}^{(6)}(M_W) & = - 1 - \Big(\frac{g_2^2}{16\pi^2}\Big)^2
\frac{\big[\gamma_2^{(0)} \big]_{Q_{2,1}^{(6)}, Q_{5,1}^{(6)}}
  \big[\gamma_2^{(0)} \big]_{Q_{5,1}^{(6)}, Q_{2,1}^{(6)}}}{2}
\log^2\frac{M_W}{\Lambda} 
\\ 
& = - 1 - \Big( \frac{g_2^2}{16\pi^2} \Big)^2
36 \log^2\frac{M_W}{\Lambda}\,.
\end{split}
\end{equation}
The mixing contributions for the other two Wilson coefficients,
$C_{3,1}^{(6)}$ and $C_{4,1}^{(6)}$, cancel. This leads to the
breaking of the original alignment, Eq.~\eqref{eq:VA:1st:alignment},
inducing a coupling to the SM vector current. The product of the large
anomalous dimensions and the square of the large logarithm
$\log(M_W/\Lambda)$, together with the coherent enhancement factor,
$A^2$, leads to
\begin{equation}
  \frac{d\sigma}{dE_R} \propto (v_T^\perp)^0 A^2 \bigg(
  \frac{\alpha_2^2}{(4\pi)^2} \bigg)^2 \bigg( 36 \log^2
  \frac{M_W}{\Lambda} \bigg)^2 \frac{1}{\Lambda^4} \sim
  \frac{10^{-18} A^2}{\text{GeV}^4} \,,
\end{equation}
resulting in the enhanced scattering rate, as shown in
Fig.~\ref{fig:VA_j1}.

It is important to realize that it is not sufficient to use the
first-order-expanded solution to the RG equations, as the effect
arises only at the second order in the mixing. While the effect
corresponds to a two-loop correction in the ``full theory'', our
method automatically captures the leading-logarithmic part of it.

\addtocontents{toc}{\SkipTocEntry}
\subsubsection{Vector -- axial-vector (third generation)}

Finally, let us consider an initial condition
\begin{equation}\label{eq:VA:3st:alignment}
-C_{2,3}^{(6)}(\Lambda) = C_{3,3}^{(6)}(\Lambda) =
C_{4,3}^{(6)}(\Lambda) \,,
\end{equation}
so that DM couples to the third generation of quarks through the operator
\begin{equation}
- Q_{2,3}^{(6)} + Q_{3,3}^{(6)} + Q_{4,3}^{(6)} = (\bar \chi
\gamma_\mu \chi) (\bar t \gamma^\mu \gamma_5 t + \bar b \gamma^\mu
\gamma_5 b)\,.
\end{equation}
This axial-vector current has a vanishingly small nuclear matrix
element, see Sec.~\ref{sec:eff:norun}.  Without mixing, the leading
contribution to the scattering rate is thus due to the renormalizable
gauge interactions, Eqs.~\eqref{eq:Higgs:penguin:scaling}
and~\eqref{eq:W:box:scaling}.

The largest contribution comes, however, from the mixing. At one loop
the top-quark Yukawa interactions induce mixing of $Q_{2,3}^{(6)}$ and
$Q_{3,3}^{(6)}$ into $Q_{16}^{(6)}$ with anomalous dimensions $6$ and
$-6$, respectively, see Eq.~\eqref{eq:yt-quark-higgs}. The
contributions add up for the axial-vector quark current, giving, for
the initial condition \eqref{eq:VA:3st:alignment},
\begin{equation}
\begin{split}
C_{16}^{(6)}(M_W) & = \frac{y_t^2}{16\pi^2} \Big( -
\big[\gamma_t^{(0)} \big]_{Q_{2,3}^{(6)}, Q_{16}^{(6)}} +
\big[\gamma_t^{(0)} \big]_{Q_{3,3}^{(6)}, Q_{16}^{(6)}} \Big)
\log\frac{M_W}{\Lambda} \\ & = - 12 \, \frac{y_t^2}{16\pi^2} \,
\log\frac{M_W}{\Lambda}\,.
\end{split}
\end{equation}
The above result takes into account the cancelations of contributions
due to the actual values of the anomalous dimensions, and neglects
numerically subleading terms. (The two-step mixing effect, described
in the previous subsection, is still present, but subleading.)

\begin{figure}
\begin{center}
\includegraphics[width=0.49\linewidth]{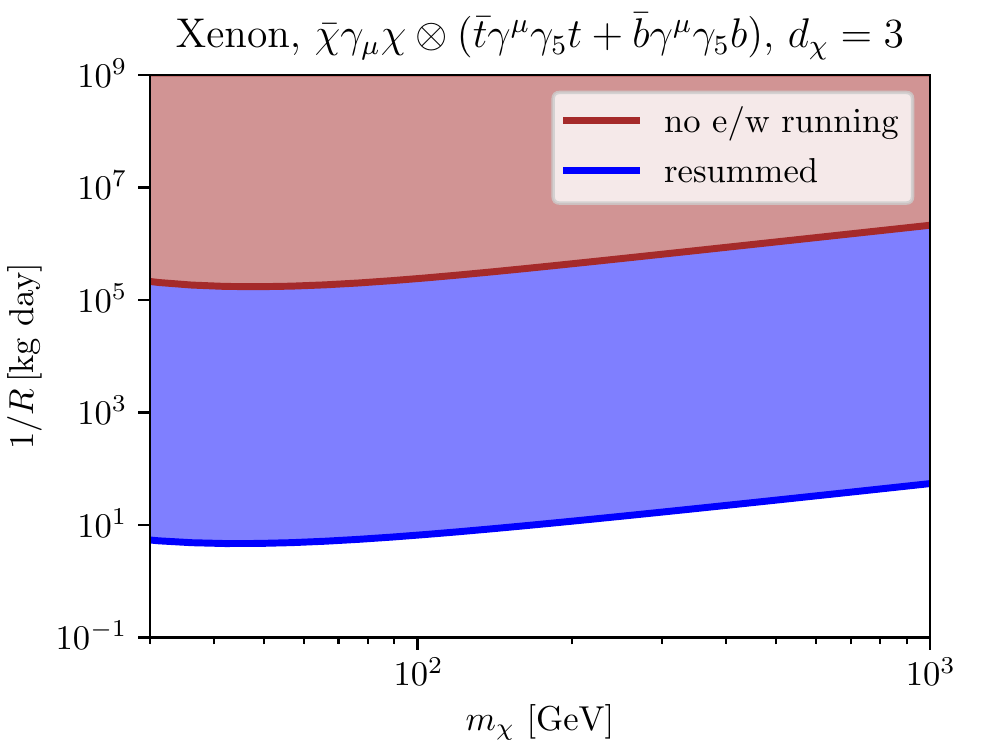}~~~~~
\includegraphics[width=0.49\linewidth]{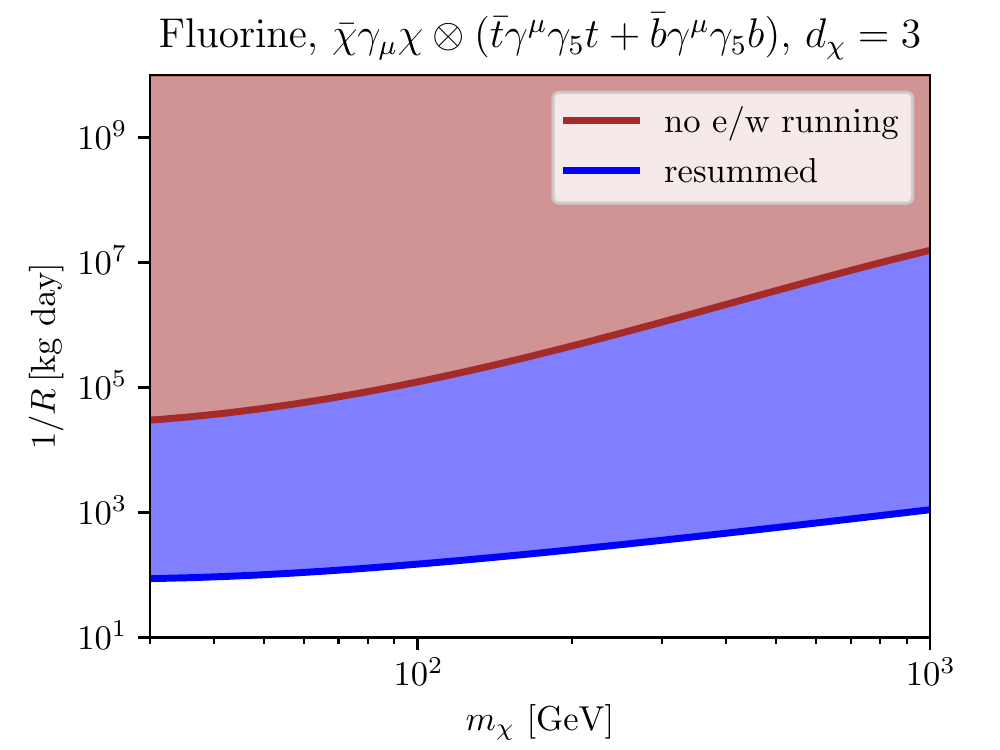}
\end{center}
\caption{ Same as Fig.~\ref{fig:A_j1_3rd}, but for $I_\chi=1,
  Y_\chi=0$ Dirac fermion DM with additional dimension-six
  interactions coupling a DM vector current to the SM axial-vector
  third-generation current, setting $\Lambda=1\,$TeV.}
         \label{fig:VA_j3}
\end{figure}

The operator $Q_{16}^{(6)}$ leads to a vector-vector interaction after
integrating out the $Z$ boson, cf. Eq.~\eqref{eq:match-6-up-1}, giving
a coherently enhanced scattering cross section of parametric size
\begin{equation}
\label{eq:Q16:induced:xsec}
  \frac{d\sigma}{dE_R} \propto (v_T^\perp)^0 A^2 \bigg(
  \frac{\alpha_t}{4\pi} \bigg)^2 \bigg( 12 \log \frac{M_W}{\Lambda}
  \bigg)^2 \frac{1}{\Lambda^4} \sim \frac{10^{-14}
    A^2}{\text{GeV}^4}\,.
\end{equation}
which is several orders of magnitude larger than the Higgs-penguin
induced one. This is illustrated in Figure~\ref{fig:VA_j1}. Included
in the numerics is the additional enhancement of the cross section by
the resummation of leading QCD logarithms below the weak scale (see
Ref.~\cite{Brod:2018ust} for details). Note that the mixing induced
effect, Eq. \eqref{eq:Q16:induced:xsec}, is independent of the weak
isospin of DM and is present even for SM-singlet
DM~\cite{Crivellin:2014qxa}.

%% file: summary.tex
\section{Summary and Conclusions}\label{sec:summary}
In this article we presented a Renormalization Group (RG) analysis of
Dark Matter (DM) interactions with the SM mediated by higher dimension
operators, up to and including dimension six. We calculated the
one-loop RG evolution of these operators, for the case of Dirac
fermion DM, from the high scale $\Lambda$ down to the weak scale, and
the matching to the tower of effective theories below the weak scale,
distinguishing the two cases, $m_\chi \sim m_Z$ and $m_\chi \ll
m_Z$. We allow for DM to be part of an electroweak multiplet.

The loop corrections are important whenever both the renormalizable
interactions and the tree-level insertions of higher dimension
operators give suppressed direct detection scattering rates. For DM
charged under the electroweak gauge group, the scattering due to
renormalizable interactions is either spin-dependent or effectively of
two-loop size. This means that the contributions from higher dimension
interactions, even if loop suppressed, can still give the leading
contribution. This is true in particular if the tree-level
contributions from higher dimension operators have small nuclear
matrix elements, while the loop-induced ones do not. In
Section~\ref{sec:eff:run} we illustrated this for three examples of DM
coupling to axial-vector quark currents, where the loop-induced
effects are especially large. Since the anomalous dimensions are
numerically large, the mixing induced effects can dominate the
scattering rate even if they are effectively of two-loop order. The RG
evolution automatically picks up the leading-logarithmic parts of such
corrections to all orders.

The computed corrections are not optional, as they arise from SM
particles running in the loop. They thus need to be included when
connecting the processes that occur at the mass of the DM (such as the
indirect detection and the LHC searches) with the processes occurring
at the low scale, e.g., direct detection scattering. The anomalous
dimensions are of two types: (i) the contributions due to Higgs
exchanges, which are present even in the case that DM is an
electroweak singlet, and (ii) the contributions that are due to the
exchanges of gauge bosons. The latter are present only if DM is part
of an electroweak multiplet.

The resulting RG evolution is implemented in the public code
\ddm~\cite{Bishara:2017nnn} and is available at
\begin{center}
\url{https://directdm.github.io}\,.
\end{center}
The code should make it relatively straightforward to use our results
when comparing indirect detection and LHC bounds with the results of
direct detection experiments, including the scattering on electrons,
that is in many cases generated already at one-loop level.

There are several directions for future work. The remaining case for
Dirac fermion DM, $m_\chi \gg m_Z$, requires the transition to Heavy
DM EFT already above the weak scale.  This will result both in a
different basis of EFT operators above the electroweak scale, as well
as changes to the anomalous dimensions. The calculations of anomalous
dimensions should also be extended to include dimension seven
operators (the full basis was presented in
Ref.~\cite{Brod:2017bsw}). Phenomenologically interesting is an
extension of our work to several multiplets, which would cover, for
instance, bino-wino-higgsino mixing in the MSSM. There are also
higher-loop contributions that the leading-logarithmic RG resummation
misses. For instance, at two-loop level there is mixing from
dimension-six operators with quarks and leptons into dipole
operators. Such contributions may be important when estimating the
dipole contributions to the dark matter scattering rates.

{\bf Acknowledgments.}  We thank F.~D'Eramo and E.~Stamou for useful
discussions.  JZ acknowledges support in part by the DOE grant
DE-SC0011784, and thanks the CERN theory group for hospitality for the
duration of his sabbatical stay, during which part of this project was
worked out. This work was also performed in part at Aspen Center for
Physics, which is supported by National Science Foundation grant
PHY-1607611. This work was partially supported by a grant from the
Simons Foundation. JB and JZ thank the Galileo Galilei Institute for
Theoretical Physics for hospitality and the INFN for partial support
during the completion of this work.

%% file: conventions.tex
%!TEX root = paper.tex

\section{Conventions and input}
\label{app:not-conv}

\subsection{Standard model in the unbroken and broken phases}
Here we collect the conventions that we use in the paper. Our
convention for the Lorentz vectors is $p^\mu=(p^0,\vec p\,)$,
$p_\mu=(p^0,-\vec p\,)$, while for the completely antisymmetric
Levi-Civita tensor $\epsilon^{\mu\nu\rho\sigma}$ we use the convention
$\epsilon^{0123}=+1$. The field-strength tensors are
\begin{align}
G_{\mu\nu}^a &=
\partial_{\mu} G_{\nu}^a - \partial_{\nu} G_{\mu}^a - g_s \, f^{abc}
\, G_{\mu}^b G_{\nu}^c\,,
\\
 W_{\mu\nu}^a &= \partial_{\mu} W_{\nu}^a -
\partial_{\nu} W_{\mu}^a + g_2 \, \epsilon^{abc} \, W_{\mu}^b
W_{\nu}^c\,,
\\
B_{\mu\nu} &= \partial_{\mu} B_{\nu} - \partial_{\nu}
B_{\mu}\,.
\end{align}
The $SU(3)$, $SU(2)$, and $U(1)$ coupling constants are $g_s$, $g_2$,
and $g_1$, while $f^{abc}$, $\epsilon^{abc}$ are the completely
antisymmetric $SU(3)$ and $SU(2)$ structure constants, respectively.

The covariant derivative acting on a fermion $f$ is, in our
convention,
\begin{equation}
D_\mu f= \Big(\partial_\mu - i g_s T^a G^a_\mu - i g_2 \tilde \tau^a
W^a_\mu + i g_1 \frac{Y_f}{2} B_\mu\Big)f\,,
\end{equation}
with $T^a, \tilde \tau^a$ the generators of $SU(3)$ and $SU(2)$,
respectively, and $Y_f$ the hypercharge of fermion $f$.  Specializing
to the DM fields we thus have
\begin{equation}\begin{split}\label{eq:ewcovder}
D_\mu \chi = \Big(\partial_\mu 
- i g_2 \tilde \tau^a
W^a_\mu + i g_1 \frac{Y_\chi}{2} B_\mu\Big)\chi\,,
\end{split}\end{equation}
with $Y_\chi$ the DM hypercharge. The $SU(2)$
generators $\tilde \tau^a$ for a general representation of weak isospin
$I_\chi$ can be chosen as
\begin{equation}\begin{split} \label{eq:su2-gen-def}
\big(\tilde \tau^1 \pm i \tilde\tau^2 \big)_{kl}  = \delta_{k,l\pm 1}
\sqrt{(I_\chi \mp l)(I_\chi \pm l + 1)} \, , \qquad 
\big( \tilde\tau^3 \big)_{kl}  = l \delta_{k,l} \, ,
\end{split}\end{equation}
with $k,l$ running over the values $-I_\chi,-I_\chi+1,\ldots , I_\chi
- 1, I_\chi$.

The Higgs Lagrangian in terms of the complex Higgs doublet $H$ reads
\begin{equation}\begin{split}
\mathcal{L}_H = \big( D_\mu H \big)^\dagger D^\mu H -
\frac{\lambda}{4} \big( H^\dagger H \big)^2 + \mu^2
H^\dagger H \, .
\end{split}\end{equation}
In the calculation of the anomalous dimensions above the electroweak
scale the Higgs mass term can be neglected as it does not affect the
UV divergences.

The Yukawa interactions are given by
\begin{equation}\begin{split}
\mathcal{L}_Y = - \sum_{k,l} \bar{Q}_{L}^k Y_{kl}^u u_{R}^l \tilde H -
\sum_{k,l} \bar{Q}_{L}^k Y_{kl}^d d_{R}^l H - \sum_{k,l} \bar{L}_{L}^k
Y_{kl}^\ell \ell_{R}^l H + \text{h.c.}\,,
\end{split}\end{equation}
with $k,l$ the generation indices, while the charge-conjugated Higgs
field is given by $\tilde H = i \sigma^2 H^*$. In the calculation of
the electroweak mixing we neglect the up, down, strange, electron, and
muon Yukawa couplings.

We further complement the Lagrangian involving the matter fields with
a gauge-fixing term. It is most convenient to perform the calculation
in a background-field gauge; the gauge-fixing Lagrangian can be taken
in analogy to the case of QCD~\cite{Abbott:1980hw, Abbott:1981ke}. We
use a generalized $R_\xi$ gauge, with gauge fixing
term~\cite{Denner:1994xt}
\begin{equation}
{\cal L}_\text{gf} = - \frac{1}{2\xi^W} \Big[ \big( \delta^{ac}
  \partial_\mu + g_2 \epsilon^{abc} \hat W_\mu^b \big) W^{c,\mu}
  \Big]^2 - \frac{1}{2\xi^B} \big( \partial_\mu B^\mu \big)^2 \,,
\end{equation}
and checked explicitly the $\xi^W$ and $\xi^B$ gauge-parameter
independence of our results.

After electroweak symmetry breaking we use the mass eigenbasis for the
gauge bosons,
\begin{equation}\begin{split}
W_\mu^\pm = \big( W_\mu^1 \mp i W_\mu^2 \big)/\sqrt{2}\, , \qquad 
\begin{pmatrix}
Z_\mu\\A_\mu
\end{pmatrix}
=
\begin{pmatrix}
c_w&s_w\\-s_w&c_w
\end{pmatrix}
\begin{pmatrix}
W_\mu^3\\B_\mu
\end{pmatrix},
\end{split}\end{equation}
where $c_w \equiv \cos\theta_w = {g_2}/{\sqrt{g_1^2 + g_2^2}}\, , s_w
\equiv \sin\theta_w$.  The electric charge is given by
\begin{equation}\begin{split}
e = \frac{g_1 g_2}{\sqrt{g_1^2 + g_2^2}} = g_2 s_w = g_1 c_w \, .
\end{split}\end{equation}
The electric charge of the components of the DM multiplet is given by
the Gell-Mann Nishijima relation, $Q_\chi = \tilde \tau^3 +
Y_\chi/2$. Defining $\tilde \tau^\pm = \tilde \tau^1 \pm i \tilde
\tau^2$, we can write the covariant derivative \eqref{eq:ewcovder} in
terms of the broken fields as
\begin{equation}\begin{split}\label{eq:ewcovderbroken}
D_\mu = \partial_\mu + i g_s T^a G^a_\mu - \frac{i}{\sqrt{2}}
\frac{e}{s_w} \big( \tilde \tau^- W_\mu^+ + \tilde \tau^+ W_\mu^-
\big) + ie Q_\chi A_\mu - \frac{ie}{s_w c_w} \big( \tilde \tau^3 -
s_w^2 Q_\chi \big) Z_\mu \, .
\end{split}\end{equation}
The Higgs doublet field after EWSB is given by
\begin{equation}\begin{split}
H(x) = 
\begin{pmatrix}
G^+(x)\\
\frac{1}{\sqrt{2}}\big(v + h(x) + iG^0(x)\big)
\end{pmatrix},
\end{split}\end{equation}
where $G^+(x)$ and $G^0(x)$ are the pseudo-Goldstone fields.

\subsection{Numerical inputs for the electroweak running}

The parameters used in our numerics for the electroweak RG evolution
are $\hat g_1$, $\hat g_2$, $\hat g_3$, $\hat y_c$, $\hat y_\tau$,
$\hat y_b$, $\hat y_t$, where the hat denotes the values in the
\text{$\overline{\rm MS}$} scheme at scale $M_Z$. All the numerical
inputs are taken from Ref.~\cite{PDG2018}. Our strategy to determine
the initial values at scale $\mu = M_Z$ is as follows. We use the
values $\sin^2\theta_w (M_Z) = 0.23122(4)$, $\alpha^{-1} (M_Z) =
127.955(10)$ to determine $\hat g_1$ and $\hat g_2$ directly via the
relation
\begin{equation}
  \sin^2\theta_w (\mu) \equiv \frac{g_1^2(\mu)}{g_1^2(\mu) +
    g_2^2(\mu)} \,.
\end{equation}
In this way we find $\hat g_1 = 0.36$, $\hat g_2 = 0.65$. The strong
coupling $\hat g_3 = 1.22$ is determined from $\alpha_s(M_Z) =
0.1181(11)$.

To determine $\hat y_\tau$ we use $m_\tau = 1.77686(12)\,$GeV, and the
relations
\begin{equation}
  y_\tau = \frac{\sqrt{2} m_\tau}{v_{\rm EW}} \,, \qquad G_F = \frac{1}{\sqrt{2}v_{\rm EW}^2} \,.
\end{equation}
Note that $G_F = 1.1663787(6) \times 10^{-5}\,$GeV$^{-2}$ is RG
invariant, and we neglect the QED running of $m_\tau$. We find $\hat
y_\tau(M_Z) = 0.010$.  We obtain $m_c(M_Z)$ by QCD running from
$m_c(m_c) = 1.275(3) \,$GeV and then convert to find $\hat y_c(M_Z) =
0.0045$. We obtain $m_b(M_Z)$ by QCD running from $m_b(m_b) =
4.18^{+0.04}_{-0.03} \,$GeV and then convert to find $\hat y_b(M_Z) =
0.018$. We obtain $m_t(M_Z)$ by converting the top-quark pole mass
$M_t = 173.0(0.4)\,$GeV to the QCD and electroweak
\text{$\overline{MS}$} scheme at scale $\mu = M_t$, and use subsequent
QCD and electroweak running from $\mu = M_t$ to $\mu = M_Z$.
%Since $y_t(M_Z)$ is given in terms of $m_t(M_Z)$, this has to be done
%iteratively.
We find $\hat y_t(M_Z) = 1.05$.
For the Higgs self-coupling $\lambda$, we take $\lambda(M_Z)\sim\lambda(M_h)=2\sqrt{2}\,M_h^2\,G_F\approx 0.52$ with $M_h=125.1$ GeV.

%% file: appcNR.tex
\section{Nonrelativistic coefficients}\label{sec:cNR}

The operators ${\cal Q}_{23,q}^{(7,0)}$ and ${\cal Q}_{25}^{(7,0)}$,
defined in Eq.~\eqref{eq:twist2:op}, lead to the following additional
contributions to the nonrelativistic coefficients:
\begin{align}
\label{eq:cNR1}
c_{1}^p & = \frac{3}{4} m_p \Big( \hat \C_{25}^{(7)} \, f_{g,p}^{(2)}
+ \sum_q \hat \C_{23,q}^{(7)} \, f_{q,p}^{(2)} \Big) \,,
\end{align}
with $f_{u,p}^{(2)} = 0.346(7)$, $f_{d,p}^{(2)} = 0.192(6)$,
$f_{s,p}^{(2)} = 0.034(3)$, and $f_{g,p}^{(2)} = 0.419(19)$, evaluated
at renormalizations scale $\mu=2\,$GeV~\cite{Hill:2014yxa}. The
coefficients for neutrons can be obtained by the exchange $p
\leftrightarrow n$, $u \leftrightarrow d$.

%% file: appWarsaw.tex
%!TEX root = paper.tex

\section{SM EFT operators}
\label{app:SMEFT}
In this appendix we provide the results for the mixing of the SM-DM
operators into the pure SM operators, restricting the discussion to
the SM operators that enter the RG running at one loop. Assuming
conservation of lepton and baryon number, only dimension-six operators
are relevant. The dimension-six effective Lagrangian is (we use the
basis in Ref.~\cite{Grzadkowski:2010es}, but with renamed operators)
\begin{equation}
{\cal L} = \sum_a \frac{C_a^{\text{\sc sm},(6)}}{\Lambda^2} S_a^{(6)} \,,
\end{equation}
where the operators involving only quark fields are
\begin{align}
S_{1,ij}^{(6)} &= (\bar Q_L^i \gamma_\mu \tau^a Q_L^i)(\bar Q_L^j \gamma^\mu 
\tau^a Q_L^j)\,, & S_{2,ij}^{(6)} &= (\bar Q_L^i \gamma_\mu Q_L^i)(\bar Q_L^j 
\gamma^\mu Q_L^j)\,, \label{eq:smeft:QQQQ}\\
S_{3,ij}^{(6)} &= (\bar Q_L^i \gamma_\mu Q_L^i)(\bar u_R^j \gamma^\mu u_R^j)\,, & S_{4,ij}^{(6)} &= (\bar Q_L^i \gamma_\mu Q_L^i)(\bar d_R^j \gamma^\mu d_R^j)\,,\\
S_{5,ij}^{(6)} &= (\bar u_R^i \gamma_\mu u_R^i)(\bar u_R^j \gamma^\mu u_R^j)\,, & S_{6,ij}^{(6)} &= (\bar u_R^i \gamma_\mu u_R^i)(\bar d_R^j \gamma^\mu d_R^j)\,,\\
S_{7,ij}^{(6)} &= (\bar d_R^i \gamma_\mu d_R^i)(\bar d_R^j \gamma^\mu d_R^j)\,. & & 
\end{align}
The operators involving only lepton fields can be chosen as
\begin{align}
S_{8,ij}^{(6)} &= (\bar L_L^i \gamma_\mu L_L^i)(\bar L_L^j \gamma^\mu L_L^j) \,, 
& S_{9,ij}^{(6)} &= (\bar L_L^i \gamma_\mu L_L^i)(\bar \ell_R^j \gamma^\mu \ell_R^j) \,,\\
S_{10,ij}^{(6)}  &= (\bar \ell_R^i \gamma_\mu \ell_R^i)(\bar \ell_R^j \gamma^\mu \ell_R^j) \,. 
& &
\end{align}
The mixed quark--lepton operators are
\begin{align}
S_{11,ij}^{(6)} &= (\bar Q_L^i \gamma_\mu \tau^a Q_L^i)(\bar L_L^j \gamma^\mu \tau^a L_L^j)\,, 
& S_{12,ij}^{(6)} &= (\bar Q_L^i \gamma_\mu Q_L^i)(\bar L_L^j \gamma^\mu L_L^j)\,, \\
S_{13,ij}^{(6)} &= (\bar Q_L^i \gamma_\mu Q_L^i)(\bar \ell_R^j \gamma^\mu \ell_R^j)\,, 
& S_{14,ij}^{(6)} &= (\bar u_R^i \gamma^\mu u_R^i)(\bar L_L^j \gamma_\mu L_L^j)\,, \\
S_{15,ij}^{(6)} &= (\bar d_R^i \gamma^\mu d_R^i)(\bar L_L^j \gamma_\mu L_L^j)\,, 
& S_{16,ij}^{(6)} & = (\bar u_R^i \gamma^\mu u_R^i)(\bar \ell_R^j \gamma_\mu \ell_R^j)\,, \\
S_{17,ij}^{(6)} &= (\bar d_R^i \gamma^\mu d_R^i)(\bar \ell_R^j \gamma_\mu 
\ell_R^j)\,, \label{eq:smeft:ddee}& & 
\end{align}
while the Higgs-fermion operators can be taken as
\begin{align}
S_{18,i}^{(6)} &= (\bar Q_L^i \gamma^\mu \tau^a Q_L^i)(H^\dagger i\lrD{}^a_\mu H)\,,
 & S_{19,i}^{(6)} &= (\bar Q_L^i \gamma^\mu Q_L^i)(H^\dagger i\lrD_\mu H)\,,\label{eq:dim6:Q15Q17:SM}\\ 
S_{20,i}^{(6)} &= (\bar u_R^i \gamma^\mu u_R^i)(H^\dagger i\lrD_\mu H) \,, 
& S_{21,i}^{(6)} &= (\bar d_R^i \gamma^\mu d_R^i)(H^\dagger i\lrD_\mu H)\,,\\
S_{22,i}^{(6)} &= (\bar L_L^i \gamma^\mu \tau^a L_L^i)(H^\dagger i\lrD{}^a_\mu H)\,, 
& S_{23,i}^{(6)} &= (\bar L_L^i \gamma^\mu L_L^i)(H^\dagger i\lrD_\mu H)\,,\\
S_{24,i}^{(6)} &= (\bar \ell_R^i \gamma^\mu \ell_R^i)(H^\dagger i\lrD_\mu H)\,.&  &
\end{align}
The remaining operator, involving only Higgs fields, is
\begin{equation}
S_{25}^{(6)} = (H^\dagger i\lrD_\mu H)(H^\dagger 
i\lrD^\mu H) \,.\label{eq:smeft:hhhh}
\end{equation}

The mixing of the SM sector into the DM-SM sector proceeds via penguin
insertions. The nonzero results for the four-fermion operators are,
for $i=j$,
\begin{equation}
\big[\gamma_1^{(0)} \big]_{S_{1\ldots 7,ii}^{(6)}\times Q_{2\ldots 4,i}^{(6)}}= Y_\chi
\begin{pmatrix}
\frac{1}{6}& 0& 0\\
 \frac{14}{9}& 0& 0\\
 \frac{4}{3}& \frac{2}{3}& 0\\
 - \frac{2}{3}& 0&\frac{2}{3}\\
 0&\frac{32}{9}& 0\\
 0& - \frac{2}{3}&\frac{4}{3}\\
 0& 0& - \frac{16}{9}\\
\end{pmatrix},
\quad
\big[\gamma_1^{(0)} \big]_{S_{12\ldots 17,ii}^{(6)}\times Q_{2\ldots 4,i}^{(6)}}= Y_\chi
\begin{pmatrix}
 - \frac{2}{3}& 0& 0\\
 - \frac{2}{3} & 0& 0\\
 0& - \frac{2}{3}& 0\\
 0& 0& - \frac{2}{3}\\
 0& - \frac{2}{3} & 0\\
 0& 0& - \frac{2}{3} 
\end{pmatrix},
\end{equation}
for mixing into operators with quark currents, and
\begin{equation}
\big[\gamma_1^{(0)} \big]_{S_{8\ldots 10,ii}^{(6)}\times Q_{10,11,i}^{(6)}}= 
Y_\chi 
\begin{pmatrix}
 - 2& 0\\
 - \frac{2}{3} & - \frac{2}{3}\\
 0& - \frac{8}{3} \\
\end{pmatrix},
\quad
\big[\gamma_1^{(0)} \big]_{S_{12\ldots 17,ii}^{(6)}\times Q_{10,11,i}^{(6)}}= 
Y_\chi 
\begin{pmatrix}
 \frac{2}{3}& 0&\\
  0&\frac{2}{3}&\\
 \frac{4}{3}& 0&\\
 - \frac{2}{3}& 0&\\
 0&\frac{4}{3}&\\
 0& - \frac{2}{3}
\end{pmatrix}.
\end{equation}
for mixing into operators with lepton currents.

The mixing proportional to $g_2$ has only a few non-vanishing entries,
given by
\begin{equation}
\begin{split}
&\big[\gamma_2^{(0)} \big]_{S_{1,ii}^{(6)} Q_{1,i}^{(6)}}=\frac{10}{3}\,,\quad
\big[\gamma_2^{(0)} \big]_{S_{2,ii}^{(6)} Q_{1,i}^{(6)}}=\big[\gamma_2^{(0)} \big]_{S_{8,ii}^{(6)} Q_{9,i}^{(6)}}=\frac{8}{3}\,,\\
&\big[\gamma_2^{(0)} \big]_{S_{11,ii}^{(6)} Q_{1,i}^{(6)}}=\frac{2}{3}\,,\quad
\big[\gamma_2^{(0)} \big]_{S_{11,ii}^{(6)} Q_{9,i}^{(6)}}=2\,.
\end{split}
\end{equation}
All the other entries are zero. 

The result for $i\neq j$ are (note that
the order of the flavor indices matters, except when the operator is
symmetric in $i$ and $j$)
\begin{align}
\big[\gamma_1^{(0)} \big]_{S_{2\dots 4,ij}^{(6)}\times Q_{2,i}^{(6)}}&=
Y_\chi\left(\frac{2}{3},\frac{4}{3},-\frac{2}{3}\right)\,, 
\\
\big[\gamma_1^{(0)} \big]_{S_{8\dots 10,ij}^{(6)}\times Q_{10\dots 11,i}^{(6)}}&=
\big[\gamma_1^{(0)} \big]_{S_{12\dots 14,ij}^{(6)}\times Q_{2\dots 3,i}^{(6)}}=
Y_\chi\begin{pmatrix}
-\frac{2}{3} & 0 \\
-\frac{2}{3} & 0 \\
0 & -\frac{2}{3}\,
\end{pmatrix},
\end{align}
as well as
\begin{align}
\big[\gamma_1^{(0)} \big]_{S_{5\dots 7,ij}^{(6)}\times Q_{3\dots 4,i}^{(6)}}&=
Y_\chi\begin{pmatrix}
\frac{4}{3} & 0 \\
-\frac{2}{3} & 0 \\
0 & -\frac{2}{3} \\
\end{pmatrix},
\quad
&\big[\gamma_1^{(0)} \big]_{S_{15\dots 17,ij}^{(6)}\times Q_{3\dots 4,i}^{(6)}}&=
Y_\chi\begin{pmatrix}
 0 & -\frac{2}{3} \\
 -\frac{2}{3} & 0 \\
 0 & -\frac{2}{3} \\
\end{pmatrix},
\\
\big[\gamma_1^{(0)} \big]_{S_{3\dots 6,ji}^{(6)}\times Q_{3\dots 4,i}^{(6)}}&=
Y_\chi\begin{pmatrix}
\frac{2}{3} & 0 \\
0 & \frac{2}{3} \\
\frac{4}{3} & 0 \\
0 & \frac{4}{3}
\end{pmatrix},\quad
&\big[\gamma_1^{(0)} \big]_{S_{14\dots 17,ji}^{(6)}\times Q_{10\dots 11,i}^{(6)}}&=
Y_\chi\begin{pmatrix}
\frac{4}{3} & 0 \\
-\frac{2}{3} & 0 \\
0 & \frac{4}{3} \\
0 & -\frac{2}{3}
\end{pmatrix},
\end{align}
and
\begin{align}
&\big[\gamma_1^{(0)} \big]_{S_{9,ji}^{(6)}\times Q_{11,i}^{(6)}}=-\frac{2Y_\chi}{3}\,,&
&\big[\gamma_1^{(0)} \big]_{S_{12\dots 13,ji}^{(6)}\times Q_{10\dots 11,i}^{(6)}}=
Y_\chi\diag\left(\frac{2}{3}, \frac{2}{3} \right)\,,
\\
&\big[\gamma_2^{(0)} \big]_{S_{1,ij}^{(6)} Q_{1,i}^{(6)}}=
\big[\gamma_2^{(0)} \big]_{S_{11,ji}^{(6)} Q_{9,i}^{(6)}}=2\,,&
&\big[\gamma_2^{(0)} \big]_{S_{11,ij}^{(6)} Q_{1,i}^{(6)}}=\frac{2}{3}\,. 
\end{align}
All the other entries are zero.

The nonzero mixings of the operators involving Higgs currents are given by
\begin{align}
&\big[\gamma_2^{(0)} \big]_{S_{18,i}^{(6)} Q_{1,i}^{(6)}}=\big[\gamma_2^{(0)} \big]_{S_{22,i}^{(6)} Q_{9,i}^{(6)}}=\frac{1}{3}\,,&
&\big[\gamma_2^{(0)} \big]_{S_{18,i}^{(6)} Q_{15}^{(6)}}=2\,,\\
&\big[\gamma_2^{(0)} \big]_{S_{22,i}^{(6)} Q_{15}^{(6)}}=\frac{2}{3}\,,&
&\big[\gamma_2^{(0)} \big]_{S_{25}^{(6)} Q_{15}^{(6)}}=\frac{2}{3}\,,
\end{align}
as well as by
\begin{align}
&\big[\gamma_1^{(0)} \big]_{S_{19,i}^{(6)}\cdots S_{21,i}^{(6)}\times Q_{2,i}^{(6)}\cdots Q_{4,i}^{(6)}}= 
\tfrac{1}{3} Y_\chi \diag\big(1,1,1\big)\,,
\\
&\big[\gamma_1^{(0)} \big]_{S_{23,i}^{(6)}, S_{24,i}^{(6)}\times Q_{10,i}^{(6)}, Q_{11,i}^{(6)}}= 
\tfrac{1}{3} Y_\chi \diag\big(1,1\big)\,,
\\
&\big[\gamma_2^{(0)} \big]_{S_{18,i}^{(6)} Q_{1,i}^{(6)}} =
\big[\gamma_2^{(0)} \big]_{S_{22,i}^{(6)} Q_{9,i}^{(6)}} = \frac{1}{3}\,,
\end{align}
and
\begin{align}
&\big[\gamma_1^{(0)} \big]_{S_{19,i}^{(6)} Q_{16}^{(6)}}=\tfrac{2}{ 3} Y_\chi\,,\quad
\big[\gamma_1^{(0)} \big]_{S_{20,i}^{(6)} Q_{16}^{(6)}}=\tfrac{4}{3} Y_\chi \,,\quad
\big[\gamma_1^{(0)} \big]_{S_{21,i}^{(6)} Q_{16}^{(6)}}=- \tfrac{2}{3} Y_\chi\,,\\
&\big[\gamma_1^{(0)} \big]_{S_{23,i}^{(6)} Q_{16}^{(6)}}=
\big[\gamma_1^{(0)} \big]_{S_{24,i}^{(6)} Q_{16}^{(6)}}=-\tfrac{2}{3}Y_\chi \,,\quad
\big[\gamma_1^{(0)} \big]_{S_{25}^{(6)} Q_{16}^{(6)}}=\tfrac{5}{6} Y_\chi \,.
\end{align}
All the other entries vanish.

The mixing of the DM-SM sector into the SM sector also proceeds only
via penguin insertions. The SM four fermion operators in
Eqs.~\eqref{eq:smeft:QQQQ}-\eqref{eq:smeft:ddee} carry two generation
indices, where the order of the indices is important.  First, we
present the anomalous dimensions proportional to $g_1^2$. The mixing
of $Q_{2,i}^{(6)}$-$Q_{4,i}^{(6)}$ into
$S_{2,ij}^{(6)}$-$S_{7,ij}^{(6)}$ is given by
\begin{align}
\big[\gamma_1^{(0)} \big]_{Q_{2,i}^{(6)}, S_{2,ij}^{(6)}} & = 
\big[\gamma_1^{(0)} \big]_{Q_{3,i}^{(6)}, S_{3,ji}^{(6)}} = 
\big[\gamma_1^{(0)} \big]_{Q_{4,i}^{(6)}, S_{4,ji}^{(6)}} = \tfrac{2}{9} Y_\chi 
d_\chi\,,\\
\big[\gamma_1^{(0)} \big]_{Q_{2,i}^{(6)}, S_{3,ij}^{(6)}} & = 
\big[\gamma_1^{(0)} \big]_{Q_{3,i}^{(6)}, S_{5,ij}^{(6)}} = 
\big[\gamma_1^{(0)} \big]_{Q_{4,i}^{(6)}, S_{6,ji}^{(6)}} = \tfrac{8}{9} Y_\chi 
d_\chi\,,\\
\big[\gamma_1^{(0)} \big]_{Q_{2,i}^{(6)}, S_{4,ij}^{(6)}} & = 
\big[\gamma_1^{(0)} \big]_{Q_{3,i}^{(6)}, S_{6,ij}^{(6)}} = 
\big[\gamma_1^{(0)} \big]_{Q_{4,i}^{(6)}, S_{7,ij}^{(6)}} = -\tfrac{4}{9} Y_\chi d_\chi\,,
\end{align}
while the remaining entries are zero. The mixing of
$Q_{5,i}^{(6)}$-$Q_{18}^{(6)}$ into $S_{1,ij}^{(6)}$-$S_{7,ij}^{(6)}$
vanishes. The mixing of $Q_{2,i}^{(6)}$-$Q_{4,i}^{(6)}$ into
$S_{12,ij}^{(6)}$-$S_{17,ij}^{(6)}$ is given by
\begin{align}
	\big[\gamma_1^{(0)} \big]_{Q_{2,i}^{(6)}, S_{12,ij}^{(6)}} & = 
	\big[\gamma_1^{(0)} \big]_{Q_{3,i}^{(6)}, S_{14,ij}^{(6)}} = 
	\big[\gamma_1^{(0)} \big]_{Q_{4,i}^{(6)}, S_{15,ij}^{(6)}} = -\tfrac{2}{3} 
	Y_\chi 
	d_\chi\,,\\
	\big[\gamma_1^{(0)} \big]_{Q_{2,i}^{(6)}, S_{13,ij}^{(6)}} & = 
	\big[\gamma_1^{(0)} \big]_{Q_{3,i}^{(6)}, S_{16,ij}^{(6)}} = 
	\big[\gamma_1^{(0)} \big]_{Q_{4,i}^{(6)}, S_{17,ij}^{(6)}} = -\tfrac{4}{3} 
	Y_\chi 
	d_\chi\,.
\end{align}
The mixing of $Q_{1,i}^{(6)}$-$Q_{4,i}^{(6)}$ into
$S_{12,ji}^{(6)}$-$S_{17,ji}^{(6)}$, with reversed indices, vanishes.
The mixing of $Q_{1,i}^{(6)}$-$Q_{4,i}^{(6)}$ into
$S_{19,i}^{(6)}$-$S_{21,i}^{(6)}$ is given by
\begin{equation}
\big[\gamma_1^{(0)} \big]_{Q_{2,i}^{(6)}, S_{19,i}^{(6)}} = 
\big[\gamma_1^{(0)} \big]_{Q_{3,i}^{(6)}, S_{20,i}^{(6)}} = 
\big[\gamma_1^{(0)} \big]_{Q_{4,i}^{(6)}, S_{21,i}^{(6)}} = \tfrac{2}{3} 
Y_\chi 
d_\chi\,.
\end{equation}
The mixing of $Q_{10,i}^{(6)}$, $Q_{11,i}^{(6)}$ into the operators
$S_{8,ij}^{(6)}$-$S_{10,ij}^{(6)}$
is given by
\begin{align}
\big[\gamma_1^{(0)} \big]_{Q_{10,i}^{(6)}, S_{8,ij}^{(6)}} &= 
\big[\gamma_1^{(0)} \big]_{Q_{11,i}^{(6)}, S_{9,ji}^{(6)}} = -\tfrac{2}{3} 
Y_\chi d_\chi\,,\\
\big[\gamma_1^{(0)} \big]_{Q_{10,i}^{(6)}, S_{9,ij}^{(6)}} &= 
\big[\gamma_1^{(0)} \big]_{Q_{11,i}^{(6)}, S_{10,ij}^{(6)}} = -\tfrac{4}{3} 
Y_\chi d_\chi\,.
\end{align}
The mixing of $Q_{10,i}^{(6)}$, $Q_{11,i}^{(6)}$ into the operators
$S_{12,ij}^{(6)}$-$S_{17,ij}^{(6)}$ is given by 
\begin{align}
\big[\gamma_1^{(0)} \big]_{Q_{10,i}^{(6)}, S_{12,ji}^{(6)}} &= 
\big[\gamma_1^{(0)} \big]_{Q_{11,i}^{(6)}, S_{13,ji}^{(6)}} = \tfrac{2}{9} 
Y_\chi d_\chi\,,\\
\big[\gamma_1^{(0)} \big]_{Q_{10,i}^{(6)}, S_{14,ji}^{(6)}} &= 
\big[\gamma_1^{(0)} \big]_{Q_{11,i}^{(6)}, S_{16,ji}^{(6)}} = \tfrac{8}{9} 
Y_\chi d_\chi\,,\\
\big[\gamma_1^{(0)} \big]_{Q_{10,i}^{(6)}, S_{15,ji}^{(6)}} &= 
\big[\gamma_1^{(0)} \big]_{Q_{11,i}^{(6)}, S_{17,ji}^{(6)}} = -\tfrac{4}{9} 
Y_\chi d_\chi\,,
\end{align}
whereas the mixing $Q_{10,i}^{(6)}$, $Q_{11,i}^{(6)}$ into the
operators $S_{12,ij}^{(6)}$-$S_{17,ij}^{(6)}$ vanishes. The mixing of
$Q_{10,i}^{(6)}$, $Q_{11,i}^{(6)}$ into the operators
$S_{23,i}^{(6)}$, $S_{24,i}^{(6)}$ is given by
\begin{equation}
\big[\gamma_1^{(0)} \big]_{Q_{10,i}^{(6)} S_{23,i}^{(6)}} =
\big[\gamma_1^{(0)} \big]_{Q_{11,i}^{(6)} S_{24,i}^{(6)}} =
\tfrac{2}{3}Y_\chi d_\chi\,.
\end{equation}
The mixing of Higgs-DM into SM is given by
\begin{equation}
\big[\gamma_1^{(0)} \big]_{Q_{16,i}^{(6)}\times S_{19,i}^{(6)}\cdots S_{25,i}^{(6)}}= 
Y_\chi d_\chi
\begin{pmatrix}
 \frac{2}{9}&\frac{8}{9}& - \frac{4}{9}& 0& - \frac{2}{3}& - \frac{4}{3}&\frac{2}{3}
\end{pmatrix}.
\end{equation}
The mixing proportional to $g_2$ has only a few non-vanishing entries,
given by
\begin{align}
\begin{split}
\big[\gamma_2^{(0)} \big]_{Q_{1,i}^{(6)} S_{1,ij}^{(6)}}&=
\big[\gamma_2^{(0)} \big]_{Q_{1,i}^{(6)} S_{11,ij}^{(6)}}=
\big[\gamma_2^{(0)} \big]_{Q_{1,i}^{(6)} S_{18,ij}^{(6)}}=
\big[\gamma_2^{(0)} \big]_{Q_{9,i}^{(6)} S_{11,ij}^{(6)}}\\
&=\big[\gamma_2^{(0)} \big]_{Q_{9,i}^{(6)} S_{22,i}^{(6)}}=
\big[\gamma_2^{(0)} \big]_{Q_{15}^{(6)} S_{18,i}^{(6)}}=
\big[\gamma_2^{(0)} \big]_{Q_{15}^{(6)} S_{22,i}^{(6)}}
=\tfrac{8}{9} {\cal J}_\chi d_\chi\,,
\end{split}
\\
\big[\gamma_2^{(0)} \big]_{Q_{9,i}^{(6)} S_{8,ij}^{(6)}}&=
\big[\gamma_2^{(0)} \big]_{Q_{15}^{(6)} S_{25}^{(6)}}
=\tfrac{2}{9} {\cal J}_\chi d_\chi\,.
\end{align}
Again, all the undisplayed entries vanish. The mixing of the SM
operators among themselves can be taken from the
literature~\cite{Jenkins:2013zja, Jenkins:2013wua, Alonso:2013hga}.

%% file: appDM.tex
%!TEX root = paper.tex

\section{Mixing in the dark sector}
\label{app:DM}
In this appendix we provide the results for the mixing of the
operators in the SM-DM sector into the pure DM operators. We write the
dimension-six effective Lagrangian as
\begin{equation}
{\cal L} = \sum_a \frac{C_a^{\text{\sc dm},(6)}}{\Lambda^2} S_a^{(6)} \,,
\end{equation}
where the relevant operators are given in Eq.~\eqref{eq:op:DM-DM}
(recall that we neglect the mixing of operators within the dark
sector). 

The mixing of DM-SM operators into DM operators is given by
\begin{equation}
\big[\gamma_1^{(0)} \big]_{Q_{2\ldots 4,i}^{(6)} \times
  D_{1}^{(6)}}= 
\big[\gamma_1^{(0)} \big]_{Q_{6\ldots 8,i}^{(6)} \times
  D_{2}^{(6)}}= 
Y_\chi 
\begin{pmatrix}
 \frac{2}{3}\\
 \frac{4}{3}\\
  - \frac{2}{3}
\end{pmatrix},
\end{equation}

\begin{equation}
\big[\gamma_1^{(0)} \big]_{Q_{10,11,i}^{(6)} \times
  D_{1}^{(6)}}= 
\big[\gamma_1^{(0)} \big]_{Q_{13,14,i}^{(6)} \times
  D_{2}^{(6)}}= 
Y_\chi 
\begin{pmatrix}
 - \frac{2}{3}\\
 - \frac{2}{3}
\end{pmatrix},
\end{equation}

\begin{equation}
\big[\gamma_1^{(0)} \big]_{Q_{16}^{(6)} \times
  D_{1}^{(6)}}= 
\big[\gamma_1^{(0)} \big]_{Q_{18}^{(6)} \times
  D_{2}^{(6)}}= 
\tfrac{1}{3}Y_\chi \,,
\end{equation}
and
\begin{align}
\big[\gamma_2^{(0)} \big]_{Q_{1,i}^{(6)} D_3 ^{(6)}}&=
\big[\gamma_2^{(0)} \big]_{Q_{5,i}^{(6)} D_4 ^{(6)}}= 2 \,,\\
\big[\gamma_2^{(0)} \big]_{Q_{9,i}^{(6)} D_3 ^{(6)}}&=
\big[\gamma_2^{(0)} \big]_{Q_{12,i}^{(6)} D_4 ^{(6)}}=
\tfrac{2}{3} \,,\\
\big[\gamma_2^{(0)} \big]_{Q_{15}^{(6)} D_3 ^{(6)}}&=
\big[\gamma_2^{(0)} \big]_{Q_{17}^{(6)} D_4 ^{(6)}}=
\tfrac{1}{3} \,,
\end{align}
while the mixing of the DM operators into the DM-SM sector is given by
\begin{align}
\big[\gamma_1^{(0)} \big]_{D_{1}^{(6)} \cdots D_{4}^{(6)} \times
  Q_{1,i}^{(6)} \cdots Q_{4,i}^{(6)}}&= 
Y_\chi 
\begin{pmatrix}
 0&\frac{2}{9} + \frac{4}{9} d_\chi&\frac{8}{9} + \frac{16}{9} d_\chi& - \frac{4}{9} - \frac{8}{9} d_\chi\\
 0& 0& 0& 0\\
 0&\frac{2}{9} {\cal J}_\chi&\frac{8}{9} {\cal J}_\chi& - \frac{4}{9} {\cal J}_\chi\\
 0& 0& 0& 0\\
\end{pmatrix},
\\
\big[\gamma_1^{(0)} \big]_{D_{1}^{(6)} \cdots D_{4}^{(6)} \times
  Q_{5,i}^{(6)} \cdots Q_{8,i}^{(6)}}&= 
Y_\chi 
\begin{pmatrix}
 0& 0& 0& 0\\
 0&\frac{2}{9} + \frac{2}{9} d_\chi&\frac{8}{9} + \frac{8}{9} d_\chi& - \frac{4}{9} - \frac{4}{9} d_\chi\\
 0& 0& 0& 0\\
 0&\frac{2}{9} {\cal J}_\chi&\frac{8}{9} {\cal J}_\chi& - \frac{4}{9} {\cal J}_\chi
\end{pmatrix},
\\
\big[\gamma_1^{(0)} \big]_{D_{1}^{(6)} \cdots D_{4}^{(6)} \times
  Q_{9,i}^{(6)} \cdots Q_{14,i}^{(6)}}&= 
Y_\chi 
\begin{pmatrix}
 0& - \frac{2}{3} - \frac{4}{3} d_\chi& - \frac{4}{3} - \frac{8}{3} d_\chi& 0& 0& 0\\
 0& 0& 0& 0& - \frac{2}{3} - \frac{2}{3} d_\chi& - \frac{4}{3} - \frac{4}{3} d_\chi\\
 0& - \frac{2}{3} {\cal J}_\chi& - \frac{4}{3} {\cal J}_\chi& 0& 0& 0\\
 0& 0& 0& 0& - \frac{2}{3} {\cal J}_\chi& - \frac{4}{3} {\cal J}_\chi
\end{pmatrix},
\\
\big[\gamma_1^{(0)} \big]_{D_{1}^{(6)} \cdots D_{4}^{(6)} \times
  Q_{15}^{(6)} \cdots Q_{18}^{(6)}}&= 
Y_\chi 
\begin{pmatrix}
 0&\frac{2}{3} + \frac{4}{3} d_\chi& 0& 0\\
 0& 0& 0&\frac{2}{3} + \frac{2}{3} d_\chi\\
 0&\frac{2}{3} {\cal J}_\chi& 0& 0\\
 0& 0& 0&\frac{2}{3} {\cal J}_\chi
\end{pmatrix},
\end{align}
and
\begin{align}
\big[\gamma_2^{(0)} \big]_{D_1 ^{(6)} Q_{1,i}^{(6)}}&=
\big[\gamma_2^{(0)} \big]_{D_2 ^{(6)} Q_{5,i}^{(6)}}=
\big[\gamma_2^{(0)} \big]_{D_1 ^{(6)} Q_{9,i}^{(6)}}=
\big[\gamma_2^{(0)} \big]_{D_2 ^{(6)} Q_{12,i}^{(6)}}\\
&=\big[\gamma_2^{(0)} \big]_{D_1 ^{(6)} Q_{15}^{(6)}}=
\big[\gamma_2^{(0)} \big]_{D_2 ^{(6)} Q_{17}^{(6)}}=
\tfrac{8}{3}\,,\\
\big[\gamma_2^{(0)} \big]_{D_3 ^{(6)} Q_{1,i}^{(6)}}&=
\big[\gamma_2^{(0)} \big]_{D_3 ^{(6)} Q_{9,i}^{(6)}}=
\big[\gamma_2^{(0)} \big]_{D_3 ^{(6)} Q_{15}^{(6)}}=
\big( \tfrac{8}{3} + \tfrac{16}{9} d_\chi \big) {\cal J}_\chi - \tfrac{8}{3}\,,\\
\big[\gamma_2^{(0)} \big]_{D_4 ^{(6)} Q_{5,i}^{(6)}}&=
\big[\gamma_2^{(0)} \big]_{D_4 ^{(6)} Q_{12,i}^{(6)}}=
\big[\gamma_2^{(0)} \big]_{D_4 ^{(6)} Q_{17}^{(6)}}=
\big( \tfrac{8}{3} + \tfrac{8}{9} d_\chi \big) {\cal J}_\chi - \tfrac{8}{3}\,.
\end{align}
All non-displayed entries vanish.

%% file: appUnphys.tex
%!TEX root = paper.tex
\section{Unphysical operators}
\label{sec:unphys}
We extract the anomalous dimensions by renormalizing off-shell Greens
functions in $d=4-2\epsilon$ dimensions.  In the intermediate stages
of the computation it is thus necessary to introduce unphysical
operators.

\subsection{Evanescent operators}
The one-loop mixing among the ``physical'' operators is not affected
by the definition of {\it evanescent} operators, i.e., operators that
are required to project one-loop Green's functions in $d=4-2\epsilon$
dimensions but vanish in $d=4$. Nevertheless, for completeness and
future reference we list below the ones we used for the one-loop
computations. The evanescent operators with quark fields are chosen as
\begin{align}
E_{1,i}^{(6)} &= (\bar\chi\gamma_\mu\gamma_\nu\gamma_\rho
\tilde\tau^a\chi)(\bar Q_L^i \gamma^\mu\gamma^\nu\gamma^\rho
\tau^a Q_L^i) - 10 Q_{1,i}^{(6)} + 6 Q_{5,i}^{(6)}\,,\\
E_{2,i}^{(6)} &= (\bar\chi\gamma_\mu\gamma_\nu\gamma_\rho\gamma_5
\tilde\tau^a\chi)(\bar Q_L^i \gamma^\mu\gamma^\nu\gamma^\rho
\tau^a Q_L^i) + 6 Q_{1,i}^{(6)} - 10 Q_{5,i}^{(6)}\,,\\
E_{3,i}^{(6)} &= (\bar\chi\gamma_\mu\gamma_\nu\gamma_\rho
\chi)(\bar Q_L^i \gamma^\mu\gamma^\nu\gamma^\rho Q_L^i) - 10
Q_{2,i}^{(6)} + 6 Q_{6,i}^{(6)}\,,\\ 
E_{4,i}^{(6)} &= (\bar\chi\gamma_\mu\gamma_\nu\gamma_\rho\gamma_5
\chi)(\bar Q_L^i \gamma^\mu\gamma^\nu\gamma^\rho Q_L^i) + 6 Q_{2,i}^{(6)}
- 10 Q_{6,i}^{(6)}\,,\\ 
E_{5,i}^{(6)} &= (\bar\chi\gamma_\mu\gamma_\nu\gamma_\rho
\chi)(\bar u_R^i \gamma^\mu\gamma^\nu\gamma^\rho u_R^i) - 10
Q_{3,i}^{(6)} - 6 Q_{7,i}^{(6)}\,,\\
E_{6,i}^{(6)} &= (\bar\chi\gamma_\mu\gamma_\nu\gamma_\rho\gamma_5
\chi)(\bar u_R^i \gamma^\mu\gamma^\nu\gamma^\rho u_R^i) - 6 Q_{3,i}^{(6)}
- 10 Q_{7,i}^{(6)}\,,\\
E_{7,i}^{(6)} &= (\bar\chi\gamma_\mu\gamma_\nu\gamma_\rho
\chi)(\bar d_R^i \gamma^\mu\gamma^\nu\gamma^\rho d_R^i) - 10
Q_{4,i}^{(6)} - 6 Q_{8,i}^{(6)}\,,\\
E_{8,i}^{(6)} &= (\bar\chi\gamma_\mu\gamma_\nu\gamma_\rho\gamma_5
\chi)(\bar d_R^i \gamma^\mu\gamma^\nu\gamma^\rho d_R^i) - 6 Q_{4,i}^{(6)}
- 10 Q_{8,i}^{(6)}\,,
\end{align}
while the evanescent operators involving lepton fields are
\begin{align}
E_{9,i}^{(6)} &= (\bar\chi\gamma_\mu\gamma_\nu\gamma_\rho
\tilde\tau^a\chi)(\bar L_L^i \gamma^\mu\gamma^\nu\gamma^\rho
\tau^a L_L^i) - 10 Q_{9,i}^{(6)} + 6 Q_{12,i}^{(6)}\,,\\
E_{10,i}^{(6)} &= (\bar\chi\gamma_\mu\gamma_\nu\gamma_\rho\gamma_5
\tilde\tau^a\chi)(\bar L_L^i \gamma^\mu\gamma^\nu\gamma^\rho
\tau^a L_L^i) + 6 Q_{9,i}^{(6)} - 10 Q_{12,i}^{(6)}\,,\\
E_{11,i}^{(6)} &= (\bar\chi\gamma_\mu\gamma_\nu\gamma_\rho
\chi)(\bar L_L^i \gamma^\mu\gamma^\nu\gamma^\rho L_L^i) - 10
Q_{10,i}^{(6)} + 6 Q_{13,i}^{(6)}\,,\\ 
E_{12,i}^{(6)} &= (\bar\chi\gamma_\mu\gamma_\nu\gamma_\rho\gamma_5
\chi)(\bar L_L^i \gamma^\mu\gamma^\nu\gamma^\rho L_L^i) + 6 Q_{10,i}^{(6)}
- 10 Q_{13,i}^{(6)}\,,\\ 
E_{13,i}^{(6)} &= (\bar\chi\gamma_\mu\gamma_\nu\gamma_\rho
\chi)(\bar \ell_R^i \gamma^\mu\gamma^\nu\gamma^\rho \ell_R^i) - 10
Q_{11,i}^{(6)} - 6 Q_{14,i}^{(6)}\,,\\
E_{14,i}^{(6)} &= (\bar\chi\gamma_\mu\gamma_\nu\gamma_\rho\gamma_5
\chi)(\bar \ell_R^i \gamma^\mu\gamma^\nu\gamma^\rho \ell_R^i) - 6 Q_{11,i}^{(6)}
- 10 Q_{14,i}^{(6)}\,.
\end{align}

\subsection{E.o.m.-vanishing operators}

The equations of motion (e.o.m.) for the $W$ and $B$ gauge-boson field
are, in our conventions,
\begin{equation}
\begin{split}
D^\nu  W^{a}_{\nu\mu} \equiv (\partial^\nu \delta^{ab} - g_2
\epsilon^{abc} W^{\nu,c}) W^{b}_{\nu\mu} = - g_2 \sum_\psi
\bar\psi \tilde\tau^a \gamma_\mu \psi - i g_2 H^\dagger
\stackrel{\leftrightarrow}{D_\mu^a} H, 
\end{split}
\end{equation}
and
\begin{equation}
\begin{split}
D^\nu B_{\nu\mu} \equiv \partial^\nu B_{\nu\mu} = g_1 \sum_\psi
\frac{Y}{2} \bar\psi \gamma_\mu \psi + i \frac{g_1}{2} H^\dagger
\stackrel{\leftrightarrow}{D}_\mu H \,,
\end{split}
\end{equation}
up to gauge-fixing and ghost terms (see Ref.~\cite{Simma:1993ky} for a
more detailed discussion of the e.o.m. in effective theories.). The
sum is over all active fermion fields. 

The following operators vanish via the e.o.m. of the gauge fields;
they contribute to the same amplitudes as the physical four-fermion
operators. Therefore, the mixing of physical operators into the
e.o.m.-vanishing operators (computed from penguin diagrams) affects
the anomalous dimensions of four-fermion operators. There are four
operators involving DM currents,
\begin{align}
P_{1}^{(6)} &= \frac{1}{g_2}(\bar\chi\gamma_\mu
\tilde\tau^a\chi)D_\nu W^{a,\nu\mu} + \sum_i \big(Q_{1,i}^{(6)} +
Q_{9,i}^{(6)}\big) + Q_{15}^{(6)} + D_3^{(6)} \,,
\\ 
\begin{split}
P_{2}^{(6)} &= \frac{1}{g_1}(\bar\chi\gamma_\mu \chi)D_\nu B^{\nu\mu}\\
& \quad - \sum_i \Big( \frac{1}{6}
Q_{2,i}^{(6)} + \frac{2}{3} Q_{3,i}^{(6)} - \frac{1}{3} Q_{4,i}^{(6)} - \frac{1}{2}
Q_{10,i}^{(6)} - Q_{11,i}^{(6)} \Big) - \frac{1}{2} Q_{16}^{(6)} -
\frac{Y_\chi}{2} D_1^{(6)} \,,
\end{split}
\\ 
P_{3}^{(6)} &= \frac{1}{g_2}(\bar\chi\gamma_\mu \gamma_5 \tilde\tau^a\chi)D_\nu
W^{a,\nu\mu} + \sum_i \big( Q_{5,i}^{(6)} + Q_{12,i}^{(6)} \big) +
Q_{17}^{(6)} + D_4^{(6)} \,,
\\ 
\begin{split}
P_{4}^{(6)} &= \frac{1}{g_1}(\bar\chi\gamma_\mu \gamma_5 \chi)D_\nu
B^{\nu\mu} \\ & \quad - \sum_i \Big( \frac{1}{6} Q_{6,i}^{(6)} + \frac{2}{3} Q_{7,i}^{(6)} -
\frac{1}{3} Q_{8,i}^{(6)} - \frac{1}{2} Q_{13,i}^{(6)} -
Q_{14,i}^{(6)} \Big) - \frac{1}{2} Q_{18}^{(6)} - \frac{Y_\chi}{2} D_2^{(6)} \,,
\end{split}
\end{align}
four operators involving quark currents,
\begin{align}
P_{5,i}^{(6)} &= \frac{1}{g_2} (\bar Q_L^i \gamma^\mu \tilde \tau^a Q_L^i)
D_\nu W^{a,\nu\mu} + Q_{1,i}^{(6)} + \sum_j \big( S_{1,ij}^{(6)} +
S_{11,ij}^{(6)}\big) + S_{18,i}^{(6)} \,,
\\
\begin{split}
P_{6,i}^{(6)} &= \frac{1}{g_1} (\bar Q_L^i \gamma^\mu Q_L^i) D_\nu
B^{\nu\mu} - \frac{Y_\chi}{2} Q_{2,i}^{(6)} \\ & \quad - \sum_j \Big( \frac{1}{6} S_{2,ij}^{(6)} + \frac{2}{3} S_{3,ij}^{(6)} - \frac{1}{3} S_{4,ij}^{(6)} - \frac{1}{2} S_{12,ij}^{(6)} - S_{13,ij}^{(6)} \Big) - \frac{1}{2} S_{19,i}^{(6)}\,,
\end{split}
\\ 
\begin{split}
P_{7,i}^{(6)} &= \frac{1}{g_1} (\bar u_R^i \gamma^\mu u_R^i) D_\nu
B^{\nu\mu} - \frac{Y_\chi}{2} Q_{3,i}^{(6)} \\ & \quad - \sum_j \Big( \frac{1}{6} S_{3,ji}^{(6)} + \frac{2}{3} S_{5,ij}^{(6)} - \frac{1}{3} S_{6,ij}^{(6)} - \frac{1}{2} S_{14,ij}^{(6)} - S_{16,ij}^{(6)} \Big) - \frac{1}{2} S_{20,i}^{(6)} \,,
\end{split}
\\ 
\begin{split}
P_{8,i}^{(6)} &= \frac{1}{g_1} (\bar d_R^i \gamma^\mu d_R^i) D_\nu
B^{\nu\mu} - \frac{Y_\chi}{2} Q_{4,i}^{(6)} \\ & \quad - \sum_j \Big( \frac{1}{6} S_{4,ji}^{(6)} + \frac{2}{3} S_{6,ji}^{(6)} - \frac{1}{3} S_{7,ij}^{(6)} - \frac{1}{2} S_{15,ij}^{(6)} - S_{17,ij}^{(6)} \Big) - \frac{1}{2} S_{21,i}^{(6)} \, ,
\end{split}
\end{align}
two operators  involving Higgs currents,
\begin{align}
P_{9}^{(6)} &= \frac{1}{g_2} [H^\dagger
  i\negmedspace\stackrel{\leftrightarrow}{D}\negthickspace{}_\mu^a \, H]D_\nu W^{a,\nu\mu} + Q_{15}^{(6)} + \sum_i \big( S_{18,i}^{(6)} + S_{22,i}^{(6)} \big) + \frac{1}{4} S_{25}^{(6)} \,,
  \\ 
  \begin{split}
P_{10}^{(6)} &= \frac{1}{g_1} (H^\dagger 
i\negmedspace\stackrel{\leftrightarrow}{D}_\mu \negmedspace H)D_\nu B^{\nu\mu} - \frac{Y_\chi}{2} Q_{16}^{(6)} \\ & \quad - \sum_i \Big( \frac{1}{6} S_{19,i}^{(6)} + \frac{2}{3} S_{20,i}^{(6)} - \frac{1}{3} S_{21,i}^{(6)} - \frac{1}{2} S_{23,i}^{(6)} - S_{24,i}^{(6)} \Big) - \frac{1}{2} S_{25}^{(6)}\,,
\end{split}
\end{align}
and three operators involving lepton currents,
\begin{align}
P_{11,i}^{(6)} &= \frac{1}{g_2} (\bar L_L^i \gamma^\mu \tilde \tau^a L_L^i) D_\nu W^{a,\nu\mu} + Q_{9,i}^{(6)} + \sum_j \Big( \frac{1}{4} S_{8,ij}^{(6)} + S_{11,ji}^{(6)} \Big) + S_{22,i}^{(6)} \,,
\\
\begin{split}
P_{12,i}^{(6)} &= \frac{1}{g_1} (\bar L_L^i \gamma^\mu L_L^i) D_\nu B^{\nu\mu} - \frac{Y_\chi}{2} Q_{10,i}^{(6)} \\ & \quad - \sum_j \Big(
  \frac{1}{6} S_{12,ji}^{(6)} + \frac{2}{3} S_{14,ji}^{(6)} - \frac{1}{3} S_{15,ji}^{(6)} - \frac{1}{2} S_{8,ij}^{(6)} - S_{9,ij}^{(6)}
\Big) - \frac{1}{2} S_{23,i}^{(6)} \,,
\end{split}
\\ 
\begin{split}
P_{13,i}^{(6)} &= \frac{1}{g_1} (\bar \ell_R^i \gamma^\mu \ell_R^i) D_\nu B^{\nu\mu} - \frac{Y_\chi}{2} Q_{11,i}^{(6)} \\ & \quad - \sum_j \Big(
  \frac{1}{6} S_{13,ji}^{(6)} + \frac{2}{3} S_{16,ji}^{(6)} - \frac{1}{3} S_{17,ji}^{(6)} - \frac{1}{2} S_{9,ji}^{(6)} - S_{10,ij}^{(6)}
\Big) - \frac{1}{2} S_{24,i}^{(6)} \,.
\end{split}
\end{align}

Several additional operators, vanishing due to the e.o.m. for the DM
fields, are needed to project all one-loop Greens functions with
insertions of the operators in Eqs.~\eqref{Q12},~\eqref{Q56}
and~Eqs.~\eqref{eq:dim6:Q15Q17},~\eqref{eq:dim6:Q16Q18}, respectively:
two dimension-five operators,
\begin{align}
P_{1}^{(5)} &= \bar\chi \slashed{D} \slashed{D} \chi\,,
&P_{2}^{(5)} &= \bar\chi \slashed{D} \slashed{D} i \gamma_5 \chi \,, \label{N12}
\end{align}
and eight dimension-six operators,
\begin{align}
P_{14}^{(6)} &= (\bar\chi \tilde\tau^a i\slashed{D} \chi)(H^\dagger \tau^a H)\,, 
& P_{15}^{(6)} &= (\bar\chi i\lDslashed{}^\dagger \tilde\tau^a \chi)(H^\dagger \tau^a H)\,,
\label{eq:dim6:P14P15}\\ 
P_{16}^{(6)} &= (\bar\chi i\slashed{D} \chi)(H^\dagger H)\,, 
& P_{17}^{(6)} &= (\bar\chi i\lDslashed{}^\dagger \chi)(H^\dagger H)\,,
\label{eq:dim6:P16P17}\\ 
P_{18}^{(6)} &= (\bar\chi \tilde\tau^a i\slashed{D} \gamma_5 \chi)(H^\dagger \tau^a H)\,, 
& P_{19}^{(6)} &= (\bar\chi i\lDslashed{}^\dagger \gamma_5 \tilde\tau^a \chi)(H^\dagger \tau^a H)\,,
\label{eq:dim6:P18P19}
\\ 
P_{20}^{(6)} &= (\bar\chi i\slashed{D} \gamma_5 \chi)(H^\dagger H)\,, 
& P_{21}^{(6)} &= (\bar\chi i\lDslashed{}^\dagger \gamma_5 \chi)(H^\dagger H)\,.
\label{eq:dim6:P20P21}
\end{align}